\documentclass[journal,english]{IEEEtran}

\usepackage[latin9]{inputenc}
\usepackage{verbatim}
\usepackage{amsthm}
\usepackage{amsmath}
\usepackage{amssymb}
\usepackage{esint,constants}
\usepackage{graphicx}
\usepackage{subfig}
\usepackage{float}
\makeatletter
\def\blfootnote{\xdef\@thefnmark{}\@footnotetext}
\makeatother

\usepackage{hyperref}	
\hypersetup{
   colorlinks=true,
   citecolor=black,%
   filecolor=black,%
   linkcolor=black,%
   urlcolor=blue
}

\makeatletter
\theoremstyle{plain}
\newtheorem{thm}{\protect\theoremname}
  \theoremstyle{plain}
  \newtheorem{lem}[thm]{\protect\lemmaname}
 \theoremstyle{definition}
  \newtheorem{rem}{Remark}
 \theoremstyle{plain}
  \newtheorem{prop}{Proposition}
 \theoremstyle{plain}
  \newtheorem{ex}{Example}

\makeatother

\usepackage{babel}
  \providecommand{\lemmaname}{Lemma}
\providecommand{\theoremname}{Theorem}
\providecommand{\theoremname}{Remark}

\newcommand{\wt} [1]{\widetilde{#1}}
\def \SU {\operatorname{U}}
\def \SV {\operatorname{V}}
\def \SUT {{\widetilde{\operatorname{U}}}}
\def \SVT {\widetilde{\operatorname{V}}}
\def \ST {\operatorname{T}}

\def \l {\left}
\def \r {\right}
\def \real {\mathbb{R}}
\def \R {\mathcal{R}} 
\def \wt {\widetilde}

\newcommand{\edit}[1]{\textcolor{black}{#1}}

\newconstantfamily{factors}{
symbol=\alpha,
}

\begin{document}

\title{Weighted Matrix Completion and Recovery\\ with Prior Subspace Information}
\author{Armin Eftekhari, Dehui Yang, and Michael B.\ Wakin}
\maketitle

\begin{abstract}
\edit{An incoherent low-rank matrix can be efficiently reconstructed after observing a few of its entries at random, and then solving a convex program that minimizes the \emph{nuclear norm}.} In many applications, in addition to these  \edit{entries}, potentially valuable prior knowledge about the column and row spaces of  the matrix is also available to the practitioner. In this paper, we incorporate this prior knowledge in matrix completion---by minimizing a weighted nuclear norm---and precisely quantify any improvements. In particular, we find \edit{in theory} that reliable prior knowledge reduces the sample complexity of matrix completion by a logarithmic factor, \edit{and the observed improvement in numerical simulations is considerably more magnified.}  We also present similar results for the closely related problem of matrix recovery from generic linear measurements.
\end{abstract}

\section{Introduction}
\label{sec:intro}

\edit{{Matrix completion}\blfootnote{{AE is with the Alan Turing Institute in London. DY and MBW are with the \edit{Department of Electrical Engineering at the Colorado School of Mines.} (Email: aeftekhari@turing.ac.edu, dyang@mines.edu, mwakin@mines.edu.) AE is supported by the Alan Turing Institute under the EPSRC grant EP/N510129/1 and also by the Turing Seed Funding grant SF019. DY and MBW are partially supported by  EPSRC grant EP/N510129/1, NSF grant CCF--1409258, and NSF CAREER grant CCF--1149225.}} is commonly defined as the problem of recovering a  low-rank matrix $M\in\mathbb{R}^{n \times n}$ from a fraction of its entries, observed on an often random index set \cite{candes2009exact,recht2011simpler}.\footnote{\edit{For simplicity, we focus on square matrices in this paper. All results can be extended with minor modifications to general rectangular matrices. Also, in an attempt to make the introduction as accessible as possible,  technical details are kept to a minimum in this section, with more details and rigorous statements of our results deferred to Sections \ref{sec:problem statement} and~\ref{sec:results}.}}  More concretely, let $r$ denote the rank of $M$. Also let $M=U_r \Sigma_r V_r^*$ be a (thin) singular value decomposition (SVD) of $M$, where $U_r,V_r\in\mathbb{R}^{n\times r}$ have orthonormal columns and the diagonal matrix $\Sigma_r\in\mathbb{R}^{r\times r}$ contains the singular values of $M$.}

In a typical low-rank matrix completion problem, each entry of $M$ is observed with a probability of $p\in(0,1]$ so that $pn^2$ entries of $M$ are revealed in expectation. Let the matrix $Y=\mathcal{R}_p(M)\in\mathbb{R}^{n\times n}$ contain the observed entries of $M$ with zeros everywhere else.  Here, $\mathcal{R}_p(\cdot)$ represents the measurement process. In fact, with overwhelming probability, $M$ can be successfully reconstructed  from the measurements $Y$ by solving the convex program
\begin{equation}
\begin{cases}
\min_X \left\Vert X \right\Vert _{*},\\
\mbox{subject to }
\mathcal{R}_p(X)=Y,
\end{cases}
\label{eq:nuc norm exact}
\end{equation}
provided that\footnote{Throughout, we often use $\lesssim$ to simplify the presentation by suppressing universal constants. }
 \begin{equation}
1 \ge p \gtrsim   \frac{\eta(M) r\log^2 n}{n} . \label{eq:uniform intro}
 \end{equation}
Above, the nuclear norm $\|X\|_*$ returns the sum of singular values of a matrix $X$. In addition,  the \emph{coherence} $\eta(M)$  measures how ``spiky''  $M$ is, as precisely defined later in Section \ref{sec:problem statement}. One can therefore expect to successfully recover $M$ from  $O(\eta(M)\cdot  rn\log^{2}n) $
uniform samples \edit{\cite{candes2010matrix,chen2013completing,chen2015incoherence}.}

\subsection{Incorporating Prior Knowledge}
\label{sec:inc prior knowledge intro}
Let $\SU_r=\mbox{span}(U_r)$ and $\SV_r=\mbox{span}(V_r)$ be the column and row spaces of $M=U_r\Sigma_r V_r^*$.\footnote{We  always reserve upright letters for subspaces.}
Suppose that we have been presented with some prior knowledge about $M$ in the form of estimates for the subspaces $\SU_r$ and $\SV_r$. More specifically, let the $r$-dimensional subspaces $\wt{\SU}_r$ and $\wt{\SV}_r$ be the initial estimates of the column and row spaces of $M$, respectively, made available to us.

\edit{As an example in the context of \emph{collaborative filtering}, $\wt{\SU}_r$ might represent the similarities among users in the ``Netflix challenge''. To be specific,} the rows and columns of the popular Netflix matrix correspond to the Netflix subscribers and available movies, respectively. The Netflix matrix is sparsely populated with the ratings assigned by its
users and the challenge is then to complete the Netflix matrix given only the ratings available, namely  given only a small fraction of its entries.
A more realistic setup for the Netflix challenge should perhaps incorporate the changes in the preferences of Netflix users over time which  are for example significantly altered by child-rearing, \edit{see \cite{koren2010collaborative} for such temporal dynamics.} Here, $\wt{\SU}_r$ might incorporate prior information about the users \edit{and for instance might be obtained by taking an SVD of the current estimate of the database matrix $M$, with the anticipation that these features and thus $M$ itself might change over time.}
Similar problems arise in tracking changes in videos or updating the Laplacian of a
graph with time-variant connectivity. \edit{See   \cite{rao2015collaborative,xu2016dynamic} for more examples.}

As another example, in  \emph{exploration seismology}, large and often incomplete matrices are acquired
and processed in order to determine the subsurface structure of an area. Each \edit{target} matrix is comprised of responses from many sources at a certain frequency recorded at many receivers, where some recordings are missing. In this context, information from adjacent frequency bands  might help enhance \edit{completion of the target matrix}.  \edit{More specifically,} one might set
$\wt{\SU}_r$  to be the column space of the estimated response matrix from an adjacent frequency band \edit{and perform the reconstruction band-by-band in a ``dynamic'' fashion similar to the collaborative filtering setup above, see \cite[Sec.~8]{aravkin2014fast} for full details. The resulting algorithm would iteratively update the estimate of $t$th slice of the data tensor based on the estimate from, say, the $t-1$th slice, and cycle through the tensor multiple times if needed.} \edit{These ideas might also be generalized to the problem of \emph{subspace tracking} from incomplete data, in which we are interested in recovering a subspace from a sequence of generic vectors  in that subspace, observed partially. This subspace tracking problem is also closely related to \emph{streaming principal component analysis} \cite{eftekhari2016snipe,mitliagkas2013memory}.}  The related literature is more carefully studied  later in Section \ref{sec:related work}.

Motivated by such scenarios, it is perhaps natural to ask:
\begin{itemize}
\item \textbf{Question:} {How should we incorporate  in matrix completion any prior knowledge about column and row spaces?}
\end{itemize}
 We approach this question with the aid of a weighted nuclear norm as follows. Let $P_{\wt{\SU}_r},P_{\wt{\SU}_r^\perp}\in\mathbb{R}^{n\times n}$  be the orthogonal projections onto the subspace $\wt{\SU}_r$ and its orthogonal complement $\wt{\SU}_r^\perp$, respectively. For some weight $w \in(0,1]$, define
\begin{equation}
Q_{\wt{\SU}_r,w} := w \cdot P_{\wt{\SU}_r} + P_{\wt{\SU}_r^\perp} \in \mathbb{R}^{n\times n}.
\end{equation}
Likewise, define $\edit{Q_{\wt{\SV}_r,w}}\in\mathbb{R}^{n\times n}$ and let us modify Program \eqref{eq:nuc norm exact} to read
\begin{equation}
\begin{cases}
\min_X \left\Vert \edit{Q_{\wt{\SU}_r,w} \cdot X \cdot   Q_{\wt{\SV}_r,w}} \right\Vert _{*},\\
\mbox{subject to }
\mathcal{R}_p(X)=Y.
\end{cases}
\label{eq:weighted nuc norm exact}
\end{equation}
In a sense, the weight $w$ reflects our uncertainty in the prior knowledge. The smaller $w$, the more confident we are that $\SU_r\approx \wt{\SU}_r$ and $\SV_r\approx \wt{\SV}_r$ and in turn the more penalty Program  \eqref{eq:weighted nuc norm exact} places on feasible matrices with column or row spaces orthogonal to $\wt{\SU}_r$ or $\wt{\SV}_r$, respectively. In contrast, when our prior information is not reliable, we might set $w=1$, in which case \edit{$Q_{\wt{\SU}_r,w}=Q_{\wt{\SV}_r,w}=I_n$} and Program \eqref{eq:weighted nuc norm exact} reduces to standard matrix completion, namely Program \eqref{eq:nuc norm exact}, thereby completely  ignoring any prior knowledge about the problem.

\begin{figure*}[t]
\begin{center}
\subfloat[\label{fig:mcmrTests06resultsAll}]{
\includegraphics[width=.33\textwidth]{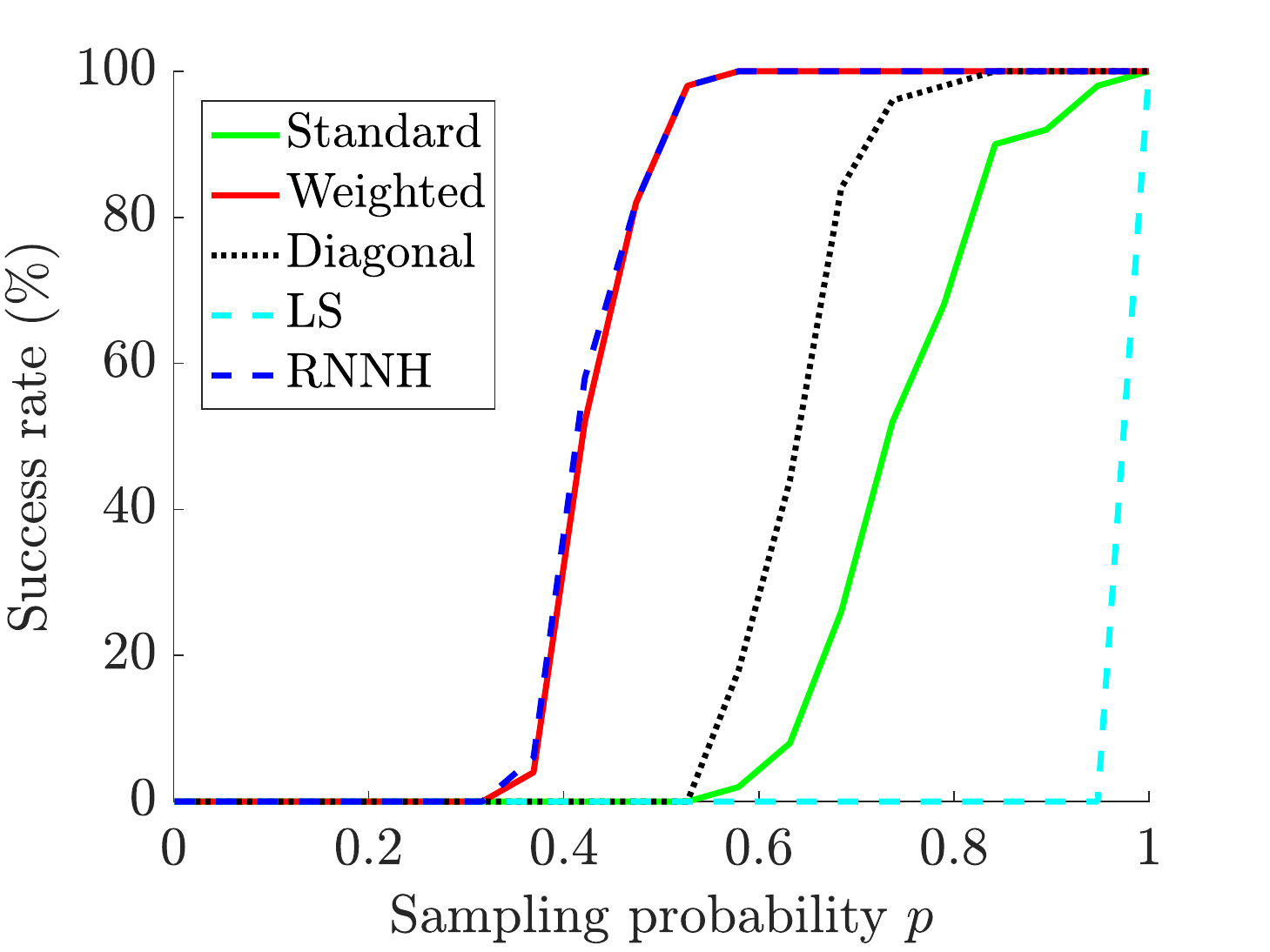}
}
\subfloat[\label{fig:mcmrTests06resultsWeighted}]{
\includegraphics[width=.33\textwidth]{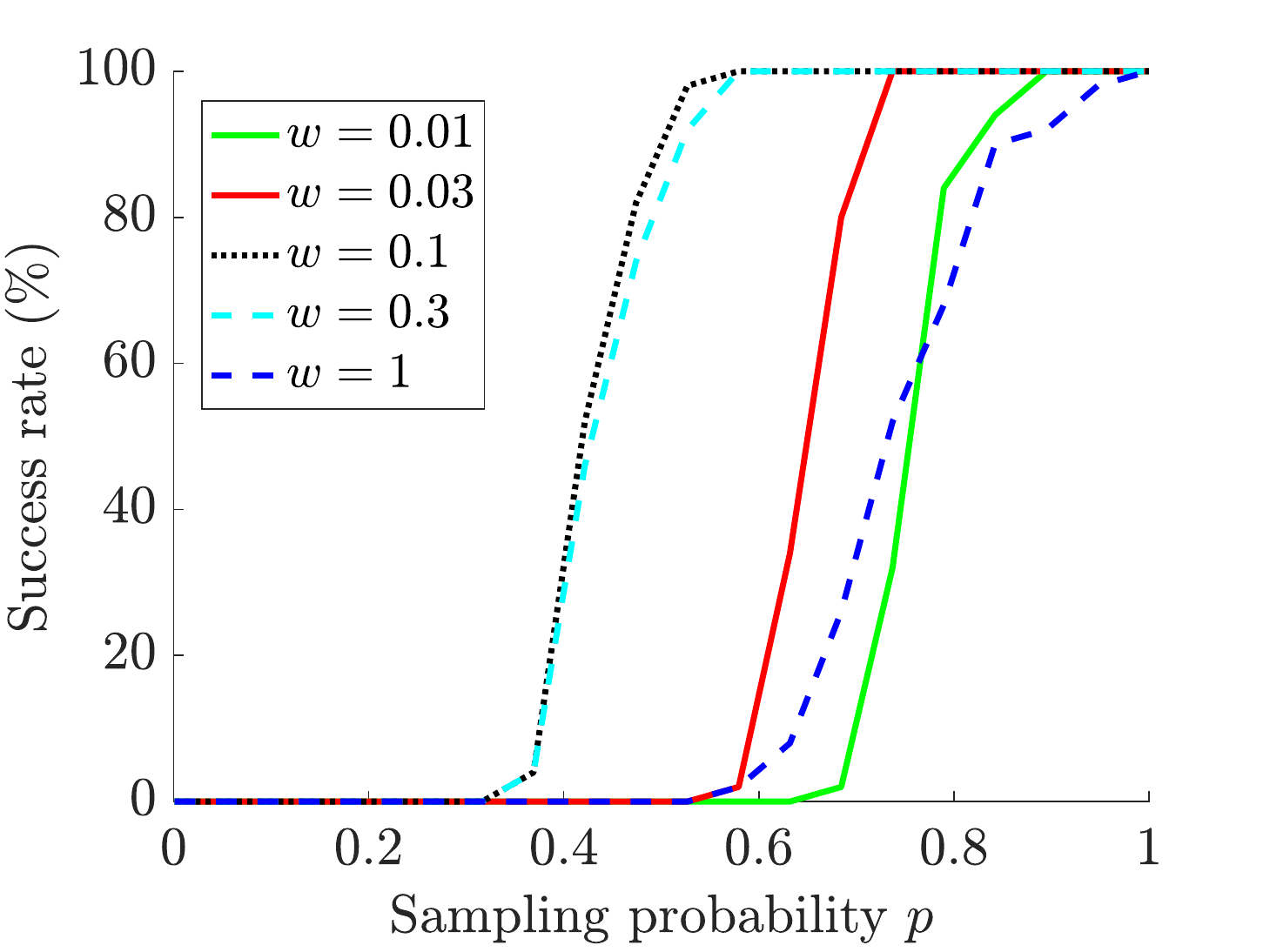}
}
\subfloat[\label{fig:mcmrTests06resultsRNNH}]{
\includegraphics[width=.33\textwidth]{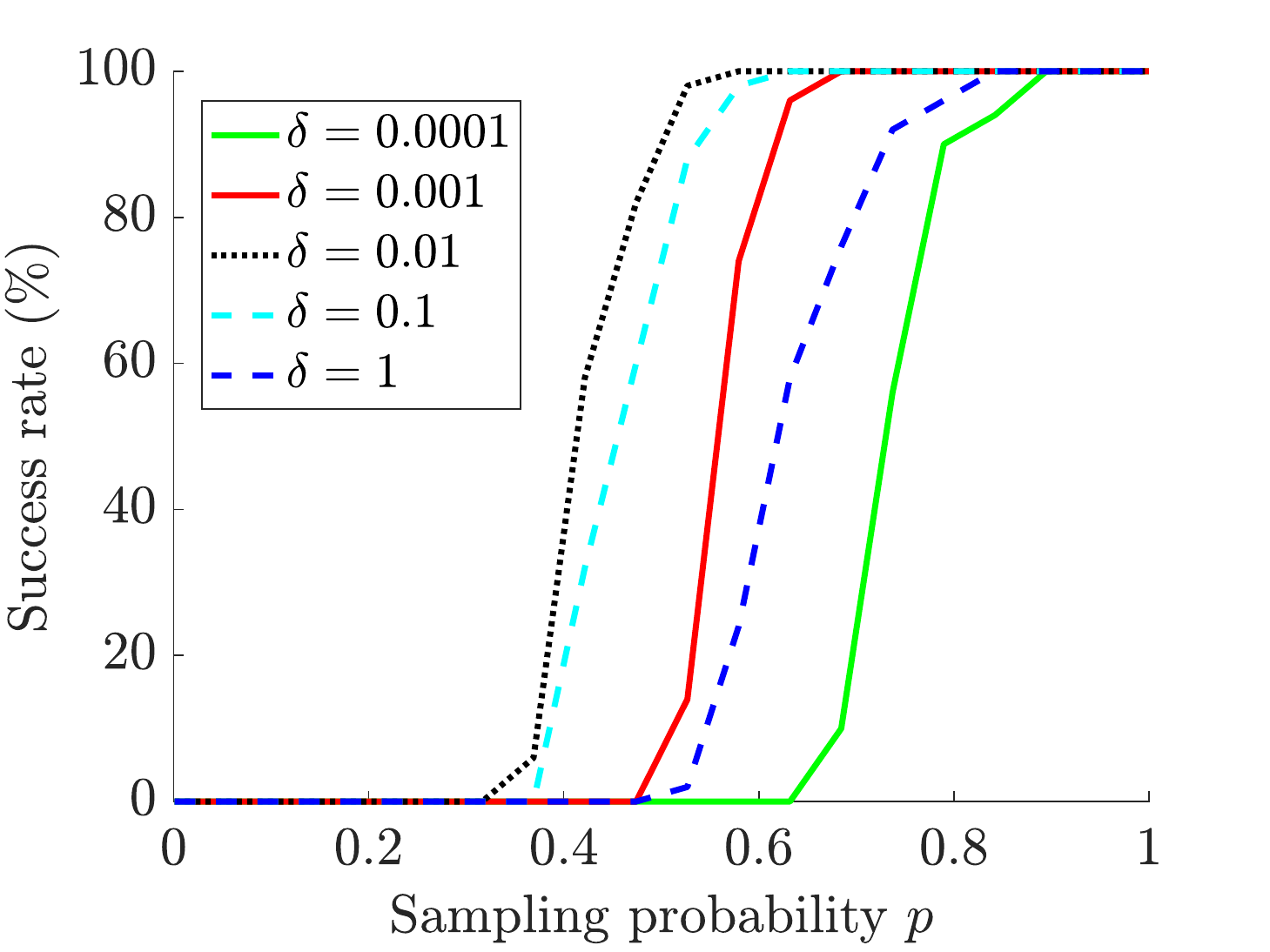}
}
\caption{\edit{Matrix completion with prior information, where the principal angles between the ground truth and prior knowledge subspaces are $u = \angle [\SU_r,\SUT_r ] = 0.099$ and $v = \angle [\SV_r,\SVT_r ] = 0.103$. (a) compares standard  matrix completion via Program \eqref{eq:nuc norm exact},  weighted matrix completion proposed in Program \eqref{eq:weighted nuc norm exact} with  $w = 0.1$, diagonal weighted matrix completion, weighted least-squares, and a weighted nuclear norm heuristic (RNNH) algorithm with the tuning parameter $\delta = 0.01$. (b) shows the performance of weighted matrix completion with various weights $w$. (c) shows RNNH with various parameters $\delta$. See Section \ref{sec:simulations} for more details. }}
\end{center}
\end{figure*}

A more general form of Program \eqref{eq:weighted nuc norm exact} is discussed in Section \ref{sec:problem statement} in greater detail. For now, let us briefly compare the two Programs \eqref{eq:nuc norm exact} and \eqref{eq:weighted nuc norm exact} in practice. \edit{We take $M$ be a square matrix of sidelength $n=20$, rank $r=4$, and norm $\|M\|_F = 1$. As prior knowledge to be used in completing $M$, we construct a perturbed version $M'$ of $M$, where $M' = M + N$ and entries of  $N\in\mathbb{R}^{n\times n}$ are independent Gaussian random variables with mean zero and variance $ 10^{-4}$. We then compute the SVD of $M'$ and let $(\wt{\SU}_r,\wt{\SV}_r)$ contain the leading $r$ left- and right-singular vectors of $M'$. Lastly we set the weight $w=0.1$. As the probability $p$ of observing each entry of $M$ varies in $(0,1]$, we solve both Programs (\ref{eq:nuc norm exact},\ref{eq:weighted nuc norm exact}) and record the results. The success rates for both programs, averaged over $50$ trials, is shown in Figure \ref{fig:mcmrTests06resultsAll}. (Program~\eqref{eq:nuc norm exact} corresponds to the solid green line, Program~\eqref{eq:weighted nuc norm exact} corresponds to the solid red line, and other lines are explained in Section~\ref{sec:simulations}.) A trial is considered successful if it recovers $M$ up to a relative error of $10^{-3}$. Observe how reliable prior knowledge, when used properly, allows for successful matrix completion from substantially fewer measurements.}

\subsection{Simplified Main Result}

\edit{One of our main results in this paper  concerns the performance of a more general form of Program \eqref{eq:weighted nuc norm exact}, which we next outline, deferring a more rigorous statement of the result to  Theorem \ref{thm:main result} in Section \ref{sec:results}.}

Consider a rank-$r$ matrix $M\in\mathbb{R}^{n\times n}$, and let  $\SU_r=\mbox{span}(M)$ and $\SV_r=\mbox{span}(M^*)$ be the column and row spaces of $M$. Let also $\eta(M)$ be the coherence of $M$,  which we briefly introduced earlier.

Suppose also that the $r$-dimensional subspaces $\SUT_r$ and $\SVT_r$, with orthonormal bases $\wt{U}_r,\wt{V}_r\in\mathbb{R}^{n\times r}$, represent our prior knowledge about the column and row spaces of $M$, respectively.
In particular, let
$u = \angle [\SU_r,\SUT_r ]$ and
$v = \angle [\SV_r,\SVT_r ]$
denote the largest principal angles between each pair of subspaces.\footnote{Principal angle between subspaces generalizes the notion of angle between lines and, for the sake of completeness, is defined later in Section \ref{sec:commons}.} We set $\theta=\max[u,v]$ for short.
Moreover, consider the sampling probability $p\in(0,1]$ and  let $Y=\mathcal{R}_p(M)$ be the matrix of measurements, defined earlier. Lastly, for weight $w\in(0,1]$, let $\widehat{M}$ be a solution of Program (\ref{eq:weighted nuc norm exact}). Then, $\widehat{M}=M$,
except with a probability of $o(n^{-19})$  and provided that\footnote{Throughout, we will occasionally use the standard ``little-$o$'' and ``big-$O$'' notation.}
\begin{align*}
& 1 \ge p \gtrsim
 \max\left[\log\left(\Cl[factors]{4p}\cdot n\right),1\right]\cdot
\frac{\eta(M) r\log n}{n}
\\
& \qquad \qquad \qquad \qquad \,\cdot \mbox{inter-coherence factor},
\end{align*}
\begin{align}
\Cl[factors]{3p}\le  \frac{1}{8},
\label{eq:p req thm simplified}
\end{align}
where the ``inter-coherence factor'' above, to be made precise later, is often not large and reflects the interaction between the coherence of prior subspaces and the true ones. Additionally,
\begin{equation*}
\Cr{4p} := {\frac{w^{4}\cos^{2}\theta+\sin^{2}\theta}{w^{2}\cos^{2}\theta+\sin^{2}\theta}},
\quad
\Cr{3p}  :=
\frac{3\sqrt{1-w^{2}}\sin \theta}{\sqrt{w^{2}\cos^{2}\theta+\sin^{2}\theta}}
.
\end{equation*}
In particular, when $w=1$, then  the prior information $(\wt{\SU}_r,\wt{\SV}_r)$ is ignored and Program \eqref{eq:weighted nuc norm exact} reduces to Program \eqref{eq:nuc norm exact}.  In this case, $\Cr{4p}=1,\Cr{3p}=0$ and \eqref{eq:p req thm simplified} reduces to
\eqref{eq:uniform intro}, save for the typically small coherence factor.\footnote{In fact, this could be easily sharpened by slightly modifying the proof to yield $p\gtrsim \eta(M) r\log^2n / n$, which precisely matches \eqref{eq:uniform intro}.
 We however opted for the more elegant but slightly loose bound presented in (\ref{eq:p req thm simplified}).}

On the other hand, when our prior knowledge is reliable, namely when $\theta$ is small, the  proper choice of $w$ in Program \eqref{eq:weighted nuc norm exact} leads to  substantial improvement over Program \eqref{eq:nuc norm exact}. \edit{For example, with $\sin \theta \lesssim 1/n$ and $w=\sqrt{\tan\theta}$, observe that $\Cr{4p}=O(\sin\theta)$, $\max\left[\log\left(\Cr{4p}\cdot n\right),1\right] = 1$, $\Cr{3p} = O(\sqrt{\sin\theta})$}, and  \eqref{eq:p req thm simplified} now reads as
\begin{equation}
 1 \ge p \gtrsim \frac{\eta(M) r\log n}{n},
\label{eq:improved result intro}
\end{equation}
save for the often small inter-coherence factor. The lower bound in \eqref{eq:improved result intro} is better than \eqref{eq:uniform intro} by a logarithmic factor \edit{and is likely near optimal, as discussed in Section \ref{sec:results}.}

\subsection{Matrix Recovery with Prior Knowledge}

Matrix completion, discussed above, is a special case of the more general \emph{matrix recovery} problem, in which the objective is to recover a matrix from  generic and often random linear measurements. This problem is reviewed in Section \ref{sec:problem statement} and our other main result, Theorem \ref{thm:main result MR} in Section \ref{sec:results}, concerns leveraging prior information in this context.

\subsection{Organization}
The rest of this paper is organized as follows. In Section \ref{sec:problem statement}, we briefly review standard low-rank matrix recovery and completion, and further motivate the use of prior knowledge in these contexts. Our main results quantify the use of prior knowledge in matrix recovery and completion, and are summarized in Section \ref{sec:results}. Section \ref{sec:simulations} offers some numerical evidence to support the theory and the related literature is highlighted in Section \ref{sec:related work}.  Technical details are postponed to Sections \ref{sec:commons}-\ref{sec:analysis of mc w pi} and appendices. In particular, Section \ref{sec:commons} collects the technical tools common to the analysis of both matrix recovery and completion. Sections \ref{sec:analysis of mr w pi} and \ref{sec:analysis of mc w pi} then contain the arguments specialized to matrix recovery and completion, respectively.

\section{Problem Statement}
\label{sec:problem statement}

Consider a matrix $M\in\mathbb{R}^{n\times n}$ and its SVD  $M=U\Sigma V^{*}$. Here,
$U,V\in\mathbb{R}^{n\times n}$ are orthonormal bases and the diagonal
matrix $\Sigma\in\mathbb{R}^{n\times n}$ collects the singular values
of $M$ in a non-increasing order,  namely $\sigma_{1}(M)\ge\sigma_{2}(M)\ge\cdots\ge\sigma_{n}(M)$. For an integer $r\le n$, let $U_r,V_r\in\mathbb{R}^{n\times r}$ comprise of the first $r$ columns of $U,V$, respectively and let $\Sigma_r\in\mathbb{R}^{r\times r}$ contain  $r$ largest singular values of $M$. Ties are broken arbitrarily. Then, $M_{r}=U_{r}\Sigma_{r}V_{r}^{*}\in\mathbb{R}^{n\times n}$ is a rank-$r$ truncation of $M$ and  we also let $M_{r^{+}}=M-M_{r}$
 denote the residual.

Suppose that we can only access $M\in\real^{n\times n}$ through a linear  operator $\mathcal{R}_m(\cdot)$ that  collects $m$ measurements from $M$.  More specifically,  let
\begin{equation}
y = \R_m(M+E) \in \real^m,
\qquad \l\| \R_m(E)\r\|_2 \le e,
\end{equation}
be the  vector of $m$ (possibly noisy) measurements. Here, $E\in\real^{n\times n}$ and $e \ge 0$ represent the noise. \emph{Matrix recovery} is \edit{then} the problem of (approximately) reconstructing $M$ from the measurement vector $y$.

The case where the entries of $M$ are randomly observed is of particular importance in practice, where  we pragmatically assume that a measurement operator $\R_p(\cdot)$ observes each entry of $M$ with a probability of $p\in(0,1]$.\footnote{\edit{Alternatively, one may observe $m$ entries of $M$ at random, with or without replacement. However, one advantage of the measurement model in \eqref{eq:def of R(Z)} is that, by assigning different probabilities to each entry, the uniform sampling operator $\R_p(\cdot)$ may easily be upgraded  for \emph{leveraged sampling}, where each entry of $M$ is observed according to its leverage (importance).} }  We set $p=m/n^2$ so that  $\R_p(M)$ contains $m$ entries of $M$, in expectation.
To be more specific, $\mathcal{R}_{p}(\cdot)$ takes
$M\in\mathbb{R}^{n\times n}$ to $\mathcal{R}_p(M)\in\mathbb{R}^{n\times n}$
defined as
\begin{equation}
\mathcal{R}_{p}(M)= \sum_{i,j=1}^{n}\frac{\epsilon_{ij}}{p}\cdot M[i,j]\cdot C_{ij},\label{eq:def of R(Z)}
\end{equation}
where  $\{\epsilon_{ij}\}_{i,j}$ is a sequence of independent Bernoulli random variables taking
one with probability of $p$ and zero otherwise.
Throughout, $C_{ij}\in\mathbb{R}^{n\times n}$
is the $[i,j]$th canonical matrix, so that  $C_{ij}[i,j]=1$ is its only nonzero entry.
We also let
\begin{equation}
Y = \mathcal{R}_p(M+E)\in\mathbb{R}^{n\times n},
\qquad
\left\|\mathcal{R}_p(E) \right\|_F\le e,
\end{equation}
be the (possibly noisy) matrix of measurements. As before, $E$ and $e$ represent the noise. \emph{Matrix completion} is the problem of (approximately) reconstructing $M$ from $Y$.

\subsection{Standard Low-Rank Matrix Recovery and Completion}

In general, both matrix recovery and completion problems are ill-posed  when $m\le n^2$ and, to rectify this issue, it is common to impose that $M$ is (nearly) low-rank. Let us briefly review both \emph{low-rank}  matrix completion  and recovery   next.

In low-rank matrix recovery, the \emph{restricted isometry property} (RIP) plays a key role by ensuring that the measurement operator preserves the geometry of the set of low-rank matrices. More specifically, for $\delta_r\in[0,1)$, we say that $\R_m(\cdot)$ satisfies  the  $(r,\delta_r)$-RIP (or simply $\delta_r$-RIP when there is no ambiguity) if
\begin{align}
& \l(1-\delta_r\r) \|X\|_F \le \l\| \R_m(X)\r\|_2 \le \l(1+\delta_r\r)\|X\|_F, \nonumber\\
\label{eq:RIP def}
\end{align}
for every  $X\in\real^{n\times n}$  with  $\mbox{rank}( X) \le r$.
It is perhaps remarkable that a ``generic'' linear operator from $\real^{n\times n}$ to $\real^m$ satisfies the RIP when the number of measurements $m$ is sufficiently large. For example, suppose that $G\in\real^{n\times n}$ is populated with independent zero-mean Gaussian random variables with variance $1/m$. Then, $\langle G,X\rangle$ collects one linear measurement from $X$. The measurement operator formed from $m$ independent copies of $\langle G,X\rangle $ is known to satisfy $\delta_r$-RIP with high probability when $m\gtrsim   rn\log n/\delta_r^2$. When $M$ is nearly low-rank (in the sense that the residual $M_{r^+}=M-M_r$ is  small) and when $\R_m(\cdot)$ satisfies the RIP, we can in fact (approximately) recover $M$ by solving the following convex program:
\begin{equation}
\begin{cases}
\min_X \left\Vert X \right\Vert _{*},\\
\mbox{subject to }\left\Vert \mathcal{R}_{m}(X)-y\right\Vert _{2}\le e .
\label{eq:MR program}
\end{cases}
\end{equation}
\edit{Above, with $\{\sigma_i(X)\}_i$ standing for the singular values of the matrix $X$, $\|X\|_*=\sum_i \sigma_i(X)$ is the nuclear norm of $X$. We also use the Frobenius norm $\|X\|_F=(\sum_i \sigma_i^2(X))^{\frac{1}{2}}$ below.} The recovery error of Program \eqref{eq:MR program} is summarized next \cite{recht2010guaranteed,fazel2008compressed}.
\begin{prop}
\label{prop:MR standard}
\textbf{\emph{(Matrix recovery)}}
For an integer $r\le n$ and matrix $M\in\real^{n\times n}$, let $M_r\in\real^{n\times n}$ be a rank-$r$ truncation of $M$ and let $M_{r^+}=M-M_r$ be the residual. Suppose  that the linear measurement operator $\R_m:\real^{n\times n}\rightarrow\real^m$ satisfies $\delta_{5r}$-RIP with  $\delta_{5r} \le 0.1$. Let also $\widehat{M}\in\real^{n\times n}$ be a solution of Program (\ref{eq:MR program}). Then it holds that
\begin{equation}
\l\|\widehat{M}-M \r\|_F \lesssim  \frac{\l\|M_{r^+} \r\|_*}{\sqrt{r}}+ e.
\end{equation}
\end{prop}

In low-rank matrix completion, on the other hand, $\R_p(\cdot)$ does \emph{not} satisfy the RIP unless $p\approx 1$, in which case nearly every entry of $M$ is observed anyway. \edit{For example, with $C_{11}$ standing for the first canonical matrix in $\real^{n\times n}$, note that $\R_p(C_{11})=0$ with a probability of $1-p$ and so $p$ must be close to one to capture the energy of $C_{11}$.}  However, $\R_p(\cdot)$ does preserve the geometry of the set of low-rank and \emph{incoherent} matrices, provided that $p$ is sufficiently large. More specifically, let $M_r=U_r\Sigma_rV_r^*$ be \edit{an} SVD of $M_r$, a rank-$r$ truncation of $M$. Then the  \emph{coherence} of $M_r$, denoted by $\eta(M_r)$ throughout, is defined as
\begin{equation}
\eta\l( M_r\r) =\frac{n}{r} \max\l[ \l\| U_r \r\|^2_{2\rightarrow\infty}, \l\| V_r \r\|^2_{2\rightarrow\infty}\r],
\label{eq:coherence def}
\end{equation}
where $\|X\|_{2\rightarrow\infty}$ returns the largest $\ell_2$ norm of the rows of $X$. It is not difficult to verify that $\eta(M_r)\in[1,\frac{n}{r}]$, and that $\eta(M_r)$ depends only on the column and row spaces of $M_r$.
When $\eta(M_r)$ is small, entries of $M_r$ tend to be \edit{roughly equal in magnitude} and we say that $M_r$ is \emph{incoherent}.
At the other extreme, when $\eta(M_r)$ is large, $M_r$ is often ``spiky'' and we say that $M_r$ is \emph{coherent}.
When  $M$ is nearly low-rank  \emph{and} incoherent, we can (approximately) reconstruct $M$ by solving the convex program
\begin{equation}
\begin{cases}
\min_X \left\Vert X \right\Vert _{*},\\
\mbox{subject to }\left\Vert \mathcal{R}_{p}(X)-Y\right\Vert _{F}\le e,
\end{cases}
\label{eq:MC program}
\end{equation}
for which the recovery error is obtained by slightly modifying Theorem 7 in \cite{candes2010matrix} to fit our setup \cite[Proposition 2]{eftekhari2016mc}.
\begin{prop}
\label{prop:MC standard}
\textbf{\emph{(Matrix completion)}}
For an integer $r\le n$ and matrix $M\in\real^{n\times n}$, let $M_r\in\real^{n\times n}$ be a rank-$r$ truncation of $M$ and let $M_{r^+}=M-M_r$ be the residual. Let $\widehat{M}\in\real^{n\times n}$ be a solution of Program (\ref{eq:MC program}). Then, except with a probability of at most $o(n^{-19})$, it holds that
\begin{equation}
\l\|\widehat{M}-M \r\|_F \lesssim  \frac{\l\|M_{r^+} \r\|_*}{\sqrt{p}}+ e\sqrt{pn},
\label{eq:no of samples vanilla}
\end{equation}
provided that
\begin{equation}
1 \ge p \gtrsim \frac{\eta\l( M_r\r) r\log^2 n}{n}.
 \label{eq:bnd on p in standard matrix completion}
\end{equation}
\end{prop}
For instance, when $M_r$ is incoherent,  say $\eta_r(M_r)\approx 1$, solving Program \eqref{eq:MC program} approximately completes $M$ after observing only $O(rn\log^2n)$ of its samples, in expectation.

\subsection{Incorporating Prior Knowledge}
\label{sec:incorporating pi}
Ideally, if the column and row spaces of a rank-$r$ matrix $M_r$ were known \emph{a priori}, only $r^2$ linear measurements of $M_r$ would  suffice for exact recovery in the absence of noise. Indeed, if rank-$r$ matrices $A_r,B_r\in\mathbb{R}^{n\times r}$ span the column and row spaces of $M_r$, then $A_r^* M_r B_r\in\mathbb{R}^{r\times r}$ contains all the necessary information to reconstruct $M_r$.

More generally, consider $M\in\real^{n\times n}$  and let $M_r$ be a rank-$r$ truncation of $M$ as before.
 If available, suppose that the $r$-dimensional subspaces $\widetilde{\SU}_r$ and $\widetilde{\SV}_r$ represent our prior knowledge about the column and row spaces of $M_r=U_r\Sigma_r V_r^*$. In order to incorporate this prior knowledge into  matrix recovery and completion, we propose the following approach. Let 
 $$
 P_{\SUT_r}\in\mathbb{R}^{n\times n} \mbox{ and } P_{\SUT_r^\perp}\in\mathbb{R}^{n\times n}
 $$
be the orthogonal projections onto the subspace $\wt{\SU}_r$ and its complement, respectively.  Likewise, we define $n\times n$ projection matrices $P_{\wt{\SV}_r}$ and $P_{\wt{\SV}_r^\perp}$. For  (left and right) weights $\lambda,\rho\in[0,1]$, set
\begin{equation*}
Q_{\SUT_r,\lambda}:=\lambda \cdot P_{\SUT_r}+P_{\SUT_r^\perp}\in\mathbb{R}^{n\times n},
\end{equation*}
\begin{equation}
Q_{\SVT_r,\rho}:=\rho \cdot P_{\SVT_r}+P_{\SVT_r^\perp}\in\mathbb{R}^{n\times n}.
\label{eq:def of Qs}
\end{equation}
In order to leverage the prior information $(\wt{\SU}_r,\wt{\SV}_r)$ in low-rank matrix recovery, we modify  Program \eqref{eq:MR program} as follows:
\begin{equation}
\begin{cases}
\min_X \left\Vert Q_{\SUT_r,\lambda}\cdot X\cdot Q_{\SVT_r,\rho}\right\Vert _{*},\\
\mbox{subject to }\left\Vert \mathcal{R}_m(X)-y\right\Vert _{2}\le e.
\end{cases}
\label{eq:p1 MR}
\end{equation}
Note that the weights $\lambda,\rho\in [0,1]$ reflect our  uncertainty (or lack of confidence) in available prior knowledge, as the following examples might help clarify. Notice also that the above setup allows for weighting column and row spaces differently by choosing $\lambda\ne \rho$. 
\begin{ex}
Consider  the rank-$r$ matrix $M_r=U_r\Sigma_r V_r^*\in\real^{n\times n}$  and suppose that $\wt{\SU}_r=\SU_r=\mbox{span}(M_r)$ and $\wt{\SV}_r=\SV_r=\mbox{span}(M^*_r)$, namely our prior knowledge about $M$ is perfectly accurate. To represent the lack of uncertainty in this knowledge, we  set $\lambda=\rho = 0$ so that $Q_{\wt{\SU}_r}= P_{{\SU}_r^\perp}$ and
$Q_{\wt{\SV}_r}= P_{{\SV}_r^\perp}$, which in turn penalizes the component of solution  orthogonal to the column and row spaces of $M_r$ in Program (\ref{eq:p1 MR}).
\end{ex}
\begin{ex}
At the other extreme, suppose that $\wt{\SU}_r$ and $\wt{\SV}_r$ are poor estimates of the true column and row spaces of $M_r=U_r\Sigma_r V_r^*$. To represent our uncertainty about  the available prior information, we might set $\lambda=\rho=1$, in which case Program (\ref{eq:p1 MR}) \edit{ignores the prior information and} simplifies to standard matrix recovery, namely Program (\ref{eq:MR program}).
\end{ex}
Similarly, for low-rank matrix completion, given the prior information $(\wt{\SU}_r,\wt{\SV}_r)$, we consider the following modification of Program \eqref{eq:MC program}:
\begin{equation}
\begin{cases}
\min_X \left\Vert Q_{\SUT_r,\lambda}\cdot X\cdot Q_{\SVT_r,\rho}\right\Vert _{*},\\
\mbox{subject to }\left\Vert \mathcal{R}_p(X)-Y\right\Vert _{F}\le e.
\end{cases}
\label{eq:p1}
\end{equation}
To what extent, does prior knowledge help (or hurt)  matrix recovery and completion? \edit{We answer this question by quantifying the performance of Programs (\ref{eq:p1 MR}) and (\ref{eq:p1}) in the next section.}

\section{Main Results}
\label{sec:results}

In Section  \ref{sec:incorporating pi}, we proposed Programs  \eqref{eq:p1 MR} and \eqref{eq:p1} in order to leverage available prior knowledge in matrix recovery and completion, respectively. Our first main result, proved in Section \ref{sec:analysis of mr w pi}, is concerned with  the performance of Program \eqref{eq:p1 MR}.

\begin{thm}
\label{thm:main result MR}
\textbf{\emph{(Matrix recovery with prior knowledge)}}
For an integer $r$ and matrix $M\in\mathbb{R}^{n\times n}$, let $M_r\in\mathbb{R}^{n\times n}$ be a  rank-$r$ truncation of $M$ and let $M_{r^+}=M-M_r$ be the residual. Let also $\SU_r = \mbox{span}(M_r)$ and $\SV_r=\mbox{span}(M_r^*)$ denote the column and row spaces of $M_r$, respectively. Additionally, let $\eta(M_r)$ be the coherence of $M_r$, see (\ref{eq:coherence def}).
Suppose that the $r$-dimensional subspaces $\SUT_r$ and $\SVT_r$ represent the prior knowledge about $\SU_r$ and $\SV_r$,  respectively.
Let
\begin{equation*}
u = \angle \l[\SU_r,\SUT_r \r],
\qquad
v = \angle \l[\SV_r,\SVT_r \r],
\end{equation*}
denote the largest principal angles between each pair of subspaces.

For an integer $m$, suppose that the linear measurement operator $\mathcal{R}_m(\cdot)$ satisfies $\delta_{32r}$-RIP with \begin{equation}
\delta_{32r} \le
\frac{0.9-\max\l[\Cl[factors]{1},\Cl[factors]{2}\r]/\sqrt{30}}{0.9+\max\l[\Cr{1},\Cr{2}\r]/\sqrt{30}},
\label{eq:req on delta thm}
\end{equation}
and  acquire the (possibly noisy) measurement vector $y=\mathcal{R}_m(M+E)\in\mathbb{R}^m$, where $\|\mathcal{R}_m(E)\|_2\le e$. Lastly, for weights $\lambda,\rho\in(0,1]$,  let $\widehat{M}$ be a solution of Program (\ref{eq:p1 MR}). Then it holds that
\begin{equation}
\l\| \widehat{M} -M\r\|_{F} \lesssim  \frac{\left\Vert M_{r^+}\right\Vert _{*}}{\sqrt{{r}}}+ e.
\label{eq:nuclear bound weighted}
\end{equation}
Above, $\Cr{1}$ and $\Cr{2}$ are set to be
\begin{equation*}
\Cr{1} :=  \sqrt{\frac{\lambda^{4}\cos^{2}u+\sin^{2}u}{\lambda^{2}\cos^{2}u+\sin^{2}u}}
+
 \sqrt{\frac{\rho^{4}\cos^{2}v+\sin^{2}v}{\rho^{2}\cos^{2}v+\sin^{2}v}},
\end{equation*}
\begin{equation}
\Cr{2} :=
 \sqrt{\frac{2(1-\lambda^{2})\sin^{2}u}{\lambda^{2}\cos^{2}u+\sin^{2}u}}
 +\sqrt{\frac{2(1-\rho^{2})\sin^{2}v}{\rho^{2}\cos^{2}v+\sin^{2}v}}
.
\end{equation}
\end{thm}
A few remarks are in order to help clarify Theorem \ref{thm:main result MR}.
\begin{rem}
\label{rem:conn to standard matrix recovery}
\textbf{{(Connection to standard low-rank matrix recovery)}}
If we set $\lambda=\rho =1$, Program (\ref{eq:p1 MR}) reduces to Program (\ref{eq:MR program}) for  standard low-rank matrix recovery, which entirely ignores the prior knowledge $(\SUT_r,\SVT_r)$. In this case, $\Cr{1}=2,\Cr{2}=0$,
 and  (\ref{eq:req on delta thm}) reads $\delta_{32r}\le 0.42$.
It is known that $\delta_{t} \le \frac{t-1}{s-1}\cdot \delta_s$ for $t\ge s>1$ \cite[Exercise 6.10]{foucart2013mathematical}. Therefore, $\delta_{32r}\le 7.75\cdot  \delta_{5r}$, so that $\delta_{5r}\le 0.05$ implies $\delta_{32r}\le 0.42$.
This bound is slightly more conservative than $\delta_{5r}\le 0.1$ in Proposition \ref{prop:MR standard},  as we made no attempts to optimize the constants.

On the other hand, even with the conservative bounds in  Theorem \ref{thm:main result MR}, we observe that Program (\ref{eq:p1 MR}) indeed outperforms  Program (\ref{eq:MR program})  when the prior knowledge is reliable (namely,  the principal angles $u,v$ are small) and when the weights $\lambda,\rho$ are selected small to reflect our confidence about the available information.  For example, suppose that $u=v=\theta$ and $\lambda=\rho=\sqrt{\tan \theta}$ which gives $\Cr{1},\Cr{2}\le 2 \sqrt{2\sin \theta}$. Then, if we repeat the calculations at the end of Section \ref{sec:body MR} to find the tightest bound here, we find that matrix recovery is successful and  (\ref{eq:nuclear bound weighted}) holds if $\delta_{5r}\le 0.75$ and $\theta \le 0.0248$.
 This requirement is substantially better than $\delta_{5r}\le 0.1$ in Proposition \ref{prop:MR standard}.
That is, the bound for Program (\ref{eq:p1 MR}) considerably improves upon the bound for Program (\ref{eq:MR program}) and it does so by leveraging the available prior information.

\begin{rem}\textbf{{(Different weights for column and row spaces)}}
Note that the formulation in Program (\ref{eq:p1 MR})
 allows for assigning different weights to the column and row spaces by selecting $\lambda\ne \rho$. This enables the user to handle  scenarios when the  uncertainty about $\widetilde{\SU}_r$ and $\widetilde{\SV}_r$ are different. For example, if one expects $u=\angle [{\SU}_r,\widetilde{\SU}_r] \approx \frac{\pi}{2}$ and $v=\angle [{\SV}_r,\widetilde{\SV}_r] \approx 0$, one might naturally choose $\lambda \approx 1$  and $\rho \approx 0$ to  best handle this scenario.
\end{rem}

\begin{rem}
\textbf{{(On choosing the weights)}}
\label{rem:choice of weights}
Ideally, the weights $\lambda,\rho\in (0,1]$ must  reflect our uncertainty (or lack of confidence) in the prior information $(\widetilde{\SU}_r,\widetilde{\SV}_r)$. To
\edit{loosen} the restriction on the isometry constant for $\R_m(\cdot)$ in (\ref{eq:req on delta thm}), inaccurate prior knowledge must be given lower influence in Program (\ref{eq:p1}) and vice versa. As a concrete example, suppose that $u=v=\theta$ and $\lambda=\rho$. Then, if $\theta\approx \frac{\pi}{2}$ for example, the prior information is obviously unreliable, and it is wise to choose $\lambda=\rho\approx 1$ so as to give less influence to $\widetilde{\SU}_r$ and $\widetilde{\SV}_r$ in Program (\ref{eq:p1}). On the contrary, if $\theta \approx 0$, the prior information is reliable and it is best to take $\lambda=\rho\approx 0$ to reflect our confidence in the prior knowledge.

More specifically, given the principal angle $u=v=\theta$ (or its estimate), one might naturally ask: What is the optimal choice of weights $\lambda=\rho$ in Program (\ref{eq:p1 MR})?

For a fixed angle $\theta \ne 0$, it is not difficult to verify that $\max[\Cr{1},\Cr{2}]$ is minimized by   the choice of $\lambda^2= \rho^2=\sqrt{\tan^4 \theta+\tan^2 \theta}- \tan^2\theta$. This choice in turn maximizes the right hand  side of (\ref{eq:req on delta thm}). In particular, when the principal angle is small ($\theta\approx 0$) this suggests the choice of $\lambda=\rho\approx \sqrt{\tan \theta}$.

\edit{Here the reader will note that the optimal weights $\lambda,\rho$ depend on the angles $u,v$ between the true subspaces and prior knowledge, which will in general be unknown. However, in some applications such as those discussed in the introduction (collaborative filtering and seismology), it is reasonable to assume that a practitioner may have some educated guess as to the accuracy of the prior information, which could be used to set the weights. In addition, our simulations in Section~\ref{sec:simulations} indicate that over a broad range of weight choices, the weighted algorithm outperforms the standard one, and so there is some robustness to the selection of these parameters.}

\end{rem}

Our second main result in this paper, proved in Section \ref{sec:analysis of mc w pi}, quantifies the performance of Program \eqref{eq:p1} for low-rank matrix completion with prior knowledge.\footnote{\edit{This theorem can be extended to general rectangular matrices $M\in\mathbb{R}^{n_{1}\times n_{2}}$ by replacing $n$ with $\max(n_1,n_2)$ except in the failure probability, where $n$ is replaced by $\min(n_1,n_2)$. The conclusion of Theorem~\ref{thm:main result MR} is unaltered.}}

\begin{thm}\label{thm:main result}
\textbf{\emph{(Matrix completion with prior knowledge)}}
Recall the first paragraph of Theorem \ref{thm:main result MR} and let $\eta(M_r)=\eta(U_rV_r^*)$ denote the coherence of $M_r$, see (\ref{eq:coherence def}).
Additionally, let $\breve{U}$ and $\breve{V}$ be orthonormal bases  for $\mbox{span}([\SU_r,\widetilde{\SU}_{r}])$ and $\mbox{span}([\SV_r,\widetilde{\SV}_{r}])$,  respectively. For $p\in(0,1]$ and recalling (\ref{eq:def of R(Z)}), acquire the (possibly noisy) measurement matrix $Y=\mathcal{R}_p(M+E)$ where $\|\mathcal{R}_p(E)\|_F\le  e$ for noise level $ e\ge 0$. Lastly, for $\lambda,\rho\in(0,1]$, let $\widehat{M}$ be a solution of Program (\ref{eq:p1}). Then it holds that
\begin{equation}
\left\Vert \widehat{M}-M\right\Vert _{F}
\lesssim \frac{  \l\|M_{r^+} \r\|_*}{\sqrt{p}} +  { e\sqrt{pn}} ,
\label{eq:err bnd thm simple}
\end{equation}
except with a probability of $o(n^{-19})$, and provided that
\begin{align*}
&1 \ge p \gtrsim \max\left[\log\left(\Cl[factors]{4}\cdot n\right),1\right]\cdot
\frac{\eta(M_r) r\log n}{n} \nonumber\\
& \qquad \quad\, \cdot\max\left[\Cl[factors]{5} \left(1+\sqrt{\frac{\eta( \breve{U}\breve{V}^* ) }{\eta(U_rV_r^* )}} \right),1\right],
\end{align*}
\begin{align}
\Cl[factors]{3}\le  \frac{1}{8},
\label{eq:p req thm}
\end{align}
where $\eta(\breve{U}\breve{V}^*)$ is the coherence of $\breve{U}\breve{V}^*$. Above, we also set
\begin{equation*}
\Cr{4}
:= \sqrt{\frac{\lambda^{4}\cos^{2}u+\sin^{2}u}{\lambda^{2}\cos^{2}u+\sin^{2}u}}
 \cdot \sqrt{\frac{\rho^{4}\cos^{2}v+\sin^{2}v}{\rho^{2}\cos^{2}v+\sin^{2}v}},
 \end{equation*}
 \begin{align*}
\Cr{5}& := \l(
\sqrt{\frac{\lambda^2 \cos^2 u+\sin^2 u}{\rho^2 \cos^2 v+\sin^2 v}} + \sqrt{\frac{\rho^2\cos^2 v+\sin^2 v}{\lambda^2 \cos^2 u+\sin^2 u}} \r)
\nonumber\\
& \qquad
\cdot
\l(
\sqrt{\lambda^4 \cos^2 u +\sin^2 u}+
\sqrt{\rho^4 \cos^2 v+\sin^2 v}
\r),
 \end{align*}
 \begin{equation*}
\Cr{3} :=
\frac{3\sqrt{1-\lambda^{2}}\sin u}{2\sqrt{\lambda^{2}\cos^{2}u+\sin^{2}u}}
+\frac{3\sqrt{1-\rho^{2}}\sin v}{2\sqrt{\rho^{2}\cos^{2}v+\sin^{2}v}}.
\end{equation*}
\end{thm}

A few remarks are in order about Theorem \ref{thm:main result}.

\begin{rem}
\textbf{{(Connection to standard low-rank matrix completion)}}
Note that, by taking $\lambda=\rho=1$, Program (\ref{eq:p1}) reduces to Program (\ref{eq:MC program}) for  standard  matrix completion, thereby ignoring any  prior information.
In this special case, $\Cr{4}=1,\Cr{5} = 4,\Cr{3} = 0$, and (\ref{eq:p req thm}) reads
\begin{equation*}
1 \ge p \gtrsim \frac{\eta(M_r) r\log^2 n}{n} \cdot \l(1+ \sqrt{\frac{\eta(\breve{U}\breve{V}^* )}{\eta(U_rV_r^*)}}\r)
 ,
\end{equation*}
which is  worse than (\ref{eq:bnd on p in standard matrix completion}) in Proposition \ref{prop:MC standard} because of the term $\eta(\breve{U}\breve{V}^*)/\eta(U_rV_r^*)$. However, employing a slightly sharper bound in Appendix \ref{sec:Estimates-of-S' norms} gives $p\gtrsim \eta r\log^2n / n$, which precisely matches (\ref{eq:bnd on p in standard matrix completion}). We opted for the looser bound in (\ref{eq:p req thm}) to keep the bound compact.

As was the case in matrix recovery, Program (\ref{eq:p1}) improves over Program (\ref{eq:MC program}) when the prior knowledge is reliable and our confidence is reflected in the small choice of weights. For example, suppose again that $u=v=\theta$ and $\lambda=\rho=\sqrt{\tan \theta}$. Then a simple calculation shows that
\begin{equation*}
\Cr{4} = \frac{2\sin\theta}{\sin\theta+\cos \theta} \le 2\sin\theta.
\end{equation*}
Therefore, if $\sin\theta \le 1/n$, the logarithmic factors in (\ref{eq:p req thm}) reduce to merely $\log n$. If also $\eta( \breve{U}\breve{V}^* ) \approx \eta(U_rV_r^* )$, then the lower bound on sampling probability in \eqref{eq:p req thm} improves over  that in standard matrix completion, where prior knowledge is not utilized, by reducing $\log^2 n$ to $\log n$.

\edit{Note also that, in the extreme case of $\theta=0$ (namely, when the row and column spaces are exactly known \emph{a priori}), one might recover $M$ from only $r^2$ samples (or equivalently $p=r^2/n^2$)  by solving a simple least-squares program. For incoherent matrices, it is natural to ask whether it is possible to obtain a theoretical result that interpolates between a total sample complexity of $O(r n \log^2 n)$ when prior information is ignored (as in standard matrix completion) and $O(r^2)$ when perfect prior subspace information is available and utilized. However, we point out that even when the prior subspace estimates are very accurate but \emph{not} perfect, the number of degrees of the freedom remains substantially greater than $O(r^2)$. We argue this by considering the dimension of the Grassmannian manifold of all $r$-dimensional subspaces of $\mathbb{R}^n$: its dimension is $r(n-r)$ and so the number of degrees of freedom required to parameterize the difference between two (even very nearby) subspaces on this manifold is $O(rn)$. In particular, for the rank-$r$ matrix $M_r=U_r \Sigma_r V_r^*$, one may verify that the elements of the tangent space to the manifold of rank-$r$  $n\times n$ matrices take the form 
$$
U_r \Delta_1 V_r^* + U_r^\perp \Delta_2 V_r^* + U_r \Delta_3^* (V_r^\perp)^*,
$$
where $\Delta_1\in\mathbb{R}^{r\times r},\Delta_2,\Delta_3\in\mathbb{R}^{n-r\times r}$ are arbitrary. That is, the tangent space at $M_r$ is a $(2n-r)r$-dimensional linear subspace. Thus, there appears to be is a discontinuity in the achievable sample complexity as a function of $\theta$: when $\theta = 0$, a specialized algorithm can succeed with $O(r^2)$ samples, but when $\theta \neq 0$, no algorithm can succeed in general without at least $O(rn)$ samples.}

\end{rem}

\begin{rem} \textbf{{(On choosing the weights)}}
\label{rem:choice of weightsrevisited}
Similar to Remark \ref{rem:choice of weights}, this remark discusses the optimal choice of weights $\lambda,\rho\in(0,1]$ now in Program (\ref{eq:p1}).
 For simplicity, again assume that $u=v=\theta$ and $\lambda=\rho$.
Then, $\Cr{5}$ and $\Cr{3}$ are non-increasing and non-decreasing in $\lambda=\rho$, respectively. However, $\Cr{4}$ is minimized with the choice of $\lambda^2=\rho^2= \sqrt{\tan^4\theta+\tan^2\theta}-\tan^2\theta$. In particular, when $\theta$ is small, $\Cr{4}$ is minimized with the choice of $\lambda=\rho\approx \sqrt{\tan \theta}$.
\end{rem}

\begin{rem} \textbf{{(Solving Programs \eqref{eq:p1 MR} and~\eqref{eq:p1})}}
\edit{As noted in~\cite{aravkin2014fast}, both $Q_{\SUT_r,\lambda}$ and $Q_{\SVT_r,\rho}$ are invertible matrices when $\lambda,\rho > 0$. Therefore the weighting matrices appearing in the objective function of~\eqref{eq:p1 MR} and~\eqref{eq:p1} can be absorbed into the constraints by minimizing the nuclear norm of $X' = Q_{\SUT_r,\lambda} \cdot X \cdot Q_{\SVT_r,\rho}$ and making the substituition   in the constraints that $X = Q_{\SUT_r,\lambda}^{-1} \cdot X' \cdot Q_{\SVT_r,\rho}^{-1}$. Therefore, standard semi-definite programming~\cite{recht2010guaranteed} or bilinear factorization methods~\cite{burer2003nonlinear} can be used to solve the weighted nuclear norm minimization problems.}
\end{rem}

\section{Simulations}
\label{sec:simulations}

\edit{This section provides some numerical characterization of the weighted matrix recovery and completion schemes. All simulations were performed using CVX~\cite{grant2008cvx}.}

\subsection{How to best use prior information?}
\label{sec:howbest}

\edit{As a baseline test, we construct a square matrix $M$ of sidelength $n=20$ and rank $r = 4$, namely  $M=U_r V_r^*$. Here, $U_r\in\mathbb{R}^{n\times r}$ spans a random $r$-dimensional subspace of $\mathbb{R}^n$, namely drawn from the uniform distribution on the Grassmannian, as it is constructed by orthogonalizing the columns of a standard random Gaussian matrix $G\in\mathbb{R}^{n\times r}$. Likewise,  $V_r \in\mathbb{R}^{n\times r}$ is constructed  from an independent copy of $G$. Finally, $M$ is normalized such that $\|M\|_F = 1$. As prior knowledge to be used in completing $M$, we construct a perturbed version $M'$ of $M$, where $M' = M + N$ and the entries of $N\in\mathbb{R}^{n\times n}$ are independent Gaussian random variables with mean zero and variance $\sigma^2 = 10^{-4}$. We then compute the SVD of $M'$ and let $(\wt{\SU}_r,\wt{\SV}_r)$ contain the leading $r$ left- and right-singular vectors of $M'$, and let $\wt{\Sigma}_r$ contain the leading $r$ singular values of $M'$. The principal angles between the ground truth and prior knowledge subspaces are $u = \angle [\SU_r,\SUT_r ] = 0.099$ and $v = \angle [\SV_r,\SVT_r ] = 0.103$.}

\edit{For various values of a sampling probability $p$, we sample without noise each element of $M$ independently with probability $p$, and we consider the matrix completion problem of filling in the unobserved entries of $M$. We compare several algorithms for solving this problem with the prior information available in $M'$:
\begin{itemize}
\item Standard matrix completion, corresponding to~\eqref{eq:MC program} with $e = 0$.
\item Weighted matrix completion, corresponding to Program~\eqref{eq:p1} with $e = 0$. For simplicity, we take the left and right weights to be equal: $\lambda=\rho=w$ for various values of $w$ that we test. This algorithm uses $(\wt{\SU}_r,\wt{\SV}_r)$ to aid in the completion of $M$.
\item A diagonal weighting algorithm proposed in~\cite[end of Sec.\ 5]{chen2013completing}. This approach takes advantage of prior information of the leverage scores of a matrix to reweight the rows and columns of the matrix in order to equalize its leverage scores with the sampling probabilities (which are uniform in our example). Specifically, this algorithm uses only the row norms of $\wt{\SU}_r$ and $\wt{\SV}_r$ to aid in the completion of $M$, see \eqref{eq:coherence def}.
\item A weighted least-squares algorithm proposed in~\cite{mohan2012iterative}. This approach appears as one iteration of an iterative, reweighted least squares (IRLS) algorithm for matrix completion. This algorithm uses all of $M'$ (rather than just its leverage scores or singular vectors) to aid in the completion of $M$. It involves a regularization parameter $\gamma$, which we fix to $0.1$ throughout this section. (Results were similar with other values of $\gamma$.)
\item A reweighted nuclear norm heuristic algorithm proposed in~\cite{mohan2010reweighted}. This approach appears as one iteration of an iterative, reweighted algorithm for matrix completion, which we discuss further in Section~\ref{sec:related work}. Specifically, to use this heuristic in a non-iterative fashion, we solve Program~\eqref{eq:fazelrw} once with weighting matrices $W_1^k$ and $W_2^k$ that are constructed not from some previous iteration but rather from $M'$ itself. The full recipe for constructing $W_1^k$ and $W_2^k$ is provided in~\cite{mohan2010reweighted}; it involves using all of $M'$. This recipe also involves a regularization parameter $\delta$, for which we test various values.
\end{itemize}
}

\edit{Figure~\ref{fig:mcmrTests06resultsAll} shows the results from the five algorithms, averaging the success probability over $50$ trials, where a successful trial is declared if $M$ is recovered up to a relative error of $10^{-3}$. The weighted matrix completion result is shown with a choice of $w=0.1$ and the reweighted nuclear norm heuristic (RNNH) result is shown with a choice of $\delta = 0.01$. The results from these two algorithms with other values of $w$ and $\delta$ are shown in Figures~\ref{fig:mcmrTests06resultsWeighted} and~\ref{fig:mcmrTests06resultsRNNH}, respectively. We see in these plots that, with proper choices of $w$ and $\delta$, both the weighted algorithm and RNNH can substantially outperform standard matrix completion. In both cases, setting $w$ or $\delta$ too large or too small can hamper the performance. Moreover, in the best case, the weighted algorithm and RNNH perform comparably to each other. However, the weighted algorithm uses slightly weaker prior information, since it depends only on $(\wt{\SU}_r,\wt{\SV}_r)$ rather than all of $M'$. The diagonal weighting algorithm also outperforms standard matrix completion, but its performance is not as strong as the weighted algorithm and RNNH, possibly because it relies on much less prior information (only the leverage scores of $M'$). The least-squares algorithm does not succeed unless $p=1$ because one iteration of a least-squares algorithm is not enough to promote low-rank structure.}

\edit{Figures~\ref{fig:mcmrTests08resultsAll}--\ref{fig:mcmrTests08resultsRNNH} show the results from a second experiment, where we use more accurate prior information. Specifically, in constructing $M'$, we set $M' = M + N$ where the entries of $N$ are independent Gaussian random variables with mean $0$ and variance $\sigma^2 = 10^{-6}$. The principal angles between the ground truth and prior knowledge subspaces are $u = \angle [\SU_r,\SUT_r ] = 0.0099$ and $v = \angle [\SV_r,\SVT_r ] = 0.0104$. We see a slight improvement in the performance of the weighted algorithm and RNNH, although the optimal choices of both $w$ and $\delta$ have decreased, as anticipated in Remarks~\ref{rem:choice of weights} and~\ref{rem:choice of weightsrevisited}. Although the best case performance is only slightly better than in the earlier experiment, we do observe a greater robustness to parameter choices: there are much wider ranges of $w$ and $\delta$ for which the algorithms perform well.}

\begin{figure*}[t]
\begin{center}
\subfloat[\label{fig:mcmrTests08resultsAll}]{
\includegraphics[width=.33\textwidth]{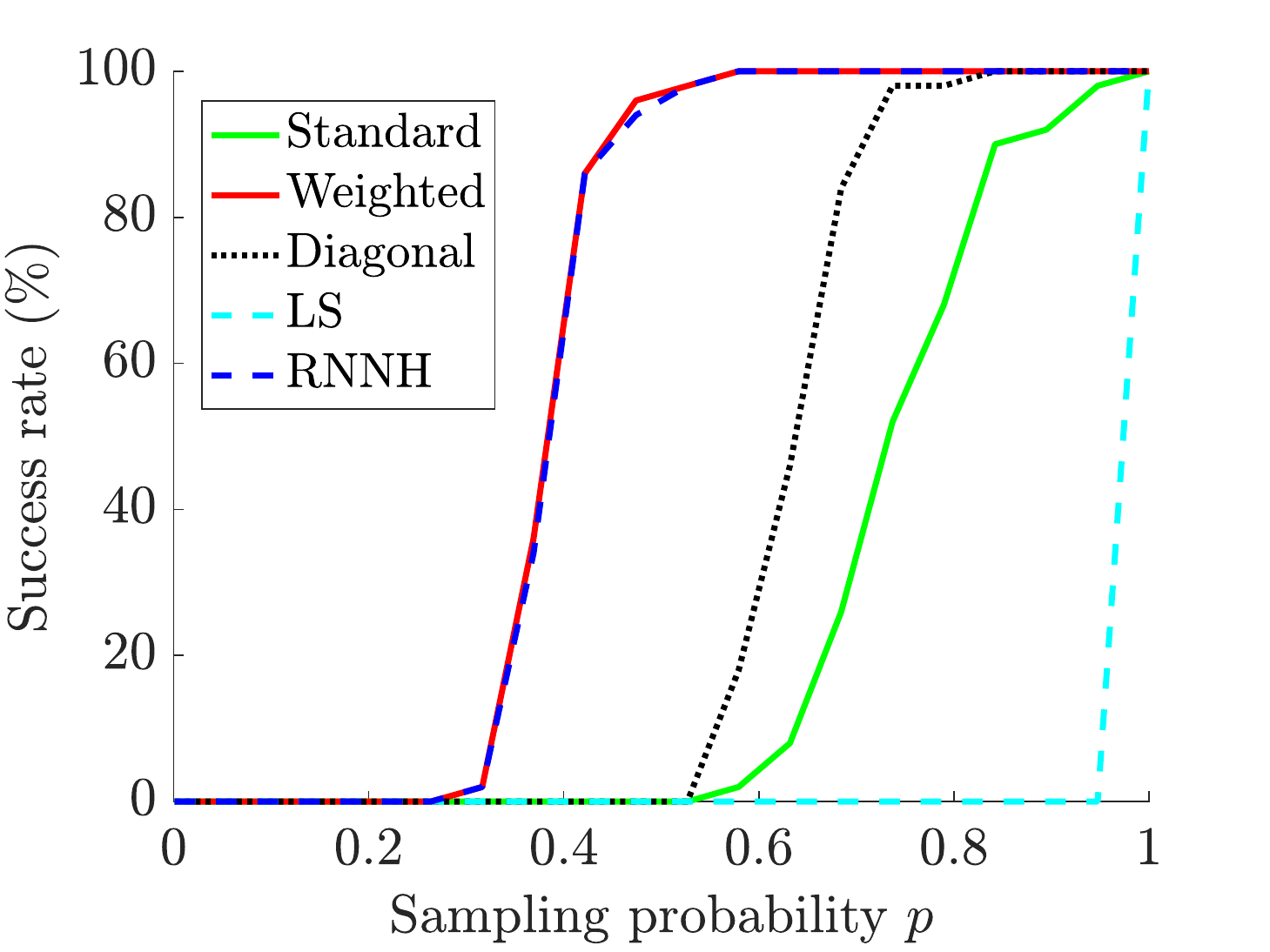}
}
\subfloat[\label{fig:mcmrTests08resultsWeighted}]{
\includegraphics[width=.33\textwidth]{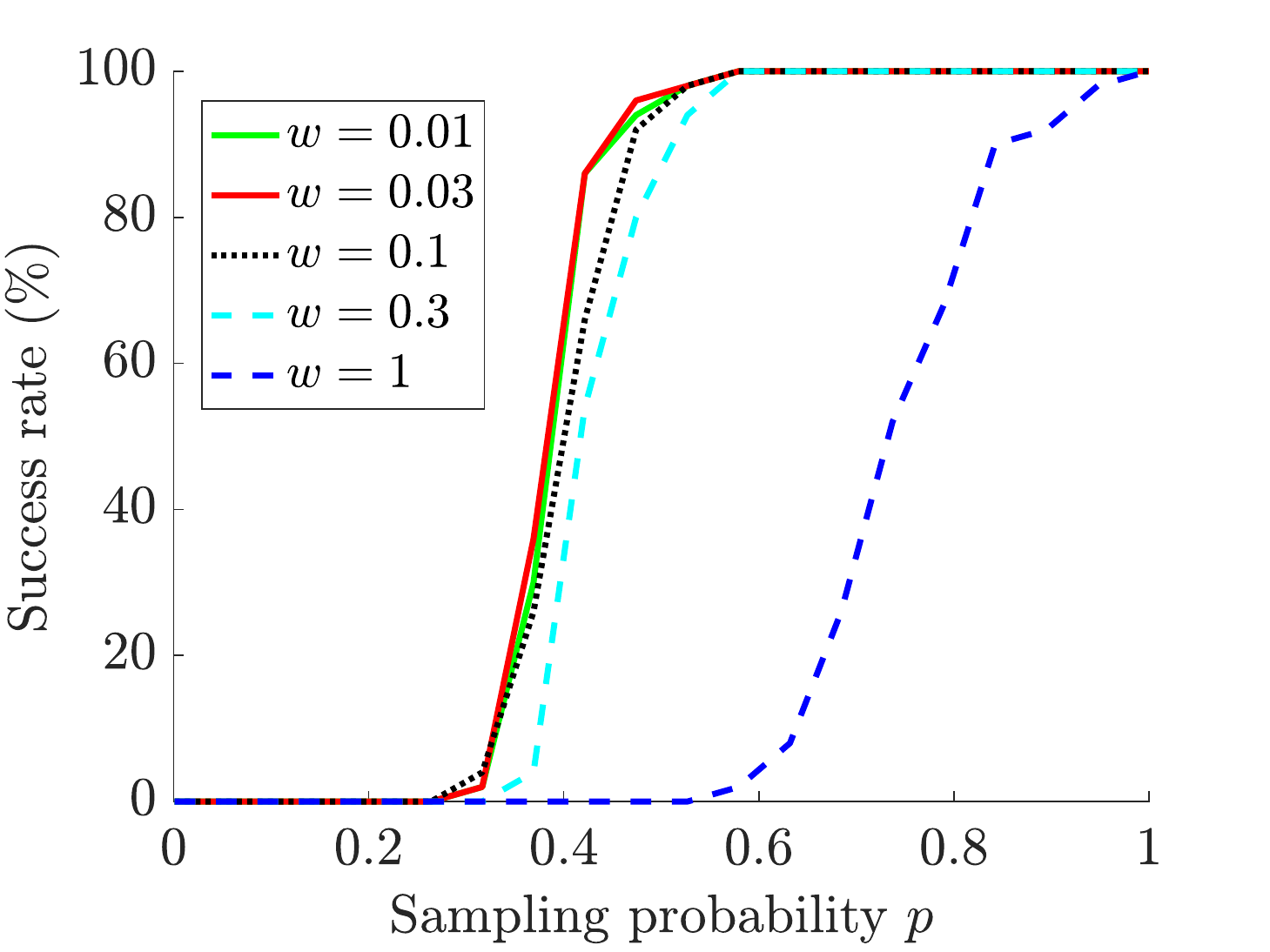}
}
\subfloat[\label{fig:mcmrTests08resultsRNNH}]{
\includegraphics[width=.33\textwidth]{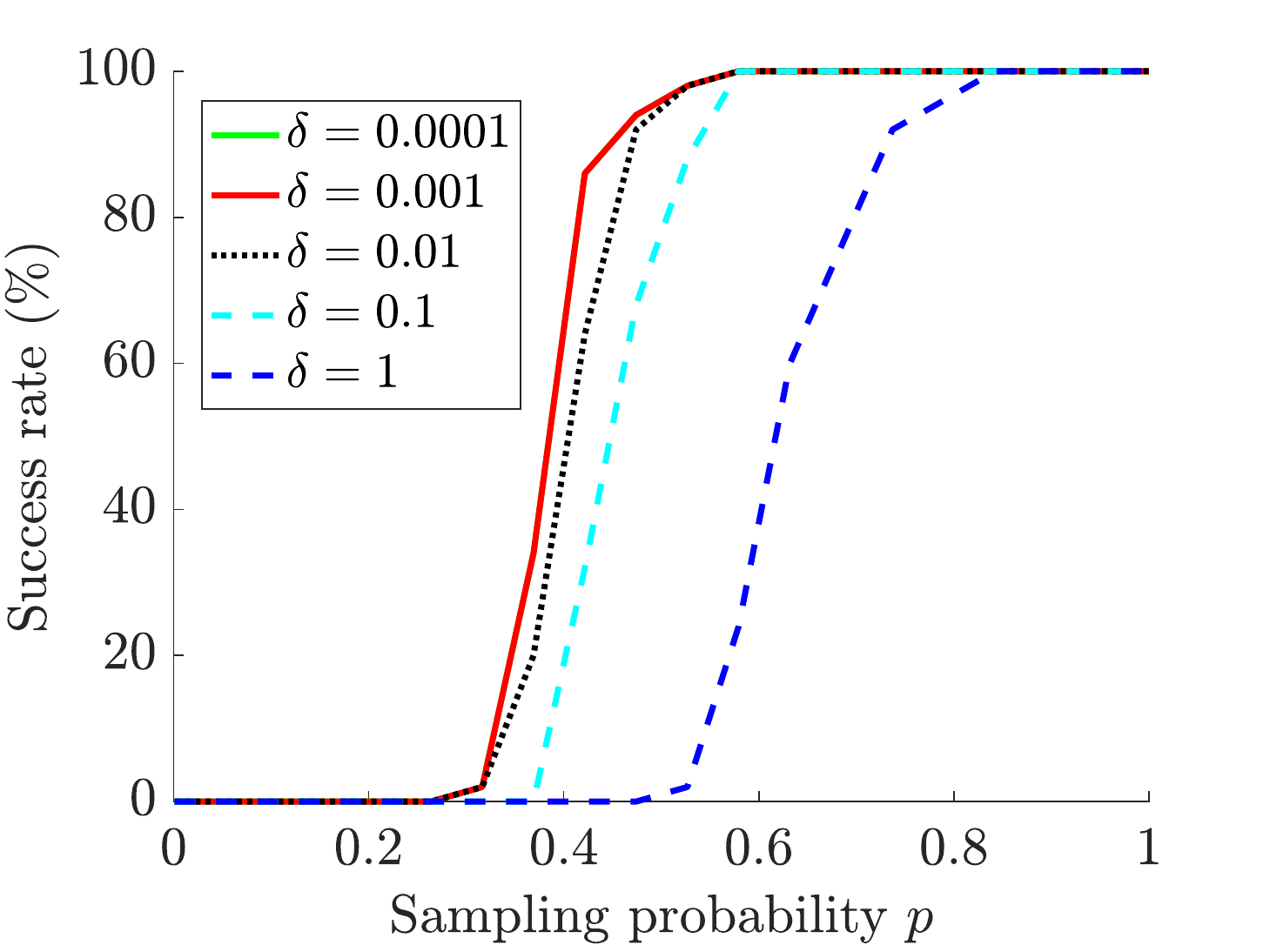}
}
\caption{\edit{Matrix completion with stronger prior information, where the principal angles between the ground truth and prior knowledge subspaces are $u = \angle [\SU_r,\SUT_r ] = 0.0099$ and $v = \angle [\SV_r,\SVT_r ] = 0.0104$. (a) Comparison of standard (unweighted) matrix completion, weighted matrix completion with $w = 0.03$, diagonal weighted matrix completion, weighted least-squares, and a weighted nuclear norm heuristic (RNNH) algorithm with $\delta = 0.001$. (b) Weighted matrix completion with various weights $w$. (c) RNNH with various parameters $\delta$.}}
\end{center}
\end{figure*}

\edit{Figures~\ref{fig:mcmrTests07resultsAll}--\ref{fig:mcmrTests07resultsRNNH} repeat this experiment with less accurate prior information. In constructing $M'$, we set $\sigma^2 = 0.0025$. The principal angles between the ground truth and prior knowledge subspaces are $u = \angle [\SU_r,\SUT_r ] = 0.512$ and $v = \angle [\SV_r,\SVT_r ] = 0.480$. Compared to the other experiments, we see a degradation in the performance of the weighted algorithm and RNNH, due to worse prior information.  While both can still outperform standard matrix completion by choosing $w$ and $\delta$ suitably large, they can also both underperform standard matrix completion with an improper choice of parameters.}

\begin{figure*}[t]
\begin{center}
\subfloat[\label{fig:mcmrTests07resultsAll}]{
\includegraphics[width=.33\textwidth]{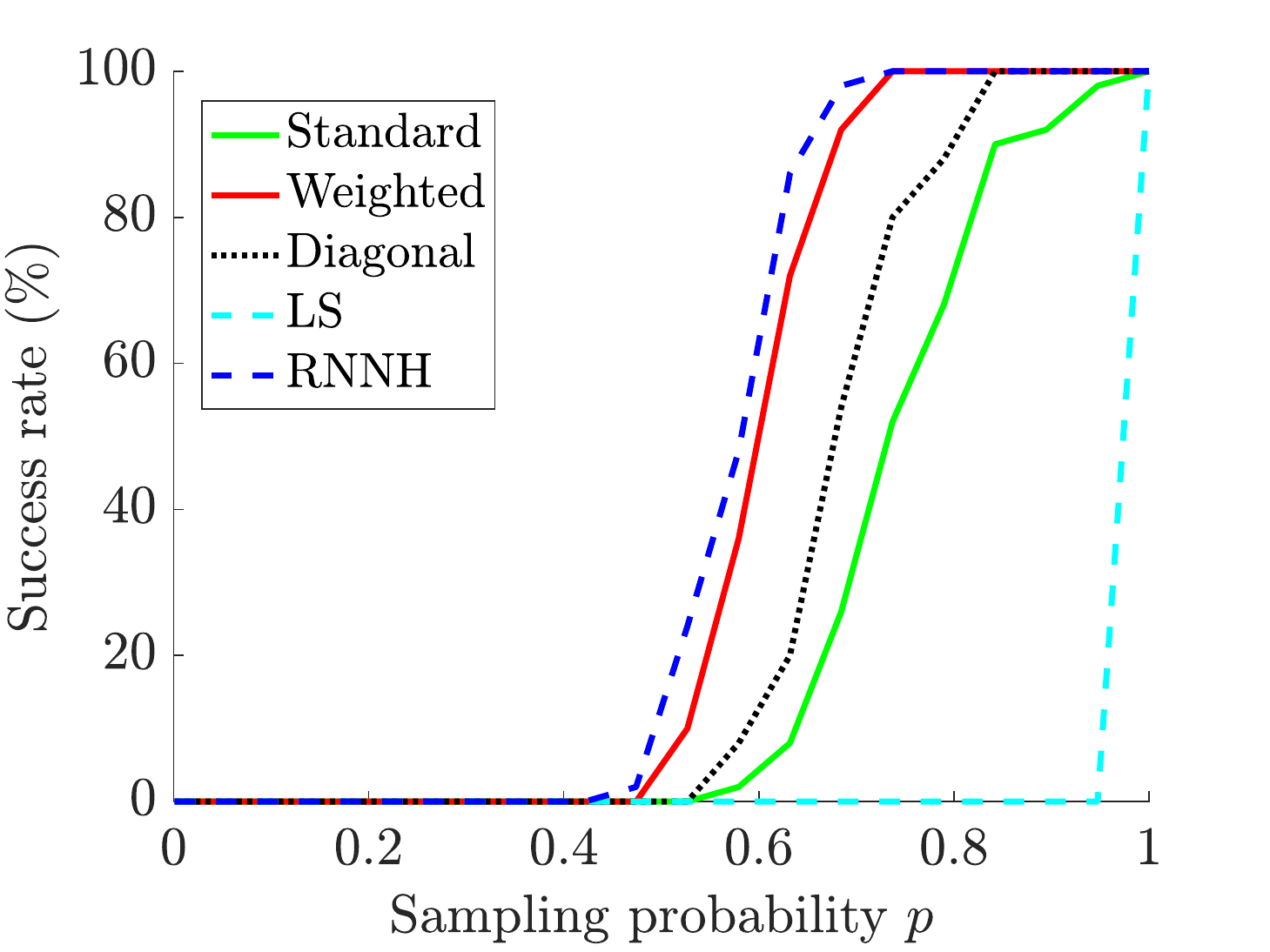}
}
\subfloat[\label{fig:mcmrTests07resultsWeighted}]{
\includegraphics[width=.33\textwidth]{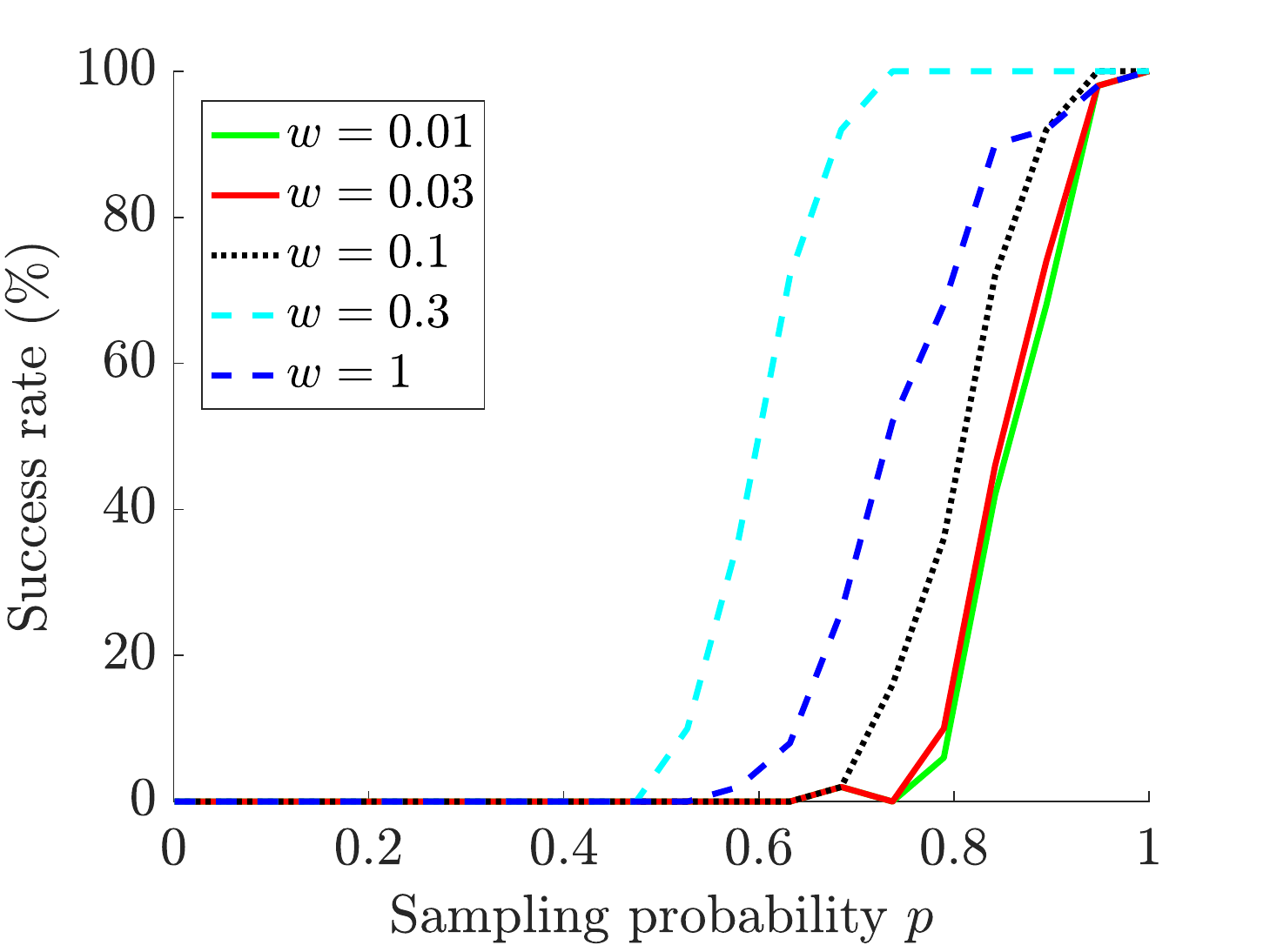}
}
\subfloat[\label{fig:mcmrTests07resultsRNNH}]{
\includegraphics[width=.33\textwidth]{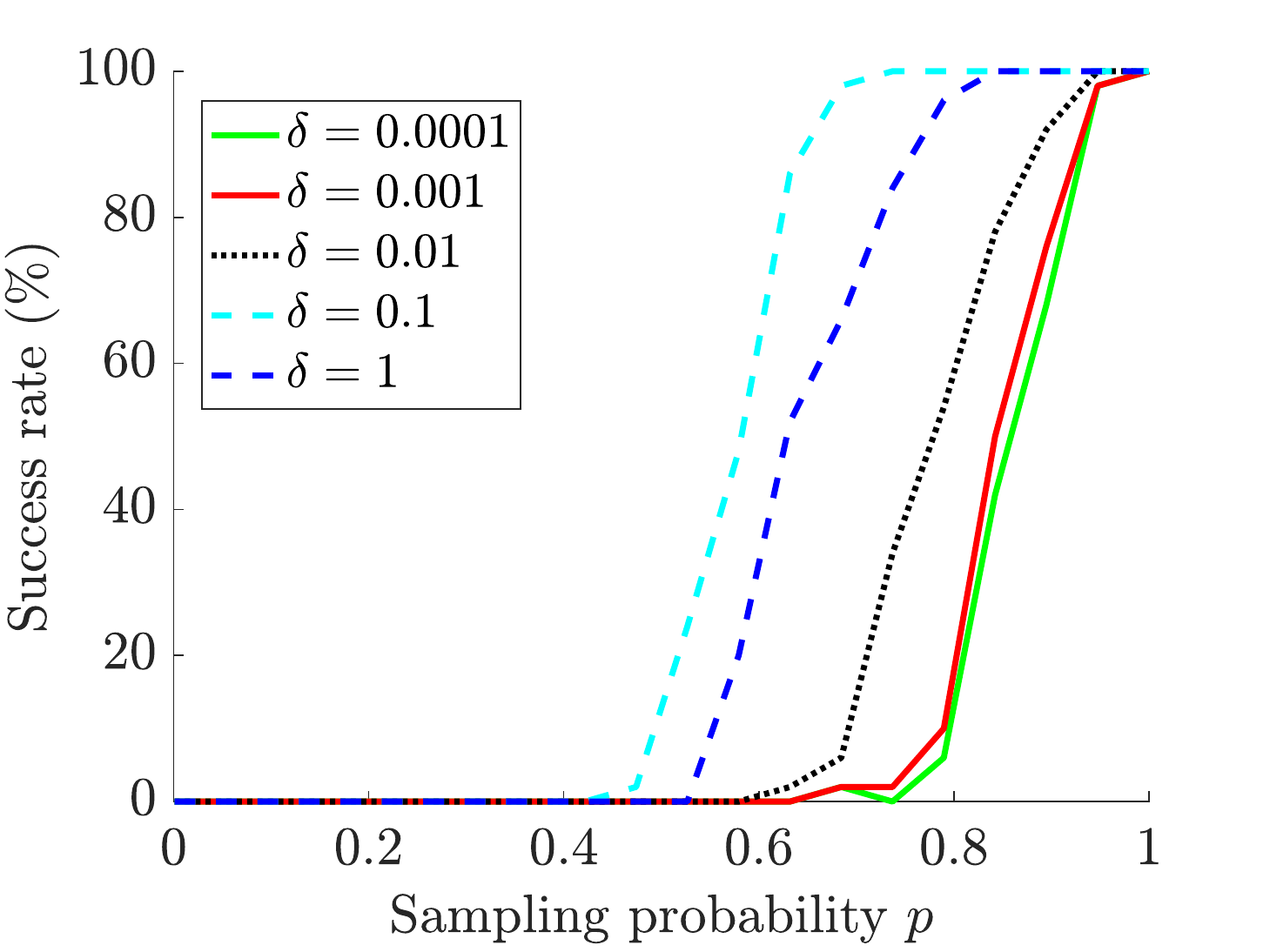}
}
\caption{\edit{Matrix completion with weaker prior information, where the principal angles between the ground truth and prior knowledge subspaces are $u = \angle [\SU_r,\SUT_r ] = 0.512$ and $v = \angle [\SV_r,\SVT_r ] = 0.480$. (a) Comparison of standard (unweighted) matrix completion, weighted matrix completion with $w = 0.3$, diagonal weighted matrix completion, weighted least-squares, and a weighted nuclear norm heuristic (RNNH) algorithm with $\delta = 0.1$. (b) Weighted matrix completion with various weights $w$. (c) RNNH with various parameters $\delta$.}}
\end{center}
\end{figure*}

\edit{Figure~\ref{fig:mcmrTests09resultsAll} repeats the experiment from Figure~\ref{fig:mcmrTests06resultsAll} but with an equivalent number of random linear measurements in place of sampling the matrix entries. The algorithms are modified to perform matrix recovery instead of matrix completion, except we omit the diagonal weighting algorithm as the leverage score weighting algorithm in~\cite{chen2013completing} was tailored to the problem of matrix completion. The relative algorithm performance is similar to other experiments.}

\begin{figure}[t]
\begin{center}
\includegraphics[width=.43\textwidth]{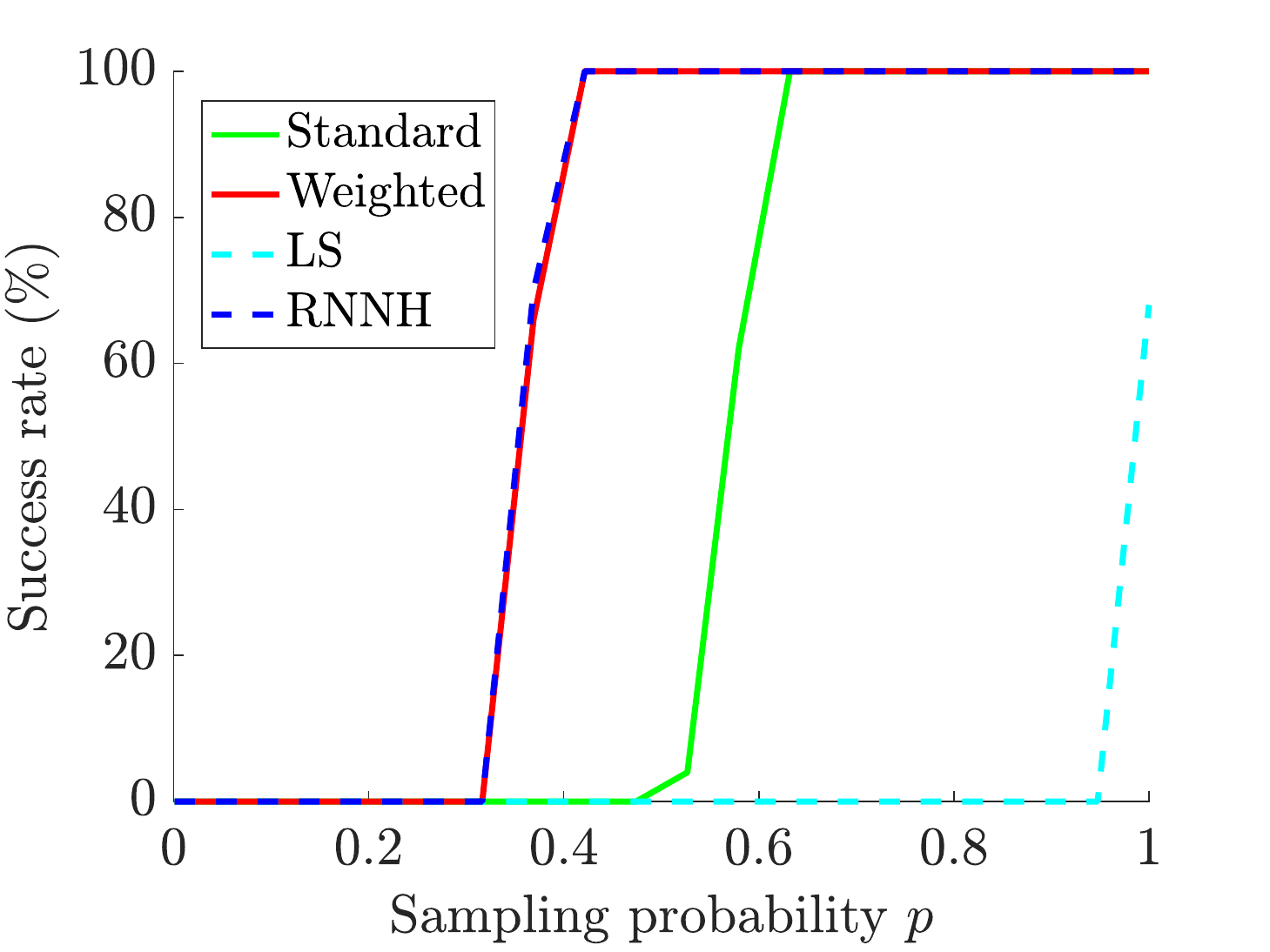}
\caption{\label{fig:mcmrTests09resultsAll} \edit{Matrix recovery from random measurements with prior information, where $u = \angle [\SU_r,\SUT_r ] = 0.119$ and $v = \angle [\SV_r,\SVT_r ] = 0.086$. The figure shows a comparison of standard (unweighted) matrix completion, weighted matrix recovery with $w = 0.3$, weighted least-squares, and a weighted nuclear norm heuristic (RNNH) algorithm with $\delta = 0.01$.}}
\end{center}
\end{figure}

\edit{Figure~\ref{fig:mcmrTests10resultsAll} repeats the matrix completion experiment from Figure~\ref{fig:mcmrTests06resultsAll} but with a rank $r=2$ matrix rather than $r=4$. All algorithms perform better, as anticipated.}

\begin{figure}[t]
\begin{center}
\includegraphics[width=.43\textwidth]{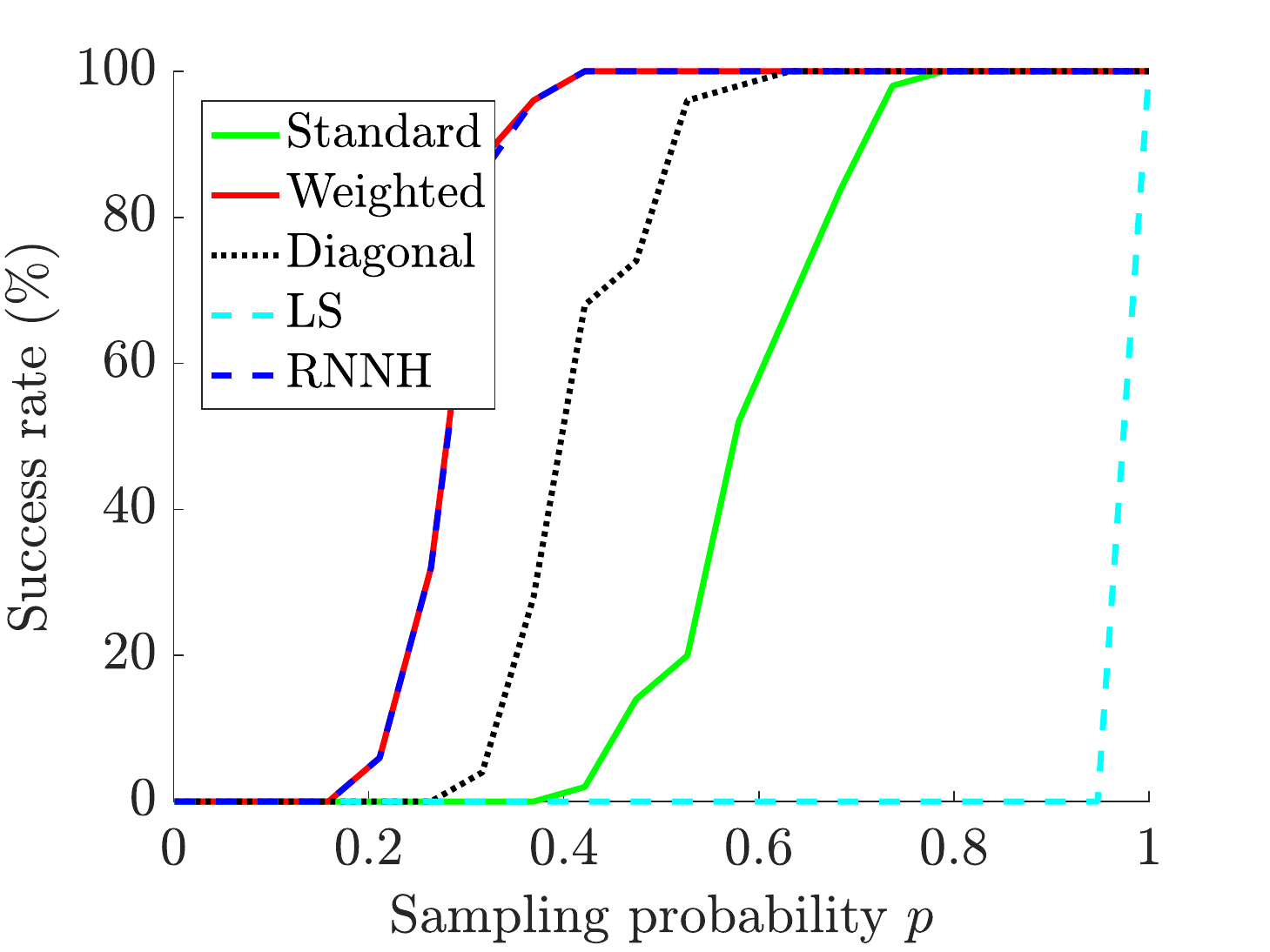}
\caption{\label{fig:mcmrTests10resultsAll} \edit{Matrix completion with prior information for a rank $r=2$ matrix. Here, $u = \angle [\SU_r,\SUT_r ] = 0.075$ and $v = \angle [\SV_r,\SVT_r ] = 0.064$. The algorithm parameters are $w = 0.1$ and $\delta = 0.01$.}}
\end{center}
\end{figure}

\edit{Figure~\ref{fig:mcmrTests11resultsAll} repeats the matrix completion experiment from Figure~\ref{fig:mcmrTests06resultsAll} but with a coherent matrix $M$ having larger leverage scores (the largest leverage score is $4.055$, as opposed to $2.211$ for the matrix used in the earlier experiment). This higher coherence hampers the performance of standard matrix completion and makes the advantages of the diagonal weighted algorithm more substantial. However, the weighted algorithm and RNNH still have the best performance in this experiment.}

\begin{figure}[t]
\begin{center}
\includegraphics[width=.43\textwidth]{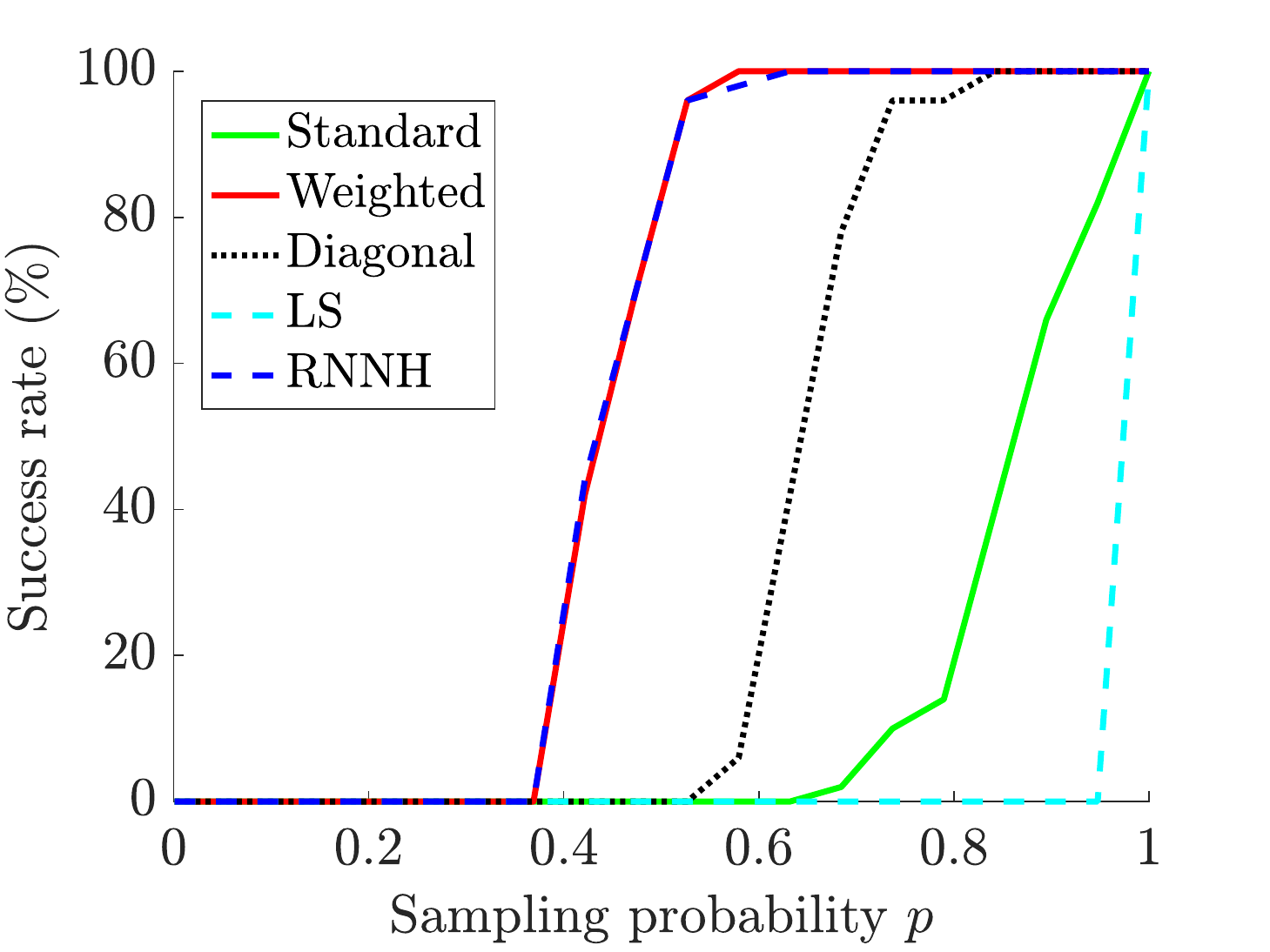}
\caption{\label{fig:mcmrTests11resultsAll} \edit{Matrix completion of a coherent matrix with prior information. Here, $u = \angle [\SU_r,\SUT_r ] = 0.096$ and $v = \angle [\SV_r,\SVT_r ] = 0.111$. The algorithm parameters are $w = 0.1$ and $\delta = 0.01$.}}
\end{center}
\end{figure}

\edit{Finally, Figure~\ref{fig:mcmrTests12resultsAll} repeats the matrix completion experiment from Figure~\ref{fig:mcmrTests06resultsAll} but with a matrix sidelength of $n=40$ rather than $n=20$.}

\begin{figure}[t]
\begin{center}
\includegraphics[width=.43\textwidth]{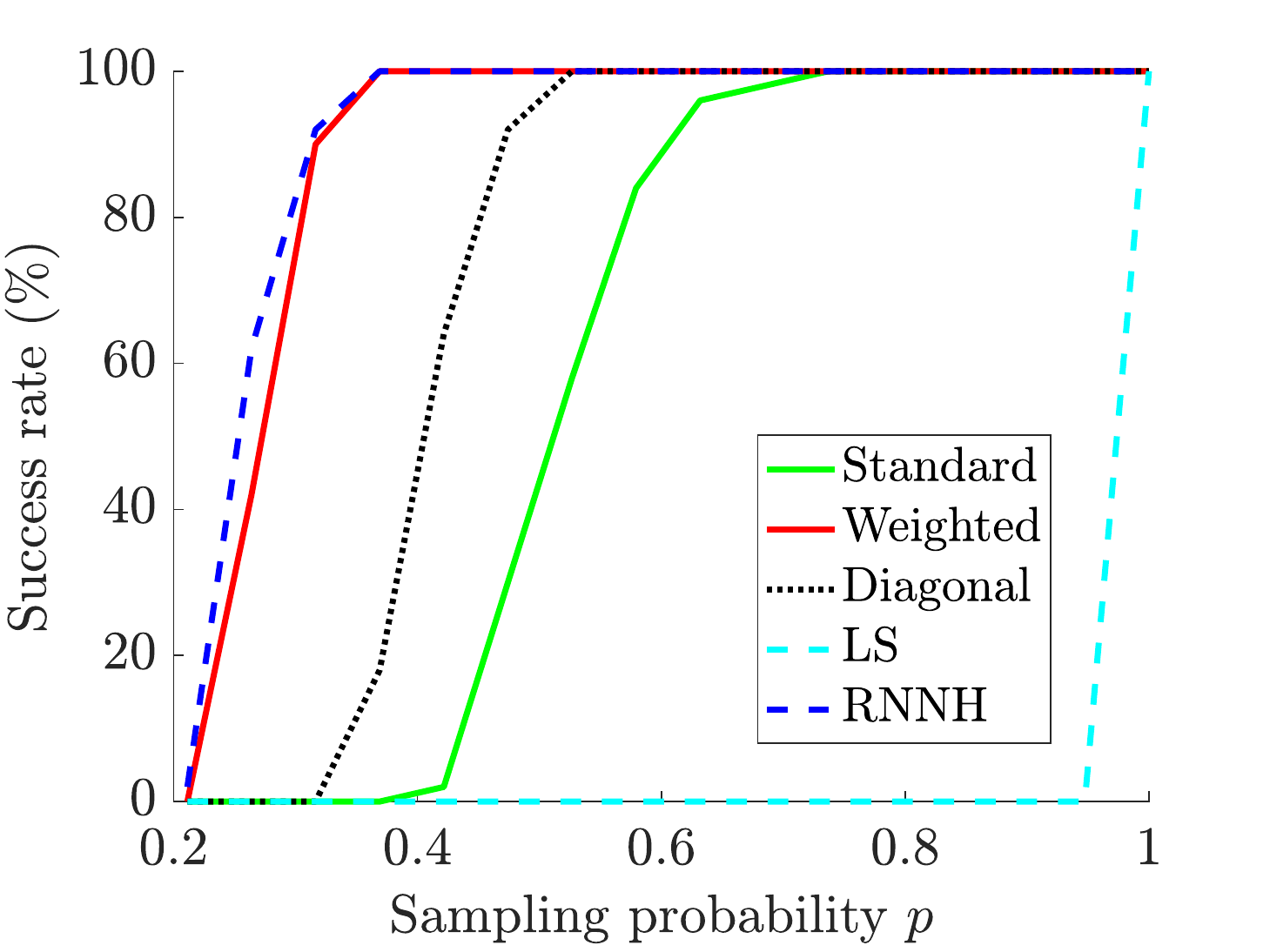}
\caption{\label{fig:mcmrTests12resultsAll} \edit{Matrix completion of a $40\times 40$ rank-$4$ matrix with prior information. Here, $u = \angle [\SU_r,\SUT_r ] = 0.139$ and $v = \angle [\SV_r,\SVT_r ] = 0.142$. The algorithm parameters are $w = 0.1$ and $\delta = 0.01$.}}
\end{center}
\end{figure}

\edit{In all experiments in this section, we see that with proper parameter choices, both the weighted algorithm and RNNH can substantially outperform other techniques. Interestingly, the weighted algorithm and RNNH generally perform similarly to each other, although they originally came from different motivations (one as a means of employing prior information in a non-iterative fashion, the other as a step in an iterative algorithm that does not require prior information). Finally, as we have noted, the weighted algorithm uses slightly weaker prior information than RNNH, since it depends only on $(\wt{\SU}_r,\wt{\SV}_r)$ rather than all of $M'$.}

\subsection{Iterative reweighting without prior information}

\edit{Section~\ref{sec:howbest} focused on the question of how to best employ prior matrix information in one iteration of a matrix completion/recovery algorithm; this is indeed the main topic of this paper.}

\edit{However, it is also possible to consider employing weighting in an iterative fashion, even when no prior information is available. In such a case, one begins by solving standard matrix completion/recovery, uses the result (call it $M'$) as ``prior'' information to set weighting matrices, and then solves a weighted matrix completion/recovery program. As discussed in Section~\ref{sec:related work}, this is the original motivation for the RNNH that was evaluated in Section~\ref{sec:howbest}. In Figure~\ref{fig:mcmrTests13resultsAll}, we compare an iterative, reweighted application of~\eqref{eq:p1} to the iterative reweighted nuclear norm minimization algorithm from~\cite{mohan2010reweighted}. We see that after one weighted iteration, the algorithm from~\cite{mohan2010reweighted} has a slight performance advantage, while after four iterations (as prescribed in~\cite{mohan2010reweighted}), the two iterative algorithms perform similarly.}

\begin{figure}[t]
\begin{center}
\includegraphics[width=.43\textwidth]{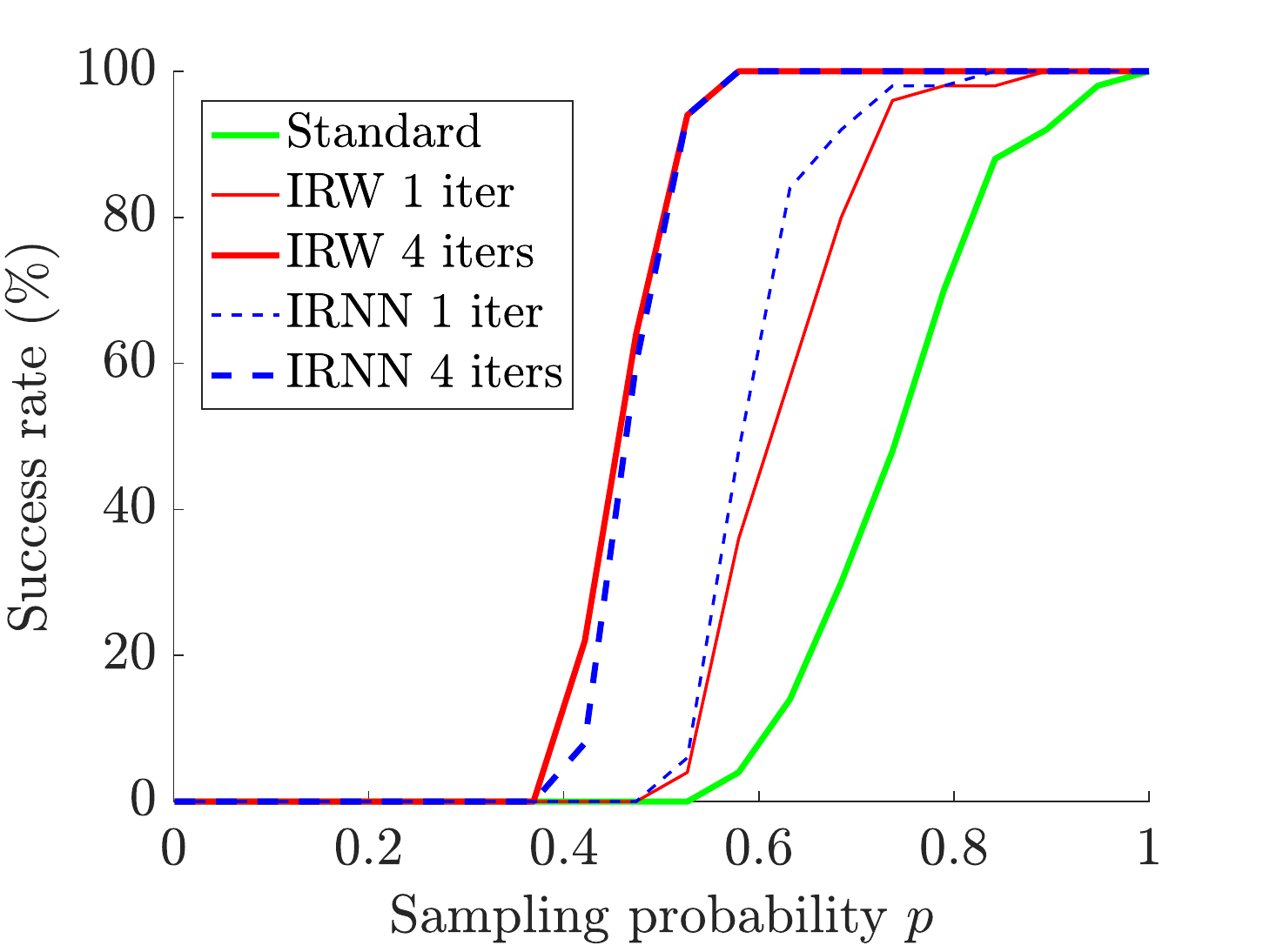}
\caption{\label{fig:mcmrTests13resultsAll} \edit{Matrix completion of a $20\times 20$ rank-$4$ matrix without prior information, through an iterative application of the weighted program~\eqref{eq:p1} (labeled IRW) and an iterative application of the RNNH (labeled IRNN). Both algorithms are initialized using standard (unweighted) nuclear norm minimization, and the results after one and four subsequent weighted iterations are shown.}}
\end{center}
\end{figure}

\edit{Overall, whether used in one iteration as a means of incorporating prior information or used in iterative fashion starting from an unweighted initialization, the weighted program in~\eqref{eq:p1} and RNNH appear to perform similarly to each other across a range of problem instances. The goal of this paper is to quantify the benefit of incorporating prior information in one iteration of~\eqref{eq:p1}. It would be interesting to develop a similar analysis for RNNH, or to study the convergence of the iterative application of~\eqref{eq:p1}.}

\section{Related Work}
\label{sec:related work}

Programs similar to \eqref{eq:p1 MR} and \eqref{eq:p1} have appeared in the literature before, and we wish to summarize here some of the related work.
In~\cite{xu2013speedup,jain2013provable}, the authors incorporate side information for matrix completion using nuclear norm minimization. However, their works differ from ours in that they assume perfect subspace information, which they use to reduce the dimension of the low-rank recovery problem (as well as the sample complexity). A later paper~\cite{chiang2015matrix} supplements this recovery program with a correction term to account for imperfect side information. Theory is again provided which allows for a reduction in sample complexity; however, this theory is limited to randomly generated matrices and uses a different characterization of subspace accuracy than the  principal angles we consider.

In \cite{mohan2010reweighted}, the authors considered the following non-convex program for rank minimization:
\begin{equation}
\begin{cases}
\min_X \log\l(\det(X)\r)\\
X\in \mbox{C}.
\end{cases}
\label{eq:Fazel}
\end{equation}
Here, the feasible set $\mbox{C}\subset\real^{n\times n}$ is assumed to be convex. (Local) linearization of the above objective function leads to a \emph{majorization-minimization} algorithm to solve Program  \eqref{eq:Fazel}, in which the $k$-th iteration takes the form of
\begin{equation}
\begin{cases}
\min_X \l\| W_1^k X W_2^k \r\|_* \\
X\in \mbox{C},
\end{cases}
\label{eq:fazelrw}
\end{equation}
for certain weight matrices $W_1^k$ and $W_2^k$. Convergence of this \emph{reweighted} algorithm to a local minimum of Program \eqref{eq:Fazel} is known. See also \cite{eftekhari2016expect} for a related problem.

In \cite{rao2015collaborative}, the authors study the following program:
\begin{equation}
\min_{X\in\mathbb{R}^{n\times n}}
\frac{1}{m} \l\|P_{\Omega} \l( X-Y \r) \r\|_F^2
+ \lambda_N
\l\| A XB \r\|_*.
\label{eq:Nikhil}
\end{equation}
Here, $\Omega \subseteq [1:n]^2$ is a random index set of size $m$, and $P_{\Omega}(X)\in\mathbb{R}^{n\times n}$ retains the entries of $X$ on the index set $\Omega$ and sets the rest to zero. In addition, $Y=P_\Omega(M+E)$ where we take  $M\in\mathbb{R}^{n\times n}$ to be rank-$r$ for simplicity, and the entries of $E\in\mathbb{R}^{n\times n}$  are independent zero-mean Gaussian random variables with variance $\sigma^2/n$. Lastly, $\lambda_N>0$ and $A,B\in\mathbb{R}^{n\times n}$  are both assumed to be invertible. Let $\widehat{M}\in\mathbb{R}^{n\times n}$ be a solution of Program \eqref{eq:Nikhil}. Then, Theorem 2 in the same reference  establishes that
\begin{equation}
\l\| \widehat{M}-M \r\|_F^2
\lesssim
\alpha_N^2 \max\l[ 1,\sigma^2\r]
\frac{r\log n}{m},
\label{eq:Nikhil's result}
\end{equation}
with high probability and provided that $\lambda_N \gtrsim \sqrt{\frac{n\log n}{m}}$. Above,
\begin{equation}
\alpha_N := n \cdot
\frac{\l\|A M B^* \r\|_{\infty}}{\l\|A M B^*\r\|_F}.
\end{equation}
By setting $A=Q_{\widetilde{\SU}_r,\lambda}$ and $B=Q_{\widetilde{\SV}_r,\rho}$ (see \eqref{eq:def of Qs}), Program \eqref{eq:Nikhil} will be equivalent to  Program \eqref{eq:p1} for the right choice of $\lambda_N$. To compare \eqref{eq:Nikhil's result} and Theorem \ref{thm:main result}, let $Q_{\widetilde{\SU}_r,\lambda}MQ^*_{\widetilde{\SV}_r,\rho}=U_{N,r}\Sigma_{N,r}V_{N,r}^*$ be the SVD of $AMB$ and note that
\begin{align}
\alpha_N^2 & = n^2 \cdot
\frac{\l\|Q_{\widetilde{\SU}_r,\lambda} M Q_{\widetilde{\SV}_r,\rho}^* \r\|_{\infty}^2}{\l\|Q_{\widetilde{\SU}_r,\lambda} M Q_{\widetilde{\SV}_r,\rho}^*\r\|_F^2} \nonumber\\
& = n^2 \cdot
\frac{\l\|U_{N,r} \Sigma_{N,r} V_{N,r}^* \r\|_{\infty}^2}{\l\|U_{N,r} \Sigma_{N,r} V_{N,r}^*\r\|_F^2} \nonumber\\
& \le n^2 \cdot \frac{ \l\| U_{N,r}\r\|_{2\rightarrow\infty}^2 \cdot \l\| \Sigma_{N,r} \r\|^2 \cdot  \l\| V_{N,r}\r\|_{2\rightarrow\infty}^2}{\l\| \Sigma_{N,r}\r\|_F^2} \nonumber\\
& =  n^2 \cdot \frac{r^2\cdot \eta^2\l( Q_{\widetilde{\SU}_r,\lambda} M Q_{\widetilde{\SV}_r,\rho}^* \r) \l\| \Sigma_{N,r}\r\|^2}{n^2 \l\| \Sigma_{N,r}\r\|_F^2} \nonumber\\
& \le  \frac{r \cdot \eta^2\l( Q_{\widetilde{\SU}_r,\lambda} M Q_{\widetilde{\SV}_r,\rho}^* \r) \l\| \Sigma_{N,r}\r\|^2}{ \sigma_{r}^2\l( \Sigma_{N,r}\r)} \nonumber\\
& = r \cdot \eta^2\l( Q_{\widetilde{\SU}_r,\lambda} M Q_{\widetilde{\SV}_r,\rho}^* \r) \kappa^2\l( Q_{\widetilde{\SU}_r,\lambda} M Q_{\widetilde{\SV}_r,\rho}^*\r),
\end{align}
where $\eta(\cdot)$ and $\kappa(\cdot)$ return the coherence and condition number of a matrix, respectively. In particular, the above inequalities hold with equality when $U_{N,r},V_{N,r}$ are columns of the Fourier basis and the condition number of $Q_{\widetilde{\SU}_r,\lambda} M Q_{\widetilde{\SV}_r,\rho}^*$ equals one. Since coherence and condition number are both never smaller than one, we conclude that the right-hand side of \eqref{eq:Nikhil's result} scales with $r^2$. This, in turn, forces $m$ (number of measurements) to scale with $r^2$. In contrast, the expected number of measurements required in Theorem \ref{thm:main result} scales linearly with $r$. We must note that \cite{rao2015collaborative} itself was preceded by \cite{negahban2012restricted} where, among other contributions, a weighted program for matrix completion was studied with diagonal $A$ and $B$ in Program  \eqref{eq:Nikhil}. A similar program was empirically studied in \cite{srebro2010collaborative} in the context of collaborative filtering.

We would like to point out that our interest in weighted matrix recovery was inspired by \cite{aravkin2014fast}, which we have previously mentioned in Section 1.1. In \cite{aravkin2014fast}, weighted nuclear norm minimization has been successfully but heuristically applied to a ``frequency continuation'' matrix completion problem in seismic signal processing.

The ideas in this work have a  precedent in \emph{compressive sensing} and, more generally,  sparse regression.
In \cite{friedlander2012recovering,mansour2011weighted}, weighted $\ell_1$ minimization was proposed in order to incorporate partial information about the support; this was actually the inspiration for the weighted matrix completion algorithm in \cite{aravkin2014fast}.
Originally, iterative reweighed $\ell_1$ minimization to enhance sparse recovery appeared in \cite{candes2008enhancing}. As our analysis in Sections \ref{sec:commons}-\ref{sec:analysis of mc w pi} indicates, extending these ideas to matrices requires different techniques and is substantially more involved.

\section*{Acknowledgements}

The authors took inspiration from \cite{aravkin2014fast}, and MBW is grateful to Felix Herrmann and the Seismic Laboratory for Imaging and Modeling (SLIM) for their hospitality during part of his sabbatical. Part of this research was conducted when AE was a graduate fellow at the Statistical and Applied Mathematical Sciences Institute (SAMSI) and later a visitor at the Institute for Computational and Experimental Research in Mathematics (ICERM). AE is grateful for their
hospitality and kindness, and also acknowledges Rachel Ward for helpful conversations about a closely related problem.

\section{Commons}
\label{sec:commons}

In this section, we collect the necessary technical tools that are common to the analysis of both Programs \eqref{eq:p1 MR} and \eqref{eq:p1}. \edit{In particular, we find canonical decompositions for $M_r$ and the estimation error that takes the prior knowledge $(\SUT_r,\SVT_r)$ into account by using standard tools from matrix analysis.}

\subsection{Canonical Decomposition}\label{sec:canonical decom}
Central to the analysis is a canonical way of decomposing $M_r$ that takes the prior knowledge $(\SUT_r,\SVT_r)$ into account.  This result is well-known and a short proof is given in Appendix \ref{sec:Proof of lemma canonical} for the sake of completeness \cite{golub2013matrix}. Throughout, the empty blocks of matrices should be interpreted as filled with zeros. Also, in our notation,  $I_a\in\real^{a\times a}$ is the identity matrix and $0_a\in\real^{a\times a}$ and $0_{a\times b}\in\real^{a\times b}$ are filled with zeros.

\begin{lem}
\label{lem:canonical}
Consider a rank-$r$ matrix $M_r\in\mathbb{R}^{n\times n}$, and let $\SU_r=\mbox{span}(M_r)$ be the column span of $M_r$. Let $\SUT_r$ be another $r$-dimensional subpsace in $\mathbb{R}^n$. Then, there exists $U_r,\wt{U}_r\in\mathbb{R}^{n\times r}$, $U_r',\wt{U}_r'\in \mathbb{R}^{n\times r}$, and $U_{n-2r}''\in\mathbb{R}^{n\times (n-2r)}$ such that
\begin{equation*}
\SU_r = \mbox{span}\l(U_r \r),
\quad
\SUT_r = \mbox{span}( \wt{U}_r ),
\end{equation*}
and
\begin{align}
B_L& :=\left[\begin{array}{ccc}
U_{r} & U'_{r} & U''_{n-2r}\end{array}\right]\in\mathbb{R}^{n\times n},
\nonumber\\
\widetilde{B}_L & :=\left[\begin{array}{ccc}
\widetilde{U}_{r} & \widetilde{U}'_{r} & U''_{n-2r}\end{array}\right]\in\mathbb{R}^{n\times n},
\label{eq:def of BL}
\end{align}
 are both orthonormal bases for $\mathbb{R}^{n}$. Moreover, it holds that
\begin{equation}
B_L^* \widetilde{B}_L=
\left[
\begin{array}{ccc}
\cos u & \sin u\\
-\sin u & \cos u\\
 &  & I_{n-2r}
\end{array}
\right],
\end{equation}
where  $u\in\mathbb{R}^{r\times r}$ is diagonal and contains the principal angles between ${\SU}_{r}$ and ${\SUT}_{r}$, in a non-increasing order: $\pi/2 \ge u_{1}\ge u_{2}\ge\cdots\ge u_{r}\ge0$.\footnote{Note that, prior to Section \ref{sec:commons}, we had used $u$ (rather than $u_1$) to denote the largest principal angle, in order to keep the notation light. From now on, we will adhere to setup of Lemma \ref{lem:canonical}.} The diagonal matrix $\cos u$ is naturally defined as
\[
\cos u :=
\l[
\begin{array}{cccc}
\cos u_1  \\
& \cos u_2 \\
& & \ddots \\
& & & \cos u_r
\end{array}
\r]
 \in\mathbb{R}^{r\times r},
\]
and $\sin u\in\mathbb{R}^{r\times r}$ is defined likewise.
A similar construction exists for  $\SV_r=\mbox{span}(M_r^*)$ and $\SVT_r$, where
we form the orthonormal bases $B_R,\wt{B}_R\in\mathbb{R}^{n\times n }$ such that
\begin{equation}
B_R^* \widetilde{B}_R=
\left[
\begin{array}{ccc}
\cos v & \sin v\\
-\sin v & \cos v\\
 &  & I_{n-2r}
\end{array}
\right].
\end{equation}
As before, the diagonal of $v\in\mathbb{R}^{r \times r}$ contains the principal angles between $\SV_r$ and $\SVT_r$ in non-decreasing order.
\end{lem}

Lemma \ref{lem:canonical} immediately implies that
\[
\widetilde{U}_{r}=B_{L}\left[\begin{array}{c}
\cos u\\
-\sin u\\
0_{(n-2r)\times r}
\end{array}\right],
\]
which, in turn, allows us to derive the following expressions for orthogonal projections onto the subspace $\SUT_r$ and its complement:
\begin{align*}
P_{\SUT_r}& = \wt{U}_r \wt{U}_r^* \nonumber\\
& = B_{L}\left[\begin{array}{ccc}
\cos^{2}u & -\sin u\cdot\cos u\\
-\sin u\cdot\cos u & \sin^{2}u\\
 &  & 0_{n-2r}
\end{array}\right]B_{L}^{*},
\end{align*}
\begin{align*}
P_{\SUT_r^\perp}
& = I_n - P_{\SUT_r} \nonumber\\
& =
B_{L}\left[\begin{array}{ccc}
\sin^{2}u & \sin u\cdot\cos u\\
\sin u\cdot\cos u & \cos^{2}u\\
 &  & I_{n-2r}
\end{array}\right]B_{L}^{*}.
\end{align*}
It also follows that
\begin{align}\label{eq:Q expression pre}
& Q_{\SUT_r,\lambda} \nonumber\\
& =\lambda\cdot P_{\SUT_r}+P_{\SUT_r^\perp}  \nonumber\\
& =B_{L}\nonumber\\
& \cdot \left[\begin{array}{ccc}
\lambda\cos^{2}u+\sin^{2}u & (1-\lambda)\sin u\cdot\cos u\\
(1-\lambda)\sin u\cdot\cos u & \lambda\sin^{2}u+\cos^{2}u\\
 &  & I_{n-2r}
\end{array}\right]B_{L}^{*},
\end{align}
where we used \eqref{eq:def of Qs}.
We next mold the above expression for $Q_{\SUT_r,\lambda}$ into one that involves
an upper-triangular matrix, as this will   prove useful shortly. Define the orthonormal basis $O_L\in\mathbb{R}^{n\times n}$ as
\begin{align*}
O_{L}
& :=\left[\begin{array}{c}
\left(\lambda\cos^{2}u+\sin^{2}u\right)\Delta_{L}^{-1} \\
(1-\lambda)\sin u\cdot\cos u\cdot\Delta_{L}^{-1} \\
  0_{(n-r)\times r} \end{array} \right.
\nonumber\\
& \qquad \qquad \left.
\begin{array}{cc}
 -(1-\lambda)\sin u\cdot\cos u\cdot\Delta_{L}^{-1} & 0_{r\times (n-r)} \\
 \left(\lambda\cos^{2}u+\sin^{2}u\right)\Delta_{L}^{-1} & 0_{r\times (n-r)}\\
0_{(n-r)\times r} &   I_{n-2r}
\end{array}\right],
\end{align*}
\begin{equation}
\Delta_{L}:=\sqrt{\lambda^{2}\cos^{2}u+\sin^{2}u}\in\mathbb{R}^{r\times r},
\label{eq:pre form of Q}
\end{equation}
where $\Delta_{L}$ is invertible because $\lambda>0$, by assumption. (It is easily verify that indeed $O_L O_L^*= I_n$.) We then rewrite \eqref{eq:Q expression pre} as
\begin{align}
& Q_{\SUT_r,\lambda} \nonumber\\
&
=B_{L}\left(O_L O_{L}^{*}\right) \nonumber\\
& \cdot \left[\begin{array}{ccc}
\lambda\cos^{2}u+\sin^{2}u & (1-\lambda)\sin u\cdot\cos u\\
(1-\lambda)\sin u\cdot\cos u & \lambda\sin^{2}u+\cos^{2}u\\
 &  & I_{n-2r}
\end{array}\right]B_{L}^{*}
\nonumber
\\
& = B_{L}O_{L}\left[\begin{array}{ccc}
\Delta_{L} & (1-\lambda^{2})\sin u\cdot\cos u\cdot\Delta_{L}^{-1}\\
 & \lambda\Delta_{L}^{-1}\\
 &  & I_{n-2r}
\end{array}\right]B_{L}^{*}\nonumber \\
 & =:B_{L}O_{L}\left[\begin{array}{ccc}
L_{11} & L_{12}\\
 & L_{22}\\
 &  & I_{n-2r}
\end{array}\right]B_{L}^{*}\nonumber \\
 & =:B_{L}O_{L}LB_{L}^{*},
 \label{eq:conn btw Q and L}
\end{align}
where $L\in\mathbb{R}^{n\times n}$ is an upper-triangular matrix with blocks $L_{11},L_{12},L_{22}\in \mathbb{R}^{r\times r}$ and defined as
\begin{align}
L & :=
\l[
\begin{array}{ccc}
L_{11} & L_{12} \\
& L_{22} \\
& & I_{n-2r}
\end{array}
\r] \nonumber\\
& =
\l[
\begin{array}{ccc}
\Delta_L & \l(1-\lambda^2\r) \sin u \cdot \cos u\cdot \Delta_L^{-1} \\
& \lambda \Delta_L^{-1} & \\
& & I_{n-2r}
\end{array}
\r]
.
\label{eq:def of Wu}
\end{align}
In the third line of \eqref{eq:conn btw Q and L}, we used the fact that $O_L O_L^* = I_n$.
Because $B_L,O_L$ are both orthonormal bases, we record
that
\begin{equation}
\left\| Q_{\SUT_r,\lambda} \right\| =\|L\|=1.
\qquad \mbox{(see \eqref{eq:def of Qs} and \eqref{eq:conn btw Q and L})}
\label{eq:norm of L}
\end{equation}
We can perform the same calculations for the row spaces and, in particular,
define $R\in\mathbb{R}^{n\times n}$ as
\begin{align}
R& :=\left[\begin{array}{ccc}
R_{11} & R_{12}\\
 & R_{22}\\
 &  & I_{n-2r}
\end{array}\right] \nonumber\\
& =\left[\begin{array}{ccc}
\Delta_{R} & (1-\rho^{2})\sin v\cdot\cos v\cdot\Delta_{R}^{-1}\\
 & \rho \Delta_{R}^{-1}\\
 &  & I_{n-2r}
\end{array}\right],\label{eq:def of R}
\end{align}
with $\Delta_{R}=\sqrt{\rho^{2}\cos^{2}v+\sin^{2}v}\in\mathbb{R}^{n\times n}$.
With these calculations in mind, for an arbitrary matrix $H\in\mathbb{R}^{n\times n}$,
we find the crucial decomposition
\begin{align}
& Q_{\SUT_r,\lambda}\cdot H\cdot Q_{\SVT_r,\rho} \nonumber\\
 & =B_{L}O_{L}L\left(B_{L}^{*}HB_{R}\right)R^{*}O_{R}^{*}B_{R}^{*}
\,\, \l(\mbox{\eqref{eq:conn btw Q and L} and } Q_{\SVT_r,\rho} =  Q_{\SVT_r,\rho}^* \r)
 \nonumber \\
 & =:B_{L}O_{L}L\overline{H}R^{*}O_{R}^{*}B_{R}^{*}
\qquad \left( \overline{H}:= B_L^*H B_R \right)
 \nonumber \\
 & =:B_{L}O_{L}L\left[\begin{array}{ccc}
\overline{H}_{11} & \overline{H}_{12} & \overline{H}_{13}\\
\overline{H}_{21} & \overline{H}_{22} & \overline{H}_{23}\\
\overline{H}_{31} & \overline{H}_{32} & \overline{H}_{33}
\end{array}\right]R^{*}O_{R}^{*}B_{R}^{*},\label{eq:H decompose}
\end{align}
with $\overline{H}_{11},\overline{H}_{22}\in\mathbb{R}^{r\times r}$
and $\overline{H}_{33}\in\mathbb{R}^{(n-2r)\times(n-2r)}$ being the
diagonal blocks of $\overline{H}$. Moreover, from Lemma~\ref{lem:canonical}, recall that
\begin{equation}
\mbox{span}\l( M_r\r) = \mbox{span} \l( U_r \r),\quad
\mbox{span}\l( M_r^*\r) = \mbox{span} \l( V_r \r),
\label{eq:M is single block}
\end{equation}
which allows us to record that
\begin{align}
& Q_{\SUT_r,\lambda}\cdot M_{r}\cdot Q_{\SVT_r,\rho} \nonumber\\
 & =B_{L}O_{L}L\left(B_{L}^{*}M_{r}B_{R}\right)R^{*}O_{R}^{*}B_{R}^{*}
\qquad \mbox{(see \eqref{eq:conn btw Q and L})}
 \nonumber \\
 & =:B_{L}O_{L}L\overline{M}_{r}R^{*}O_{R}^{*}B_{R}^{*}
\qquad \left( \overline{M}_r:=B_L^* M_r B_R \right)
 \nonumber \\
 & =:B_{L}O_{L}L\left[\begin{array}{cc}
\overline{M}_{r,11}\\
 & 0_{n-r}\\
\end{array}\right]R^{*}O_{R}^{*}B_{R}^{*}
\qquad \mbox{(see \eqref{eq:M is single block})}
\nonumber \\
 & =B_{L}O_{R}\left[\begin{array}{cc}
L_{11}\overline{M}_{r,11}R_{11}\\
 & 0_{n-r}
\end{array}\right]\nonumber\\
& \qquad \qquad \qquad \cdot O_{R}^{*}B_{R}^{*}.
\qquad \l(\mbox{\eqref{eq:def of Wu} and } R_{11}=R_{11}^* \r)
\label{eq:M decompose}
\end{align}
In the last line above, we
benefited from the fact that, by construction, both $L$ and $R$
are upper-triangular matrices.
Note also that $\overline{M}_{r,11}=U_r^* M_r V_r$, as defined above, is \emph{not} necessarily diagonal.  For future reference, the following useful inequalities are proved in Appendix \ref{sec:proof of Lemma prop of L}.
\begin{lem}\label{lem:props of L}
With $L$ and its blocks $L_{11},L_{12},L_{22}$ defined in (\ref{eq:def of Wu}), it holds that
\[
\left\Vert L_{11}\right\Vert = \left\| \Delta_L \right\| \le\sqrt{\lambda^{2}\cos^{2}u_1+\sin^{2}u_1},
\]
\begin{align*}
\|L_{12}\|
& \le  \frac{\left(1-\lambda^{2}\right)\sin u_1}{\sqrt{\lambda^{2}\cos^{2}u_1+\sin^{2}u_1}} \le 1-\lambda^2,
\end{align*}
\begin{align*}
\l\|I_{r}-L_{22}\r\|
& \le\frac{\sqrt{1-\lambda^{2}}\sin u_1}
{\sqrt{\lambda^{2}\cos^{2}u_1+\sin^{2}u_1}}
\le \sqrt{1-\lambda^2},
\end{align*}
\begin{align*}
\left\Vert \left[\begin{array}{cc}
L_{11} & L_{12}\end{array}\right]\right\Vert ^{2}  & \le \frac{\lambda^{4} \cos^2 u_1 +\sin^{2}u_{1}}{\lambda^{2} \cos^2 u_1+\sin^{2}u_{1}}\le 1,
\end{align*}
\begin{align}
\left\Vert \left[\begin{array}{cc}
L_{22}-I_{r} & L_{12}\end{array}\right]\right\Vert ^{2} & \le \frac{2(1-\lambda^{2})\sin^{2}u_{1}}{\lambda^{2} \cos^2 u_1 +\sin^{2}u_{1}},
\label{eq:b1}
\end{align}
where $u_1$ is the largest principal angle between $r$-dimensional subspaces $\SU_r$ and $\SUT_r$. Similar bounds hold for $R$ and its  blocks, $R_{11},R_{12},R_{22}$.
\end{lem}

\subsection{Support}

Let $M_r$ be a rank-$r$ truncation of $M\in\real^{n\times n}$ (obtained via SVD) and consider the decomposition
\[
M=M_{r}+M_{r^{+}}=U_{r}\overline{M}_{r,11}V_{r}^{*}+M_{r^{+}},
\]
where $U_{r},V_{r}\in\mathbb{R}^{n\times r}$ (with orthonormal columns) span column and row spaces of $M_r$,
and $\overline{M}_{11}\in\mathbb{R}^{r\times r}$ is rank-$r$ but
not necessary diagonal. Let $\SU_r=\mbox{span}(U_r)=\mbox{span}(M_r)$  and $\SV_r=\mbox{span}(V_r)=\mbox{span}(M_r^*)$. Then, the {\em support} of $M_{r}\in\mathbb{R}^{n\times n}$
is the linear subspace $\ST\subset\mathbb{R}^{n\times n}$ defined as
\begin{equation}
\ST=\left\{ Z\in\mathbb{R}^{n\times n}\,:\,Z=P_{\SU_r}\cdot Z+Z\cdot P_{\SV_r}-P_{\SU_r}\cdot Z\cdot P_{\SV_r}\right\},\label{eq:supp}
\end{equation}
where $P_{\SU_r},P_{\SV_r}\in\real^{n\times n}$ are orthogonal projection onto $\SU_r,\SV_r$, respectively.
For the record, the orthogonal projection onto $\ST$ and its complement
$\ST^{\perp}$ take $Z\in\mathbb{R}^{n\times n}$ to
\[
\mathcal{P}_{\ST}(Z)=P_{\SU_r}\cdot Z+Z\cdot P_{\SV_r}-P_{\SU_r}\cdot Z\cdot P_{\SV_r},
\]
\[
\mathcal{P}_{\ST^{\perp}}(Z)=P_{\SU_r^{\perp}}\cdot Z\cdot P_{\SV_r^\perp},
\]
respectively. As suggested above, throughout we reserve the calligraphic font for matrix operators.
Note that, using Lemma \ref{lem:canonical},  we can express $\ST$ equivalently as
\begin{align}
\ST
 & =\left\{ Z\in\mathbb{R}^{n\times n}\,:\,Z=B_{L} \overline{Z}
 B_{R}^{*},\quad
 \overline{Z}=
 \left[\begin{array}{cc}
\overline{Z}_{11} & \overline{Z}_{12}
\nonumber\\
\overline{Z}_{21} & 0_{n-r}
\end{array}\right]
 \right\}
  \\
 & =:B_{L}\cdot\overline{\ST}\cdot B_{R}^{*},
\label{eq:T n Tbar}
\end{align}
where, to be clear, the new subspace $\overline{\ST}\subset\mathbb{R}^{n\times n}$ is the support of $\overline{M}_r := B_L^*M_rB_R$ and
is defined as
\begin{equation}
\overline{\ST}=\left\{ \overline{Z}\in\mathbb{R}^{n\times n}\,:\,\overline{Z}=\left[\begin{array}{cc}
\overline{Z}_{11} & \overline{Z}_{12}\\
\overline{Z}_{21} & 0_{n-r}
\end{array}\right]
\right\}.\label{eq:T bar}
\end{equation}
Also note that, for arbitrary
\[
\overline{Z}=\left[\begin{array}{cc}
\overline{Z}_{11} & \overline{Z}_{12}\\
\overline{Z}_{21} & \overline{Z}_{22}
\end{array}\right]\in\mathbb{R}^{n\times n},
\]
with $\overline{Z}_{11}\in\mathbb{R}^{r\times r},\overline{Z}_{22}\in\mathbb{R}^{(n-r)\times(n-r)}$,
the orthogonal projection onto $\overline{\ST}$ and its complement  simply take $\overline{Z}$ to
\begin{equation*}
\mathcal{P}_{\overline{\ST}}\left(\overline{Z}\right)=\left[\begin{array}{cc}
\overline{Z}_{11} & \overline{Z}_{12}\\
\overline{Z}_{21} & 0_{n-r}
\end{array}\right],
\end{equation*}
\begin{equation}
\mathcal{P}_{\overline{\ST}^{\perp}}\left(\overline{Z}\right)=\left[\begin{array}{cc}
0_{r}\\
 & \overline{Z}_{22}
\end{array}\right],\label{eq:PT PTperp}
\end{equation}
respectively.
Lastly, we record the following connection: For arbitrary $Z\in\mathbb{R}^{n\times n}$ and with $\overline{Z}=B_L^* Z B_R$, we have that
\begin{equation*}
\mathcal{P}_{\ST}(Z)=B_{L}\cdot\mathcal{P}_{\overline{\ST}}\left(\overline{Z}\right)\cdot B_{R}^{*},
\end{equation*}
\begin{equation}
\mathcal{P}_{\ST^{\perp}}(Z)=B_{L}\cdot\mathcal{P}_{\overline{\ST}^{\perp}}\left(\overline{Z}\right)\cdot B_{R}^{*}.
\label{eq:connection btw projections}
\end{equation}

\section{Analysis for Matrix Recovery}
\label{sec:analysis of mr w pi}
In this section, we study Program \eqref{eq:p1 MR} in detail and eventually prove Theorem \ref{thm:main result MR}. First, in Section \ref{sec:null space prop}, we establish a variant of the well-known \emph{nullspace property} for Program \eqref{eq:p1 MR} which loosely states that the recovery error in Program \eqref{eq:p1 MR} is concentrated along the subspace $\ST$, namely the support of $M_r$. \edit{Using this property, we then complete the proof of Theorem~\ref{thm:main result MR} in Section~\ref{sec:body MR} by breaking down the error into smaller components to which we can apply the RIP.}

\subsection{Nullspace Property}
\label{sec:null space prop}

For solution $\widehat{M}$,  let $H:=\widehat{M}-M$ be the error.
By feasibility of $M$ and optimality of $\widehat{M}=M+H$ in Program \eqref{eq:p1 MR},
we have that
\begin{equation}
\left\Vert Q_{\wt{\SU}_r,\lambda}\left(M+H\right)Q_{\widetilde{\SV}_r,\rho}\right\Vert _{*}\le\left\Vert Q_{\widetilde{\SU}_r,\lambda}M Q_{\widetilde{\SV}_r,\rho}\right\Vert _{*}.\label{eq:optimality cond}
\end{equation}
Recall that $M=M_{r}+M_{r^+}$. With the decomposition  of $M_r$ in (\ref{eq:M decompose})
at hand, the right-hand side above is then bounded follows:
\begin{align}
& \left\Vert Q_{\widetilde{\SU}_r,\lambda}MQ_{\widetilde{\SV}_r,\rho}\right\Vert _{*} \nonumber\\
& \le\left\Vert Q_{\widetilde{\SU}_r,\lambda}M_{r}Q_{\widetilde{\SV}_r,\rho}\right\Vert _{*}+\left\Vert Q_{\widetilde{\SU}_r,\lambda} M_{r^+} Q_{\widetilde{\SV}_r,\rho}\right\Vert _{*}
\nonumber
\\
 & =\left\Vert B_{L}O_{L}^{*}\left[\begin{array}{cc}
L_{11}\overline{M}_{r,11}R_{11}
\\
   & 0_{n-r}
\end{array}\right]O_{R}^{*}B_{R}^{*}\right\Vert _{*} \nonumber\\
& \qquad \qquad +\left\Vert Q_{\widetilde{\SU}_r,\lambda}M_{r^+}Q_{\widetilde{\SV}_r,\rho}\right\Vert _{*}
\qquad \mbox{(see \eqref{eq:M decompose})}
\nonumber
\\
 & =\left\Vert \left[\begin{array}{cc}
L_{11}\overline{M}_{r,11}R_{11}\\
   & 0_{n-r}
\end{array}\right]\right\Vert _{*}
+ \left\Vert Q_{\widetilde{\SU}_r,\lambda}M_{r^+}Q_{\widetilde{\SV}_r,\rho}\right\Vert _{*}
\nonumber\\
& = \l\|   L\overline{M} R^* \r\|_* +
\left\Vert Q_{\widetilde{\SU}_r,\lambda}M_{r^+}Q_{\widetilde{\SV}_r,\rho}\right\Vert _{*},
\qquad \mbox{(see \eqref{eq:M decompose})}
\label{eq:1st opt bnd}
\end{align}
where the first inequality above uses $M=M_r+M_{r^+}$ and the triangle inequality. The second identity uses the rotational invariance of the nuclear norm.
In the last line, we used the fact that $B_L,B_R,O_L,O_R$ are all orthonormal bases.
Using the decomposition of $H$ in (\ref{eq:H decompose}),
the left hand side  of \eqref{eq:opt cond} can also be bounded as follows:
\begin{align*}
 & \left\Vert Q_{\widetilde{\SU}_r,\lambda}\left(M+H\right)Q_{\widetilde{\SV},\rho}\right\Vert _{*}\nonumber \\
 & \ge\left\Vert Q_{\widetilde{\SU}_r,\lambda}\left(M_{r}+H\right)Q_{\widetilde{\SV}_r,\rho}\right\Vert _{*} -\left\Vert Q_{\widetilde{\SU}_r,\lambda}M_{r^+}Q_{\widetilde{\SV}_r,\rho}\right\Vert _{*}
 \nonumber \\
 & =\left\Vert B_{L}O_{L}L\left(\overline{M}_{r}+\overline{H}\right)R^{*}O_{R}^{*}B_{R}^{*}\right\Vert _{*}-\left\Vert Q_{\widetilde{\SU}_r,\lambda}M_{r^+}Q_{\widetilde{\SV}_r,\rho}\right\Vert _{*}
 \nonumber \\
 & =\left\Vert L\overline{M}_{r}R^{*}+L\overline{H}R^{*}\right\Vert _{*}-\left\Vert Q_{\widetilde{\SU}_r,\lambda}M_{r^+}Q_{\widetilde{\SV}_r,\rho}\right\Vert _{*}
 \nonumber \\
 & =\left\Vert L\overline{M}_{r}R^{*}+L\mathcal{P}_{\overline{\ST}}\left(\overline{H}\right)R^{*}+L\mathcal{P}_{\overline{\ST}^{\perp}}\left(\overline{H}\right)R^{*}\right\Vert _{*} \nonumber\\
& \qquad  -\left\Vert Q_{\widetilde{\SU}_r,\lambda}M_{r^+}Q_{\widetilde{\SV}_r,\rho}\right\Vert _{*}\nonumber \\
 & =\left\Vert L\overline{M}_{r}R^{*}+L\mathcal{P}_{\overline{\ST}}\left(\overline{H}\right)R^{*}+L\mathcal{P}_{\overline{\ST}^{\perp}}\left(\overline{H}\right)R^{*}
\right. \nonumber\\
& \qquad \qquad \left.
 -\mathcal{P}_{\overline{\ST}^{\perp}}\left(\overline{H}\right)+\mathcal{P}_{\ST^{\perp}}\left(\overline{H}\right)\right\Vert _{*}-\left\Vert Q_{\widetilde{\SU}_r,\lambda}M_{r^+}Q_{\widetilde{\SV}_r,\rho}\right\Vert _{*}.
 \end{align*}
Above, the second line uses $M=M_r+M_{r^+}$ and the the triangle inequality. The third line uses \eqref{eq:H decompose} and \eqref{eq:M decompose}. The fourth line uses the rotational invariance of the nuclear norm.  We continue by writing that
\begin{align*}
& \left\Vert Q_{\widetilde{\SU}_r,\lambda}\left(M+H\right)Q_{\widetilde{\SV},\rho}\right\Vert _{*} \nonumber\\
 & =\left\Vert L\overline{M}_{r}R^{*}+L\mathcal{P}_{\overline{\ST}}\left(\overline{H}\right)R^{*}+L\left[\begin{array}{ccc}
0_{r}\\
 & \overline{H}_{22} & \overline{H}_{23}\\
 & \overline{H}_{32} & \overline{H}_{33}
\end{array}\right]R^{*}
\right. \nonumber\\
& \qquad \qquad \left. -\left[\begin{array}{ccc}
0_{r}\\
 & \overline{H}_{22} & \overline{H}_{23}\\
 & \overline{H}_{32} & \overline{H}_{33}
\end{array}\right]+\mathcal{P}_{\overline{\ST}^{\perp}}\left(\overline{H}\right)\right\Vert _{*}
\nonumber\\
& \qquad \qquad
-\left\Vert Q_{\widetilde{\SU}_r,\lambda}M_{r^+}Q_{\widetilde{\SV}_r,\rho}\right\Vert _{*}
\quad \mbox{(see \eqref{eq:PT PTperp})}
\nonumber \\
 & =:\left\Vert L\overline{M}_{r}R^{*}+L\mathcal{P}_{\overline{\ST}}\left(\overline{H}\right)R^{*}+L\overline{H}'R^{*}-\overline{H}'+\mathcal{P}_{\overline{\ST}^{\perp}}\left(\overline{H}\right)\right\Vert _{*}
 \nonumber\\
& \qquad  -\left\Vert Q_{\widetilde{\SU}_r,\lambda}M_{r^+}Q_{\widetilde{\SV}_r,\rho}\right\Vert _{*}
\qquad \mbox{(see \eqref{eq:def of H'})}
 \nonumber \\
 & \ge\left\Vert L\overline{M}_{r}R^{*}+\mathcal{P}_{\overline{\ST}^{\perp}}\left(\overline{H}\right)\right\Vert _{*}-\left\Vert L\mathcal{P}_{\overline{\ST}}\left(\overline{H}\right)R^{*}\right\Vert _{*}
\nonumber\\
& \qquad  -\left\Vert L\overline{H}'R^{*}-\overline{H}'\right\Vert _{*}-\left\Vert Q_{\widetilde{\SU}_r,\lambda}M_{r^+}Q_{\widetilde{\SV},\rho}\right\Vert _{*},
\end{align*}
where the last line uses the triangle inequality. To be concrete, above we defined
\begin{equation}
\overline{H}':=\left[\begin{array}{ccc}
0_{r}\\
 & \overline{H}_{22} & \overline{H}_{23}\\
 & \overline{H}_{32} & 0_{n-2r}
\end{array}\right]\in\mathbb{R}^{n\times n}.\label{eq:def of H'}
\end{equation}
Consequently,
\begin{align}
& \left\Vert Q_{\widetilde{\SU}_r,\lambda}\left(M+H\right)Q_{\widetilde{\SV},\rho}\right\Vert _{*} \nonumber \\
 & =\left\Vert \left[\begin{array}{ccc}
L_{11}\overline{M}_{r,11}R_{11}\\
 & \overline{H}_{22} & \overline{H}_{23}\\
 & \overline{H}_{32} & \overline{H}_{33}
\end{array}\right]\right\Vert _{*}-\left\Vert L\mathcal{P}_{\overline{\ST}}\left(\overline{H}\right)R^{*}\right\Vert _{*} \nonumber\\
& \qquad\qquad  -\left\Vert L\overline{H}'R^{*}-\overline{H}'\right\Vert _{*}-\left\Vert Q_{\widetilde{\SU}_r,\lambda}M_{r^+} Q_{\widetilde{\SV}_r,\rho}\right\Vert _{*}
\nonumber \\
 & =\left\Vert \left[\begin{array}{ccc}
L_{11}\overline{M}_{r,11}R_{11}\\
 & 0_{r}\\
 &  & 0_{n-2r}
\end{array}\right]\right\Vert _{*}\nonumber\\
& \qquad +\left\Vert \left[\begin{array}{ccc}
0_{r}\\
 & \overline{H}_{22} & \overline{H}_{23}\\
 & \overline{H}_{32} & \overline{H}_{33}
\end{array}\right]\right\Vert _{*}-\left\Vert L\mathcal{P}_{\overline{\ST}}\left(\overline{H}\right)R^{*}\right\Vert _{*}
\nonumber\\
& \qquad -\left\Vert L\overline{H}'R^{*}-\overline{H}'\right\Vert _{*}   -\left\Vert Q_{\widetilde{\SU}_r,\lambda}M_{r^+}Q_{\widetilde{\SV}_r,\rho}\right\Vert _{*}  \nonumber\\
 & = \left\Vert L\overline{M}R^{*}\right\Vert _{*}+\left\Vert \mathcal{P}_{\overline{\ST}^{\perp}}\left(\overline{H}\right)\right\Vert _{*}-\left\Vert L\mathcal{P}_{\overline{\ST}}\left(\overline{H}\right)R^{*}\right\Vert _{*}
\nonumber\\
&  \qquad  -\left\Vert L\overline{H}'R^{*}-\overline{H}'\right\Vert _{*}-\left\Vert Q_{\widetilde{\SU}_r,\lambda}M_{r^+}Q_{\widetilde{\SV}_r,\rho}\right\Vert _{*}.
 \label{eq:le 0}
\end{align}
Above, we used \eqref{eq:M decompose} and \eqref{eq:PT PTperp} twice.
 In the second identity in (\ref{eq:le 0}),
we used the fact that $\|A+B\|_{*}=\|A\|_{*}+\|B\|_{*}$ whenever
both column and row spaces of $A$ are orthogonal to those of $B$, namely when  $A^*B=AB^*=0$.
Now combining (\ref{eq:optimality cond}) with the bounds in \eqref{eq:1st opt bnd} and \eqref{eq:le 0}  yields that
\begin{align}
\left\Vert \mathcal{P}_{\overline{\ST}^{\perp}}\left(\overline{H}\right)\right\Vert _{*} & \le\left\Vert L\mathcal{P}_{\overline{\ST}}\left(\overline{H}\right)R^{*}\right\Vert _{*}+\left\Vert L\overline{H}'R^{*}-\overline{H}'\right\Vert _{*}\nonumber\\
&\qquad \qquad  +2\left\Vert Q_{\widetilde{\SU}_r,\lambda}M_{r^+}Q_{\widetilde{\SV}_r,\rho}\right\Vert _{*}.\label{eq:interm}
\end{align}
We next simplify the terms  in the above inequality.  
 First, notice that
\begin{align}
& \left[
\begin{array}{ccc}
0_{r} & &\\
 & L_{22} &\\
 &  & I_{n-2r}
\end{array}
\right]
\mathcal{P}_{\overline{\ST}}\left(\overline{H}\right)
\left[
\begin{array}{ccc}
0_r & &\\
 & R_{22} &\\
 &  & I_{n-2r}
\end{array}
\right]
\nonumber\\
& =
\left[
\begin{array}{ccc}
0_{r} & &\\
 & L_{22} &\\
 &  & I_{n-2r}
\end{array}
\right]
\l[
\begin{array}{ccc}
\overline{H}_{11} & \overline{H}_{12} & \overline{H}_{13} \\
\overline{H}_{21} & 0_r & \\
\overline{H}_{31} & & 0_{n-2r}
\end{array}
\r]
\nonumber\\
& \qquad \qquad \cdot \left[
\begin{array}{ccc}
0_r & &\\
 & R_{22} &\\
 &  & I_{n-2r}
\end{array}
\right]
\qquad \mbox{(see \eqref{eq:PT PTperp})}
\nonumber\\
& = 0_n,
\label{eq:null term}
\end{align}
which, in turn, allows us to simplify the first norm on the right-hand side of \eqref{eq:interm} as follows:
\begin{align*}
 & \left\Vert L\mathcal{P}_{\overline{\ST}}\left(\overline{H}\right)R^{*}\right\Vert _{*}\nonumber \\
 & =\left\Vert \left[\begin{array}{ccc}
L_{11} & L_{12}\\
 & L_{22}\\
 &  & I_{n-2r}
\end{array}\right]\mathcal{P}_{\overline{\ST}}\left(\overline{H}\right)\left[\begin{array}{ccc}
R_{11}\\
R_{12}^{*} & R_{22}\\
 &  & I_{n-2r}
\end{array}\right] \right.
\nonumber\\
&  \left.  -
\left[\begin{array}{ccc}
0_{r}\\
 & L_{22}\\
 &  & I_{n-2r}
\end{array}\right]\mathcal{P}_{\overline{\ST}}\left(\overline{H}\right)\left[\begin{array}{ccc}
0_{r}\\
 & R_{22}\\
 &  & I_{n-2r}
\end{array}\right]\right\Vert _{*},
\end{align*}
where we used \eqref{eq:null term}. Then we continue by writing that
\begin{align*}
& \left\Vert L\mathcal{P}_{\overline{\ST}}\left(\overline{H}\right)R^{*}\right\Vert _{*} \nonumber\\
 & = \left\Vert \left[\begin{array}{ccc}
L_{11} & L_{12}\\
 & 0_{r}\\
 &  & 0_{n-2r}
\end{array}\right]\mathcal{P}_{\overline{\ST}}\left(\overline{H}\right)R^*
\right. \nonumber\\
&  \left.
+
 \left[\begin{array}{ccc}
0_{r}\\
 & L_{22}\\
 &  & I_{n-2r}
\end{array}\right]\mathcal{P}_{\overline{\ST}}\left(\overline{H}\right)\left[\begin{array}{ccc}
R_{11}\\
R_{12}^{*} & 0_{r}\\
 &  & 0_{n-2r}
\end{array}\right]\right\Vert _{*},
\end{align*}
in which we applied \eqref{eq:polarization}. Consequently,
\begin{align}
& \left\Vert L\mathcal{P}_{\overline{\ST}}\left(\overline{H}\right)R^{*}\right\Vert _{*} \nonumber \\
 & \le\left\Vert \left[\begin{array}{ccc}
L_{11} & L_{12}\\
 & 0_{r}\\
 &  & 0_{n-2r}
\end{array}\right]\mathcal{P}_{\overline{\ST}}\left(\overline{H}\right)R^*\right\Vert _{*}
\nonumber\\
&  +\left\Vert \left[\begin{array}{ccc}
0_{r}\\
 & L_{22}\\
 &  & I_{n-2r}
\end{array}\right]\mathcal{P}_{\overline{\ST}}\left(\overline{H}\right)\left[\begin{array}{ccc}
R_{11}\\
R_{12}^{*} & 0_{r}\\
 &  & 0_{n-2r}
\end{array}\right]\right\Vert _{*}
\nonumber \\
 & \le\left\Vert \left[\begin{array}{cc}
L_{11} & L_{12}\end{array}\right]\right\Vert \left\Vert \mathcal{P}_{\overline{\ST}}\left(\overline{H}\right)\right\Vert _{*}\|R\|
\nonumber\\
& \qquad +\max\left[\|L_{22}\|,1\right]\left\Vert \mathcal{P}_{\overline{\ST}}\left(\overline{H}\right)\right\Vert _{*}\left\Vert \left[\begin{array}{cc}
R_{11} & R_{12}\end{array}\right]\right\Vert
\nonumber \\
 & \le\left\Vert \left[\begin{array}{cc}
L_{11} & L_{12}\end{array}\right]\right\Vert \left\Vert \mathcal{P}_{\overline{\ST}}\left(\overline{H}\right)\right\Vert _{*}
\nonumber\\
& \qquad \qquad +\left\Vert \mathcal{P}_{\overline{\ST}}\left(\overline{H}\right)\right\Vert _{*}\left\Vert \left[\begin{array}{cc}
R_{11} & R_{12}\end{array}\right]\right\Vert
\nonumber \\
 & =\left(\left\Vert \left[\begin{array}{cc}
L_{11} & L_{12}\end{array}\right]\right\Vert +\left\Vert \left[\begin{array}{cc}
R_{11} & R_{12}\end{array}\right]\right\Vert \right)\left\Vert \mathcal{P}_{\overline{\ST}}\left(\overline{H}\right)\right\Vert _{*}.\label{eq:leg 1}
\end{align}
The second inequality above uses the fact that $\|AB\|_* \le \|A\|\cdot  \|B\|_*$ for all conforming matrices $A,B$.
The last inequality uses \eqref{eq:def of Wu}, \eqref{eq:norm of L}, and  the observation that  $\l\| L_{22}\r\| \le \|L\|$.  In the second inequality, we also used the so-called polarization identity
\begin{equation}
AZC-BZD=(A-B)ZC+BZ(C-D),
\label{eq:polarization}
\end{equation}
for conforming matrices $A,B,C,D,Z$.
The second norm on the right-hand side of (\ref{eq:interm}) may also be bounded
as follows:
\begin{align*}
 & \left\Vert L\overline{H}'R^{*}-\overline{H}'\right\Vert _{*}\nonumber \\
 & =\left\Vert \left[\begin{array}{ccc}
L_{11} & L_{12}\\
 & L_{22}\\
 &  & I_{n-2r}
\end{array}\right]\overline{H}'\left[\begin{array}{ccc}
R_{11}\\
R_{12}^{*} & R_{22}\\
 &  & I_{n-2r}
\end{array}\right]
\right. \nonumber\\
& \qquad \left. -\left[\begin{array}{ccc}
L_{11}\\
 & I_{r}\\
 &  & I_{n-2r}
\end{array}\right]\overline{H}'\left[\begin{array}{ccc}
R_{11}\\
 & I_{r}\\
 &  & I_{n-2r}
\end{array}\right]\right\Vert _{*}
\nonumber \\
 & =\left\Vert \left[\begin{array}{ccc}
0_{r} & L_{12}\\
 & L_{22}-I_{r}\\
 &  & 0_{n-2r}
\end{array}\right]\overline{H}'R^{*}
\right. \nonumber\\
& \left. \,
- \left[\begin{array}{ccc}
L_{11}\\
 & I_{r}\\
 &  & I_{n-2r}
\end{array}\right]\overline{H}'\left[\begin{array}{ccc}
0_{r}\\
R_{12}^{*} & R_{22}-I_{r}\\
 &  & 0_{n-2r}
\end{array}\right]\right\Vert _{*},
\end{align*}
Above, the first identity  uses \eqref{eq:def of Wu}. and \eqref{eq:def of H'}.  The second identity employs \eqref{eq:polarization}. We continue by applying the triangle inequality to find that
\begin{align*}
& \left\Vert L\overline{H}'R^{*}-\overline{H}'\right\Vert _{*}\nonumber \\
 & \le\left\Vert \left[\begin{array}{ccc}
0_{r} & L_{12}\\
 & L_{22}-I_{r}\\
 &  & 0_{n-2r}
\end{array}\right]\overline{H}'R^{*}\right\Vert _{*}
\nonumber\\
&  +\left\Vert \left[\begin{array}{ccc}
L_{11}\\
 & I_{r}\\
 &  & I_{n-2r}
\end{array}\right]\overline{H}'\left[\begin{array}{ccc}
0_{r}\\
R_{12}^{*} & R_{22}-I_{r}\\
 &  & 0_{n-2r}
\end{array}\right]\right\Vert _{*},
\end{align*}
 and, consequently,
 \begin{align}
 & \left\Vert L\overline{H}'R^{*}-\overline{H}'\right\Vert _{*}\nonumber \\
 & \le\left\Vert \left[\begin{array}{cc}
L_{12}^* & L_{22}-I_{r}\end{array}\right]\right\Vert \left\Vert \overline{H}'\right\Vert _{*}\|R\|\nonumber\\
& \qquad +\max\left[\|L_{11}\|,1\right]\left\Vert \overline{H}'\right\Vert _{*}\left\Vert \left[\begin{array}{cc}
R_{12}^* & R_{22}-I_{r}\end{array}\right]\right\Vert
 \nonumber \\
 & \le\left\Vert \left[\begin{array}{cc}
L_{12}^* & L_{22}-I_{r}\end{array}\right]\right\Vert \left\Vert \overline{H}'\right\Vert _{*} \nonumber\\
&\qquad +\left\Vert \overline{H}'\right\Vert _{*}\left\Vert \left[\begin{array}{cc}
R_{12}^* & R_{22}-I_{r}\end{array}\right]\right\Vert
\qquad
\nonumber \\
 & =\left(\left\Vert \left[\begin{array}{cc}
L_{12} & L_{22}-I_{r}\end{array}\right]\right\Vert +\left\Vert \left[\begin{array}{cc}
R_{12} & R_{22}-I_{r}\end{array}\right]\right\Vert \right)\left\Vert \overline{H}'\right\Vert _{*}.
\label{eq:leg 2}
\end{align}
The first inequality above uses $\|AB\|_*\le \|A\| \cdot \|B\|_*$. The second inequality applies \eqref{eq:def of Wu}, \eqref{eq:norm of L},  and the fact that  $\l\| L_{11}\r\| \le \|L\|$. The last line benefits from the fact that  $L_{12}=L_{12}^* $.  By substituting (\ref{eq:leg 1}) and (\ref{eq:leg 2}) back into (\ref{eq:interm}),
we arrive at the following inequality:
\begin{align}
 & \left\Vert \mathcal{P}_{\overline{\ST}^{\perp}}\left(\overline{H}\right)\right\Vert _{*}\nonumber \\
 & \le\left(\left\Vert \left[\begin{array}{cc}
L_{11} & L_{12}\end{array}\right]\right\Vert
+\left\Vert \left[\begin{array}{cc}
R_{11} & R_{12}\end{array}\right]\right\Vert \right)\left\Vert \mathcal{P}_{\overline{\ST}}\left(\overline{H}\right)\right\Vert _{*}
\nonumber\\
& \, +\left(\left\Vert \left[\begin{array}{cc}
L_{12} & L_{22}-I_{r}\end{array}\right]\right\Vert +\left\Vert \left[\begin{array}{cc}
R_{12} & R_{22}-I_{r}\end{array}\right]\right\Vert \right)\left\Vert \overline{H}'\right\Vert _{*}\nonumber \\
 & \qquad+2\left\Vert Q_{\widetilde{\SU}_r,\lambda}M_{r^+}Q_{\widetilde{\SV}_r,\rho}\right\Vert _{*}\nonumber \\
 & \le \l( \sqrt{\frac{\lambda^{4}\cos^{2}u_1+\sin^{2}u_1}{\lambda^{2}\cos^{2}u_1+\sin^{2}u_1}}
+
 \sqrt{\frac{\rho^{4}\cos^{2}v_1+\sin^{2}v_1}{\rho^{2}\cos^{2}v_1+\sin^{2}v_1}}
 \r) \nonumber\\
& \qquad \cdot  \left\Vert \mathcal{P}_{\overline{\ST}}\left(\overline{H}\right)\right\Vert _{*}  +
 \Bigg(
 \sqrt{\frac{2(1-\lambda^{2})\sin^{2}u_1}{\lambda^{2}\cos^{2}u_1+\sin^{2}u_1}}
 \nonumber\\
&  +\sqrt{\frac{2(1-\rho^{2})\sin^{2}v_1}{\rho^{2}\cos^{2}v_1+\sin^{2}v_1}}
\Bigg)
 \cdot \left\Vert \overline{H}'\right\Vert _{*}+2\left\Vert Q_{\widetilde{\SU}_r,\lambda}M_{r^+}Q_{\widetilde{\SV},\rho}\right\Vert _{*}
 \nonumber \\
 & =: {\Cr{1}}(u_1,v_1,\lambda,\rho)\left\Vert \mathcal{P}_{\overline{\ST}}\left(\overline{H}\right)\right\Vert _{*}+\Cr{2} (u_1,v_1,\lambda,\rho)\left\Vert \overline{H}'\right\Vert _{*} \nonumber\\
&\qquad  +2\left\Vert Q_{\widetilde{\SU}_r,\lambda}M_{r^+}Q_{\widetilde{\SV}_r,\rho}\right\Vert _{*}.\label{eq:new ni}
\end{align}
In the second inequality above, Lemma \ref{lem:props of L} is applied.
Some additional manipulation of  \eqref{eq:new ni} is in order.
First, owing to \eqref{eq:connection btw projections} and the rotational invariance of the nuclear norm, it holds that
\begin{equation*}
\left\Vert \mathcal{P}_{\ST}\left(H\right)\right\Vert _{*}=\left\Vert \mathcal{P}_{\overline{\ST}}\left(\overline{H}\right)\right\Vert _{*},
\end{equation*}
\begin{equation}
\left\Vert \mathcal{P}_{\ST^{\perp}}\left(H\right)\right\Vert _{*}=\left\Vert \mathcal{P}_{\overline{\ST}^{\perp}}\left(\overline{H}\right)\right\Vert _{*}.
\label{eq:manipulation 1}
\end{equation}
If we also  define the linear subspace $\widetilde{\ST}\subset \ST^{\perp}\subset\mathbb{R}^{n\times n}$
as
\begin{equation}
\widetilde{\ST}:=\left\{ Z\in\mathbb{R}^{n\times n}\,:\, Z=B_{L}\left[\begin{array}{ccc}
0_{r}\\
 & \overline{Z}_{22} & \overline{Z}_{23}\\
 & \overline{Z}_{32} & 0_{n-2r}
\end{array}\right]B_{R}^{*}
\r\},
\label{eq:T tilde}
\end{equation}
then we may write that
\begin{equation}
\left\Vert \overline{H}'\right\Vert _{*}=\left\Vert B_{L}\overline{H}'B_{R}^{*}\right\Vert _{*}=\left\Vert \mathcal{P}_{\widetilde{\ST}}(H)\right\Vert _{*},
\label{eq:manipulation 2}
\end{equation}
by rotational invariance of the nuclear norm and  in light of \eqref{eq:def of H'}.
Putting these all together, we may rewrite  \eqref{eq:new ni} as
\begin{align}
\label{eq:nsp final}
\left\Vert \mathcal{P}_{\ST^{\perp}}\left(H\right)\right\Vert _{*}
& \le\Cr{1}\left\Vert \mathcal{P}_{\ST}\left(H\right)\right\Vert _{*}+\Cr{2}\left\Vert \mathcal{P}_{\widetilde{\ST}}(H)\right\Vert _{*}
\nonumber\\
& +2\left\Vert Q_{\widetilde{\SU}_r,\lambda}M_{r^+}Q_{\widetilde{\SV}_r,\rho}\right\Vert _{*}
\quad \mbox{(see \eqref{eq:manipulation 1} and \eqref{eq:manipulation 2})}
\nonumber\\
& \le \Cr{1}\left\Vert \mathcal{P}_{\ST}\left(H\right)\right\Vert _{*}+\Cr{2}\left\Vert \mathcal{P}_{\widetilde{\ST}}(H)\right\Vert _{*}+2\left\Vert M_{r^+}\right\Vert _{*},
\end{align}
with $\Cr{1},\Cr{2}$ as defined in \eqref{eq:new ni}. The last line above uses the inequality $\|AB\|_*\le \|A\|\cdot \|B\|_*$  and  \eqref{eq:norm of L}.
Note that \eqref{eq:nsp final} might be interpreted as an analog of the nullspace property in standard matrix recovery \cite{recht2010guaranteed}. In particular, suppose that $M=M_r$ is rank-$r$ and $\lambda=\rho=1$ so that Program \eqref{eq:p1 MR} reduces to  Program \eqref{eq:MR program}. Then, in turn, \eqref{eq:nsp final} reduces to $\|\mathcal{P}_{\ST^\perp}(H)\|_* \le 2 \|\mathcal{P}_{\ST}(H)\|_*$ which is by a factor of two worse than the standard nullspace property.\footnote{The extra factor of two is likely an artifact of using the polarization identity.} With \eqref{eq:nsp final} at hand, we are now prepared to prove Theorem \ref{thm:main result MR}.

\subsection{Body of the Analysis}
\label{sec:body MR}

Let $\mathcal{P}_{\ST^{\perp}}(H)=U'\Sigma'(V')^{*}$ be the SVD
of $\mathcal{P}_{\ST^{\perp}}(H)$ with $U',V'\in\mathbb{R}^{n\times(n-r)}$ and where  the diagonal matrix $\Sigma'\in\mathbb{R}^{(n-r)\times(n-r)}$ contains
the singular values of $P_{\ST^{\perp}}(H)$, in a non-increasing order.
We partition the singular values into groups of size $r'$ as follows, with integer
$r'$ to be set later. Using  MATLAB's matrix notation, we form
\[
\Sigma_{i}=\Sigma'[(i-1)r'+1:ir',(i-1)r'+1:ir']\in\mathbb{R}^{r'\times r'},
\]
\[
U_{i}:=U'[:,(i-1)r'+1:ir']\in\mathbb{R}^{n\times r'},
\]
\[
V_{i}:=V'[:,(i-1)r'+1:ir']\in\mathbb{R}^{n\times r'},
\]
\[
H_i := U'_i \Sigma_i \l(V_i'\r)^*\in \real^{n\times n},
\]
for $i\ge1$.\footnote{The four last blocks might be smaller  than others.}
To unburden the notation, we also set $H_0 = \mathcal{P}_{\ST}(H)$. This
setup allows us to decompose the error  $H$ as
\begin{equation}
H=\mathcal{P}_{\ST}(H)+\mathcal{P}_{\ST^{\perp}}(H)=\sum_{i\ge0}H_i.\label{eq:deco}
\end{equation}
Note that both row and column spans of $H_i$ and
$H_j$ are orthogonal to one another when $i\ne j$,  namely
\begin{equation}
H_i^* H_j  =
H_i H_j^*=
0_n,
\qquad  i\ne j.
\end{equation}
On the other hand, by feasibility of both $M$ and $\widehat{M}$ in Program \eqref{eq:p1 MR}, we find the so-called tube constraint:
\begin{align}
\left\Vert \mathcal{R}_m(H)\right\Vert _{F}
& = \left\Vert \mathcal{R}_m(\widehat{M}-M)\right\Vert _{F} \nonumber\\
& \le \l\| \R_m ( \widehat{M} ) -y \r\|_*
+ \l\| \R_m\l( M \r) -y \r\|_*
 \nonumber\\
& \le 2e.
\qquad \mbox{(see Program \eqref{eq:p1 MR})}
\label{eq:feasibility}
\end{align}
Using  (\ref{eq:deco}), \eqref{eq:feasibility}, and the triangle inequality,  we may then write that
\begin{equation}
\left\Vert \mathcal{R}_m(H_0+H_1)\right\Vert _{F}\le\sum_{i\ge2}\left\Vert \mathcal{R}_m(H_i)\right\Vert _{F}+2e.
\label{eq:pre RIP}
\end{equation}
Recall \eqref{eq:RIP def} and  suppose that the measurement operator $\mathcal{R}_m(\cdot)$ satisfies $\delta_{r''}$-RIP   with integer $r''\ge 2r+r'$ to be set later. By construction,
\begin{equation*}
\mbox{rank}\l(H_0+H_1\r)
= \mbox{rank}\l(\mathcal{P}_{\ST}(H)+H_1\r)
\le 2r+r'\le r'',
\end{equation*}
\begin{equation*}
\mbox{rank}(H_i)\le r'\le r'', \quad i\ge 1,
\end{equation*}
 and, therefore, \eqref{eq:pre RIP} and the RIP together imply  that
\begin{align}
(1-\delta_{r''})\left\Vert H_0+H_1\right\Vert _{F} & \le(1+\delta_{r''})\sum_{i\ge2}\left\Vert H_i \right\Vert _{F}+2e
\nonumber
\\
 & \le\frac{1+\delta_{r''}}{\sqrt{r'}}\sum_{i\ge1}\left\Vert H_i \right\Vert _{*}+2e
\nonumber
 \\
 & =\frac{1+\delta_{r''}}{\sqrt{r'}}\left\Vert \sum_{i\ge 1}H_i \right\Vert _{*}+2e
\nonumber\\
  & =\frac{1+\delta_{r''}}{\sqrt{r'}}\left\Vert \mathcal{P}_{\ST^{\perp}}(H)\right\Vert _{*}+2e,
  \label{eq:key RIP use}
\end{align}
where, in the second line, we used the fact that  $\left\Vert H_{i+1} \right\Vert _{F}\le\frac{1}{\sqrt{r'}}\left\Vert H_i \right\Vert _{*}$ for every $i\ge 1$,  which itself follows directly from the non-increasing order of the singular values in $\Sigma'$ and the fact that $\mbox{rank}(H_i )\le r'$ for every $i \ge 1$. The last line uses the fact that $\|A+B\|_*=\|A\|_*+\|B\|_*$ when $\mbox{span}(A)\perp \mbox{span}(B)$ and $\mbox{span}(A^*)\perp \mbox{span}(B^*)$.
 Then, invoking the nullspace property (\ref{eq:nsp final})  below,  we find that
\begin{align*}
 \left\Vert H_0+H_1\right\Vert _{F}&  \le \frac{1+\delta_{r''}}{1-\delta_{r''}}  \sqrt{\frac{1}{r'}} \l\|\mathcal{P}_{\ST^\perp} (H)\r\|_*+ \frac{2e}{1-\delta_{r''}}\\
 & \le\frac{1+\delta_{r''}}{1-\delta_{r''}} \sqrt{\frac{1}{r'}}  \left(\Cr{1}\left\Vert \mathcal{P}_{\ST}(H)\right\Vert _{*}+\Cr{2}\left\Vert \mathcal{P}_{\widetilde{\ST}}(H)\right\Vert _{*} \right. \nonumber\\
& \qquad \qquad \qquad \left. +2\left\Vert M_{r^+}\right\Vert _{*}\right)
+\frac{2e}{1-\delta_{r''}},
 \end{align*}
 where the first and second inequalities use \eqref{eq:key RIP use} and \eqref{eq:nsp final}, respectively.
We now continue by writing that
 \begin{align}
& \left\Vert H_0+H_1\right\Vert _{F} \nonumber\\
 & \le\frac{1+\delta_{r''}}{1-\delta_{r''}} \sqrt{\frac{1}{r'}}  \left(\Cr{1}\left\Vert \mathcal{P}_{\ST}(H)\right\Vert _{*}+\Cr{2}\left\Vert \mathcal{P}_{\widetilde{\ST}}(H)\right\Vert _{*} \right. \nonumber\\
& \qquad \qquad \qquad \left. +2\left\Vert M_{r^+}\right\Vert _{*}\right)
+\frac{2e}{1-\delta_{r''}}
\qquad \mbox{(see \eqref{eq:nsp final})} \nonumber\\
 & \le\frac{1+\delta_{r''}}{1-\delta_{r''}}\sqrt{\frac{2r}{r'}}\cdot\left(\Cr{1}\left\Vert \mathcal{P}_{\ST}(H)\right\Vert _{F}+\Cr{2}\left\Vert \mathcal{P}_{\widetilde{\ST}}(H)\right\Vert _{F}\right)
 \nonumber\\
& \qquad  +2\cdot \frac{1+\delta_{r''}}{1-\delta_{r''}}\sqrt{\frac{1}{r'}}\left\Vert M_{r^+}\right\Vert _{*}+\frac{2e}{1-\delta_{r''}}
 \nonumber\\
 & =\frac{1+\delta_{r''}}{1-\delta_{r''}}\sqrt{\frac{2r}{r'}}\cdot\left(\Cr{1}\left\Vert H_0 \right\Vert _{F}+\Cr{2}\left\Vert \mathcal{P}_{\widetilde{\ST}}(H)\right\Vert _{F}\right)
\nonumber\\
& \qquad  +2\cdot \frac{1+\delta_{r''}}{1-\delta_{r''}}\sqrt{\frac{1}{r'}}\left\Vert M_{r^+} \right\Vert _{*}+\frac{2e}{1-\delta_{r''}}
 \nonumber\\
 & \le\frac{1+\delta_{r''}}{1-\delta_{r''}}\sqrt{\frac{2r}{r'}}\cdot\left(\Cr{1}\left\Vert H_0 \right\Vert _{F}+\Cr{2}\left\Vert H_1 \right\Vert _{F}\right)
\nonumber\\
& \qquad  +2\cdot \frac{1+\delta_{r''}}{1-\delta_{r''}}\sqrt{\frac{1}{r'}}\left\Vert M_{r^+}\right\Vert _{*}+\frac{2e}{1-\delta_{r''}}
 \nonumber\\
 & \le\frac{1+\delta_{r''}}{1-\delta_{r''}}\sqrt{\frac{2r}{r'}}\cdot\max\left[\Cr{1},\Cr{2}\right]\cdot \l(\left\Vert H_0\r\|_F + \l\|H_1\right\Vert _{F}\r)
\nonumber\\
& \qquad  +2\cdot \frac{1+\delta_{r''}}{1-\delta_{r''}}\sqrt{\frac{1}{r'}}\left\Vert M_{r^+}\right\Vert _{*}+\frac{2e}{1-\delta_{r''}}
\nonumber\\
 & \le \frac{1+\delta_{r''}}{1-\delta_{r''}}\sqrt{\frac{2r}{r'}}\cdot\max\left[\Cr{1},\Cr{2}\right]\cdot \sqrt{2\left\Vert H_0\r\|_F^2 + 2\l\|H_1\right\Vert _{F}^2}
\nonumber\\
& \qquad  +2\cdot \frac{1+\delta_{r''}}{1-\delta_{r''}}\sqrt{\frac{1}{r'}}\left\Vert M_{r^+}\right\Vert _{*}+\frac{2e}{1-\delta_{r''}}
 \nonumber\\
 & = 2\cdot \frac{1+\delta_{r''}}{1-\delta_{r''}}\sqrt{\frac{r}{r'}}\cdot\max\left[\Cr{1},\Cr{2}\right]\cdot \l\| H_0+H_1\r\|_F
\nonumber\\
& \qquad  +2\cdot \frac{1+\delta_{r''}}{1-\delta_{r''}}\sqrt{\frac{1}{r'}}\left\Vert M_{r^+}\right\Vert _{*}+\frac{2e}{1-\delta_{r''}},
 \label{eq:pre rearrange RIP}
\end{align}
which, as justified presently, holds as long as $r'\ge 2r$. Above, the second inequality holds because, by \eqref{eq:supp}, $\mbox{rank}(\mathcal{P}_{{\ST}}(H))\le 2r$ and, by  (\ref{eq:T tilde}),  $ \mbox{rank}(\mathcal{P}_{\widetilde{\ST}}(H))\le 2r$. The first identity there uses  the fact that $H_0 = \mathcal{P}_{\ST}(H)$.
The third inequality above follows because $\widetilde{\ST},\ST_{1}\subset \ST^{\perp}$
and $H_1$, by construction, is a rank-$r'$
 truncation of $\mathcal{P}_{\ST^{\perp}}(H)$. Therefore, as long as $r'\ge 2r\ge\mbox{rank}(\mathcal{P}_{\widetilde{\ST}}(H))$, we have that $\|\mathcal{P}_{\widetilde{\ST}}(H)\|_F\le \|H_1 \|_F$, as claimed above. Also,
the last inequality uses the fact that $a+b\le \sqrt{2a^2+2b^2}$ for scalars $a,b$, and
 the last identity  holds because $H_0=\mathcal{P}_{\ST}(H)$, $\mbox{span}(H_1)\subseteq \ST^\perp$, and consequently $\langle H_0,H_1\rangle = \mbox{trace}(H_0^* H_1)=0$.
After rearranging the terms in \eqref{eq:pre rearrange RIP}, we find that if
\begin{equation}
2\cdot \frac{1+\delta_{r''}}{1-\delta_{r''}}\cdot \sqrt{\frac{r}{r'}}  \max\l[\Cr{1},\Cr{2}\r] \le 0.9,
\end{equation}
or equivalently if
\begin{equation}
\label{eq:req on RIP cte}
\delta_{r''} \le
\frac{0.9-2\max\l[\Cr{1},\Cr{2}\r]\sqrt{\frac{r}{r'}}}{0.9+2\max\l[\Cr{1},\Cr{2}\r]\sqrt{\frac{r}{r'}}},
\end{equation}
then the following holds:
\begin{equation}
\left\Vert H_0+H_1 \right\Vert _{F}\le 20\cdot  \frac{1+\delta_{r''}}{1-\delta_{r''}}\sqrt{\frac{1}{r'}}\left\Vert M_{r^+}\right\Vert _{*}+ \frac{20 e}{1-\delta_{r''}}.
\label{eq:MR final leg 1}
\end{equation}
On the other hand, note that
\begin{align}
& \left\Vert \sum_{i\ge2}H_i \right\Vert _{F} \nonumber\\
& \le \sum_{i\ge 2}\left\Vert H_i\right\Vert _{F}
\qquad \mbox{(triangle inequality)}
\nonumber
\\
& \le \frac{\l\| \mathcal{P}_{\ST^\perp}(H)\r\|_*}{\sqrt{r'}}
\qquad \mbox{(implicit in \eqref{eq:key RIP use})}\nonumber\\
& \le 2\sqrt{\frac{r}{r'}} \cdot \max[\Cr{1},\Cr{2}] \cdot \l\| H_0+H_1\r\|_F
\nonumber\\
& \qquad + \frac{2}{\sqrt{r'}} \l\| M_{r^+} \r\|_*
\qquad \mbox{(implicit in \eqref{eq:pre rearrange RIP})}\nonumber\\
 & \le
 2\sqrt{\frac{r}{r'}} \cdot \max[\Cr{1},\Cr{2}]
 \cdot  \Bigg(20\cdot  \frac{1+\delta_{r''}}{1-\delta_{r''}}\sqrt{\frac{1}{r'}}\left\Vert M_{r^+}\right\Vert _{*} \nonumber\\
&\qquad \qquad \qquad \qquad \quad   + \frac{20e}{1-\delta_{r''}} \Bigg) + \frac{2}{\sqrt{r'}} \l\|M_{r^+} \r\|_*,
\label{eq:MR final leg 2}
\end{align}
where the last inequality uses \eqref{eq:MR final leg 1}.
Lastly, \eqref{eq:MR final leg 1} and \eqref{eq:MR final leg 2} together imply that
\begin{align*}
& \left\Vert \widehat{M}-M\right\Vert _{F} \nonumber\\
&  =\|H\|_{F}
\nonumber\\
& \le\left\Vert H_0+H_1\right\Vert _{F}+\left\Vert \sum_{i\ge2}H_i \right\Vert _{F}
\qquad \mbox{(see \eqref{eq:deco})}
\\
 & \le
\l( 1+2\sqrt{\frac{r}{r'}} \cdot \max[\Cr{1},\Cr{2}]\r)
\nonumber\\
&  \cdot  \l(20\cdot  \frac{1+\delta_{r''}}{1-\delta_{r''}}\sqrt{\frac{1}{r'}}\left\Vert M_{r^+}\right\Vert _{*}+ \frac{20 e}{1-\delta_{r''}} \r)
 + \frac{2}{\sqrt{r'}}\l\| M_{r^+} \r\|_*,
\end{align*}
provided that $r''\ge 2r+r'$ and as long as  \eqref{eq:req on RIP cte} is met. The last inequality above uses \eqref{eq:MR final leg 2}.
This completes the proof of Theorem \ref{thm:main result MR} after taking $r'=30r$ and $r''=32r$.

\section{Analysis for Matrix Completion}
\label{sec:analysis of mc w pi}

In this section, we analyze Program \eqref{eq:p1} and eventually prove Theorem \ref{thm:main result}. In fact, we prove a stronger result based on \emph{leveraged sampling}, of which Theorem \ref{thm:main result} is a special case. Let us begin with recalling the definition of  \emph{leverage scores} of a matrix. \edit{We begin with recalling the definition of leverage scores of a matrix and then leveraged random sampling, of which the uniform random sampling in Theorem \ref{thm:main result} is a special case. Then, in Section~\ref{sec:meas operator section}, we study certain isometric properties of the leveraged sampling framework that are necessary for our analysis. After that, as is standard in convex analysis, we introduce the corresponding dual certificate of Program \eqref{eq:p1} in Section~\ref{sec:meas bodya section}, the existence of which allows us to quantify the performance of Program \eqref{eq:p1} and complete the proof of Theorem \ref{thm:main result}. The  construction of the dual certificate is deferred to Appendix~\ref{sec:dual cert}.}

\subsection{Leverage Scores}
For a rank-$r$ matrix $M_r\in\real^{n\times n}$, let $\SU_r=\mbox{span}(U_r)=\mbox{span}(M_r)$ and $\SV_r=\mbox{span}(V_r)=\mbox{span}(M_r^*)$ be the column and row spaces of $M_r$ with orthonormal bases $U_r,V_r\in\mathbb{R}^{n\times r}$, respectively.
The leverage score corresponding to the $i$th row of $M_{r}$
is defined as
\begin{equation}
\mu_{i}=\mu_{i}\left(\SU_{r}\right):=\frac{n}{r}\left\Vert U_{r}[i,:]\right\Vert _{2}^{2},\qquad i\in [1:n],\label{eq:mus}
\end{equation}
where $U_r[i,:]$ is the $i$th row of $U_r$.
Similarly,
 the leverage score corresponding to the $j$th column of $M_{r}$ is defined as
\begin{equation}
\nu_{j}=\nu_{j}\left(\SV_{r}\right):=\frac{n}{r}\left\Vert V_{r}[j,:]\right\Vert _{2}^{2},\qquad j\in [1:n].\label{eq:nus}
\end{equation}
As our notation above suggests, leverage scores of a subspace are indeed independent of the choice of the orthonormal basis for subspace. In particular, notice that the coherence of a matrix is simply the largest leverage score of its column and row spans (see \eqref{eq:coherence def}), namely
\begin{equation}
\eta(M_r)  =
 \max_i \mu_i\l( \SU_r\r) \vee \max_j \mu_j\l( \SV_r\r),
 \label{eq:relation btw coh and lev}
\end{equation}
where $a\vee b= \max[a,b]$ is the shorthand for maximum.
 We
also assign leverage scores to   subspaces $\breve{\SU}= \mbox{span}([\SU_r,\widetilde{\SU}_{r}])$
and $\breve{\SV}= \mbox{span}([\SV_r,\widetilde{\SV}_{r}])$ :
\begin{equation*}
\breve{\mu}_{i}=\mu_{i}( \breve{\operatorname{U}}_{r})
,\qquad i\in[1:n],
\end{equation*}
\begin{equation}
\breve{\nu}_{j}=\nu_{i}( \breve{\operatorname{V}}_{r})
,\qquad j\in[1:n].
\label{eq:mu nu tildes}
\end{equation}
Throughout, we will continue using $\{\mu_{i},\nu_{i}\}$ and $\{\breve{\mu}_{i},\breve{\nu}_{i}\}$
as shorthand to ease the notation.
To facilitate the calculations later, let us define
the  $n\times n$ diagonal matrix
\begin{equation}
\mu
=
\left[\begin{array}{cccc}
\mu_{1}\\
 & \mu_{2}\\
 &  & \ddots\\
 &  &  & \mu_{n}
\end{array}\right].\label{eq:lev matrices}
\end{equation}
The $n\times n$ matrices $\nu,\breve{\mu},\breve{\nu}$ are
defined similarly using $\{\nu_{i},\breve{\mu}_{i},\breve{\nu}_{i}\}$,
respectively. Recall the $n\times r$ matrices $U_r,U'_r,V_r,V'_r$  constructed in Lemma \ref{lem:canonical} and denote with $\|A\|_{2\rightarrow\infty}$  the largest $\ell_{2}$ norm
of the rows of a matrix $A$.
Assuming that $\mu_i,\nu_i\ne 0$ for all $i$, the  relations  below (which follow directly from earlier definitions)
will prove useful later on:
\[
\left\Vert \left(\frac{\mu r}{n}\right)^{-\frac{1}{2}}U_r\right\Vert _{2\rightarrow\infty}=1,\qquad\left\Vert \left(\frac{\nu r}{n}\right)^{-\frac{1}{2}}V_r\right\Vert _{2\rightarrow\infty}=1,
\]
\begin{equation*}
\left\Vert \left(\frac{\mu r}{n}\right)^{-\frac{1}{2}}U'_r\right\Vert _{2\rightarrow\infty}\le \sqrt{2\max_{i}\frac{\breve{\mu}_{i}}{\mu_{i}}},
\end{equation*}
\begin{equation}
\left\Vert \left(\frac{\nu r}{n}\right)^{-\frac{1}{2}}V'_r\right\Vert _{2\rightarrow\infty}\le \sqrt{2\max_{j}\frac{\breve{\nu}_{j}}{\nu_{j}}}.\label{eq:useful eqs}
\end{equation}
To be complete, let us verify the third relation above. Letting $\breve{U}$ denote an orthonormal basis for $\breve{\SU}$, we write that
\begin{align*}
& \l\| \l(\frac{\mu r}{n}\r)^{-\frac{1}{2}} U'_r  \r\|_{2\rightarrow\infty}
\nonumber\\
& = \max_i \frac{\l\|U'_r[i,:]  \r\|_2}{\l\|U_r[i,:]\r\|_2}
\qquad \mbox{(see \eqref{eq:lev matrices} and then \eqref{eq:mus})} \nonumber\\
& \le \max_i \frac{\l\|\breve{U}[i,:]  \r\|_2}{\l\|U_r[i,:]\r\|_2}
\nonumber\\
& = \sqrt{ \max_i \frac{\breve{\mu}_i
\cdot \mbox{dim}( \breve{\SU} )}{{\mu}_i
r }}
\qquad \mbox{(see \eqref{eq:mus})}
\nonumber\\
& \le
\sqrt{\max_i \frac{\breve{\mu}_i \cdot 2r}{\mu_i r }},
\qquad \l(\mbox{dim}(\breve{\SU}) \le 2r  \r)
\end{align*}
where the third line above holds because $\SU_r' \subset \breve{\SU}$, by construction in the proof of Lemma \ref{lem:canonical}.

\subsection{Measurement Operator}
\label{sec:meas operator section}

Next, we slightly modify the measurement operator in \eqref{eq:def of R(Z)} to gain more versatility. Throughout Section \ref{sec:analysis of mc w pi}, for probabilities $\{p_{ij}\}_{i,j=1}^n\subset (0,1]$, we assume that $\mathcal{R}_{p}(\cdot)$ takes
$M\in\mathbb{R}^{n\times n}$ to $\mathcal{R}_p(M)\in\mathbb{R}^{n\times n}$
defined as
\begin{equation}
\mathcal{R}_{p}(M)= \sum_{i,j=1}^{n}\frac{\epsilon_{ij}}{p_{ij}}\cdot M[i,j]\cdot C_{ij},\label{eq:def of R(Z) appendix}
\end{equation}
where each $\epsilon_{ij}$ is a Bernoulli random variable that takes
$1$ with a probability of $p_{ij}$ (and $0$ otherwise).
Moreover, $\{\epsilon_{ij}\}$ are independent. Recall also  that $C_{ij}\in\mathbb{R}^{n\times n}$
is the $[i,j]$th canonical matrix. Throughout Section \ref{sec:analysis of mc w pi}, we will assume that $\{p_{ij}\}\subset [l,h]$ for some $0<l\le h\le 1$. In particular, note that we retrieve the measurement operator in \eqref{eq:def of R(Z)} by setting $p_{ij}=l=h=p$ for every $i,j$.

Through $\mathcal{R}_{p}(\cdot)$,
we measure $M$. In particular, for noise level $e\ge 0$, let $Y=\R_p(M+E)$ with $\|\R_p(E)\|_F\le e$ be the (possibly noisy) matrix of measurements. To (approximately) complete $M$ given the measurement matrix $Y$ and prior knowledge about column/row spaces of $M$, we solve
\begin{equation}
\begin{cases}
\min_X \left\Vert Q_{\SUT_r,\lambda}\cdot X\cdot Q_{\SVT_r,\rho}\right\Vert _{*},\\
\mbox{subject to }\left\Vert \mathcal{R}_p(X)-Y\right\Vert _{F}\le e,
\end{cases}
\label{eq:p1 appendix}
\end{equation}
where $Q_{\wt{\SU}_r,\lambda},Q_{\wt{\SV}_r,\rho}\in\real^{n\times n}$ encapsulate our prior knowledge about $M$ and were defined in \eqref{eq:def of Qs}. In the rest of Section \ref{sec:analysis of mc w pi}, we analyze Program \eqref{eq:p1 appendix} with $\R_p(\cdot)$ defined in \eqref{eq:def of R(Z) appendix}. Theorem \ref{thm:main result} will follow as a special case, as explained later.

Understanding the properties of the measurement operator is
imperative to the development of  supporting theory. To list these
properties, let us introduce the following norms which, respectively, measure the (weighted) largest entry and largest $\ell_2$ norm of the rows of a matrix \cite{chen2013completing}:
For a matrix $Z\in\mathbb{R}^{n\times n}$, we set
\begin{align}
& \|Z\|_{\mu(\infty)}\nonumber\\
& =\left\Vert \left(\frac{\mu r}{n}\right)^{-\frac{1}{2}}\cdot Z\cdot\left(\frac{\nu r}{n}\right)^{-\frac{1}{2}}\right\Vert _{\infty}\nonumber\\
&
=\max_{i,j}\sqrt{\frac{n}{\mu_{i}r}}\cdot|Z[i,j]|\cdot\sqrt{\frac{n}{\nu_{j}r}},
\qquad \mbox{(see \eqref{eq:lev matrices})}
\label{eq:mu inf norm}
\end{align}
where $\|A\|_{\infty}$ returns the largest entry of matrix $A$ in
magnitude.
 Moreover, for $Z\in\mathbb{R}^{n\times n}$, we let
\begin{align}
\label{eq:mu inf 2 norm}
& \|Z\|_{\mu(\infty,2)}
\nonumber\\
& =\left\Vert \left(\frac{\mu r}{n}\right)^{-\frac{1}{2}}\cdot Z\right\Vert _{2\rightarrow\infty}\vee\left\Vert \left(\frac{\nu r}{n}\right)^{-\frac{1}{2}}\cdot Z^{*}\right\Vert _{2\rightarrow\infty}
\nonumber\\
 & =\left(\max_{i}\sqrt{\frac{n}{\mu_{i}r}}\left\Vert Z[i,:]\right\Vert _{2}\right)\vee\left(\max_{j}\sqrt{\frac{n}{\nu_{j}r}}\left\Vert Z[:,j]\right\Vert _{2}\right),
\end{align}
return the largest $\ell_{2}$ norm of the columns and rows of $Z$
after reweighting. %
Above, $\|A\|_{2\rightarrow\infty}$ returns the largest $\ell_{2}$ norm
of the rows of matrix $A$.

Establishing the following results is a standard practice in the use
of large deviation bounds, stated here without proof from \cite{chen2013completing}.
\begin{lem}
\label{lem:near isometry 1} \emph{\textbf{ \cite[Lemma 9]{chen2013completing}}} For probabilities $\{p_{ij}\}\subset(0,1]$, consider
the measurement operator $\mathcal{R}_{p}(\cdot)$ defined in (\ref{eq:def of R(Z) appendix}).
Let the subspace $\ST$, defined in (\ref{eq:supp}), be the support
of $M_{r}\in\mathbb{R}^{n\times n}$ and let $\mathcal{P}_{\ST}(\cdot)$
be the orthogonal projection onto $\ST$. Then,
except with a probability of at most $n^{-20}$, it holds that
\[
\left\Vert \left(\mathcal{P}_{\ST}-\mathcal{P}_{\ST}\circ\mathcal{R}_{p}\circ\mathcal{P}_{\ST}\right)(\cdot)\right\Vert _{F\rightarrow F}\le\frac{1}{2},
\]
provided that
\begin{equation}\label{eq:no of samples}
 \frac{\left(\mu_{i}+\nu_{j}\right)r\log n}{n} \lesssim p_{ij} \le 1
 ,
\qquad \forall i,j\in[1:n].
\end{equation}
Above, $\|\mathcal{A}(\cdot)\|_{F\rightarrow F}=\sup_{\|X\|_{F}\le1}\|\mathcal{A}(X)\|_{F}$
is the operator norm of the linear map $\mathcal{A}(\cdot)$, and $(\mathcal{A} \circ \mathcal{B})(\cdot) = \mathcal{A}(\mathcal{B}(\cdot))$ stands for composition of operators $\mathcal{A}(\cdot)$ and $\mathcal{B}(\cdot)$.
\end{lem}

\begin{lem}
\label{lem:any Z}\emph{\textbf{\cite[Lemma 10]{chen2013completing}}} Consider the same setup as in Lemma \ref{lem:near isometry 1}
and fix a matrix $Z\in\mathbb{R}^{n\times n}$. Except with a probability of at most $n^{-20}$, it holds that
\[
\left\Vert \left(\mathcal{I}-\mathcal{R}_{p}\right)(Z)\right\Vert \lesssim\|Z\|_{\mu(\infty)}+\|Z\|_{\mu(\infty,2)},
\]
provided that (\ref{eq:no of samples})  holds. Here, $\mathcal{I}(\cdot)$ is the identity operator, so that $\mathcal{I}(Z)=Z$ for any $Z$. In particular, multiplying the far-left side of (\ref{eq:no of samples}) by a factor of $\Delta^2\ge 1$ will divide  the right-hand side above by a factor of $\Delta$.
\end{lem}

\begin{lem}
\label{lem:inf two bnd} \emph{\textbf{\cite[Lemma 11]{chen2013completing}}} Consider the same setup as Lemma \ref{lem:near isometry 1}
and fix a matrix $Z\in \ST\subset \mathbb{R}^{n\times n}$ (i.e., $\mathcal{P}_{\ST}(Z)=Z$). Then, except with a probability of at most $n^{-20}$, it holds that
\begin{align*}
& \left\Vert \left(\mathcal{P}_{\ST}-\mathcal{P}_{\ST}\circ\mathcal{R}_{p}\circ\mathcal{P}_{\ST}\right)(Z)\right\Vert _{\mu(\infty,2)}
\nonumber\\
&
\qquad \le\frac{1}{2}\|Z\|_{\mu(\infty)}+\frac{1}{2}\|Z\|_{\mu(\infty,2)},
\end{align*}
as long as (\ref{eq:no of samples}) holds.\footnote{Lemmas 11 and 12 in \cite{chen2013completing} are in fact  more general than the statements here.}
\end{lem}

\begin{lem}
\label{lem:inf bound} \emph{\textbf{\cite[Lemma 12]{chen2013completing}}} Consider the same setup as in Lemma \ref{lem:near isometry 1}
and fix a matrix $Z\in \ST\subset\mathbb{R}^{n\times n}$. Then, except with a probability of at most $n^{-20}$, it holds that
\[
\left\Vert \left(\mathcal{P}_{\ST}-\mathcal{P}_{\ST}\circ\mathcal{R}_{p}\circ\mathcal{P}_{\ST}\right)(Z)\right\Vert _{\mu(\infty)}\le\frac{1}{2}\|Z\|_{\mu(\infty)},
\]
as long as (\ref{eq:no of samples}) holds.
\end{lem}

At times, we will find it more convenient to work with the closely related operator $\overline{\mathcal{R}}_{p}(\cdot)$  that takes $\overline{Z}\in\mathbb{R}^{n\times n}$
to $\overline{\mathcal{R}}_p(\overline{Z})\in \mathbb{R}^{n\times n}$,  where
\begin{equation}
 \overline{\mathcal{R}}_{p}(\overline{Z})=B_{L}^{*}\cdot\mathcal{R}_{p}\left(B_L \overline{Z}B_{R}^{*}\right)\cdot B_{R},\label{eq:def of RUV}
\end{equation}
in which we applied \eqref{eq:def of BL} and \eqref{eq:def of R(Z) appendix}.
Corresponding to the measurement operator $\mathcal{R}_{p}(\cdot)$,
we also define the orthogonal projection $\mathcal{P}_{p}(\cdot)$ that
projects onto the support of $\mathcal{R}_{p}(\cdot)$. More specifically, $\mathcal{P}_p(\cdot)$ takes
 $Z\in\mathbb{R}^{n\times n}$ to $\mathcal{P}_p(Z)\in\mathbb{R}^{n\times n}$ defined as
\begin{equation}
\mathcal{P}_{p}(Z)=\sum_{i,j}\epsilon_{ij}Z[i,j]\cdot C_{ij}.\label{eq:P}
\end{equation}
Similarly, $\overline{\mathcal{P}}_{p}(\cdot)$   is the orthogonal
xprojection  that takes $\overline{Z}\in \mathbb{R}^{n\times n}$ to $\overline{\mathcal{P}}_{p}(\overline{Z})\in \mathbb{R}^{n\times n}$, defined as
\begin{equation}
\overline{\mathcal{P}}_{p}(\overline{Z})=B_{L}^{*}\cdot\mathcal{P}_{p}\left(B_{L}\overline{Z}B_{R}^{*}\right)\cdot B_{R}.\label{eq:P bar}
\end{equation}
We will only $\mathcal{P}_{p}(\cdot)$ and $\overline{\mathcal{P}}_{p}(\cdot)$
once. Below, we collect a few basic properties of all these operators which, for the sake of completeness, are proved in Appendix \ref{sec:proof of props of Rbar}.

\begin{lem}\label{lem:props of Rbar}
For an arbitrary  $Z\in\mathbb{R}^{n\times n}$, with { $\overline{Z}=B_{L}^{*}ZB_{R}\in\mathbb{R}^{n\times n}$}, and for the operators $\mathcal{R}_p(\cdot)$, $\overline{\mathcal{R}}_p(\cdot)$, $\mathcal{P}_p(\cdot)$, $\overline{\mathcal{P}}_p(\cdot)$ defined above, it holds that
\begin{equation}
\left\langle \overline{Z},\overline{\mathcal{R}}_{p}(\overline{Z})\right\rangle =\left\langle Z,\mathcal{R}_{p}(Z)\right\rangle,
\label{eq:aux 8.9}
\end{equation}
\begin{equation}
\left\Vert \overline{\mathcal{R}}_{p}(\overline{Z})\right\Vert _{F}
  =\left\Vert \mathcal{R}_{p}(Z)\right\Vert _{F}.\label{eq:aux 9}
\end{equation}
Additionally, if $\{p_{ij}\}\subset[l,h]$ with $0<l\le h\le 1$, it holds true that
\begin{equation}
\left( \overline{\mathcal{R}}_{p} \circ \overline{\mathcal{R}}_{p} \r)
(\cdot)
\succcurlyeq
\overline{\mathcal{R}}_{p}(\cdot).\label{eq:aux 11}
\end{equation}
\begin{equation}
\left\Vert \overline{\mathcal{R}}_{p}(\cdot)\right\Vert _{F\rightarrow F}
=
\left\Vert {\mathcal{R}}_{p}(\cdot)\right\Vert _{F\rightarrow F}
\le l^{-1}.\label{eq:aux 10}
\end{equation}
Above, for operators $\mathcal{A}(\cdot)$ and $\mathcal{B}(\cdot)$,
$\mathcal{A}(\cdot)\succcurlyeq\mathcal{B}(\cdot)$ means that $\left\langle Z,\mathcal{A}(Z)\right\rangle \succcurlyeq\left\langle Z,\mathcal{B}(Z)\right\rangle $
for any matrix $Z$.
Lastly,
\begin{equation}
\left\Vert \overline{\mathcal{P}}_{p}(\overline{Z})\right\Vert_{F}
\le  h \l\|\overline{\mathcal{R}}_{p}(\overline{Z})\right\Vert _{F}.\label{eq:P n R}
\end{equation}
\end{lem}
We are now in position to study Program \eqref{eq:p1 appendix} in more detail.

\subsection{Body of the Analysis}
\label{sec:meas bodya section}

Assume for now that $M=M_r$ is rank-$r$ and that $e=0$, namely noise is absent. Extending to noise and nearly low-rank matrices is straightforward, as described later.
For solution $\widehat{M}$,  let $H:=\widehat{M}-M$ be the error. In Program \eqref{eq:p1 appendix}, by feasibility of $M$ and optimality of $\widehat{M}=M+H$,
we may write that
\begin{equation}
\left\Vert Q_{\SUT_r,\lambda} (M+H) Q_{\SVT_r,\rho}
\right\Vert _{*}\le
\left\Vert Q_{\SUT_r,\lambda} M Q_{\SVT_r,\rho}
\right\Vert _{*}.\label{eq:opt cond}
\end{equation}
The right-hand side above can itself be bounded
as
\begin{align}
& \left\Vert Q_{\SUT_r,\lambda} M Q_{\SVT_r,\rho}
\right\Vert _{*} \nonumber\\
&
= \left\Vert Q_{\SUT_r,\lambda} M_r Q_{\SVT_r,\rho}
\right\Vert _{*}
\nonumber\\
 & \le \left\Vert B_{L}O_{L}L\overline{M}_{r}R^{*}O_{R}^{*}B_{R}^{*}\right\Vert _{*}
 \nonumber\\
 & =\left\Vert L\overline{M}_{r}R^{*}\right\Vert _{*}
\qquad \mbox{(rotational invariance)}
\nonumber \\
 & =\left\Vert \left[\begin{array}{cc}
L_{11}\overline{M}_{r,11}R_{11}\\
 & 0_{n-r}
\end{array}\right]\right\Vert _{*},
\qquad \mbox{(see \eqref{eq:M decompose})}
\label{eq:upp bnd MC}
\end{align}
where the third \edit{line} applies  \eqref{eq:M decompose} and  then \eqref{eq:norm of L}.
 Similarly, the left-hand side of (\ref{eq:opt cond}) can be bounded
from below as follows:
\begin{align}
\label{eq:low bnd MC}
& \left\Vert Q_{\SUT_r,\lambda} \left(M+H \right) Q_{\SVT_r,\rho}\right\Vert _{*} \nonumber\\
& = \left\Vert Q_{\SUT_r,\lambda} \l( M_{r}+H \right) Q_{\SVT_r,\rho}\right\Vert _{*}
\nonumber\\
 & \ge \left\Vert B_{L}O_{L}L\left(\overline{M}_{r}+\overline{H}\right)R^{*}O_{R}^{*}B_{R}^{*}\right\Vert _{*}-\left\Vert  M_{r^+} \right\Vert _{*}
 \nonumber\\
 & =\left\Vert L\left(\overline{M}_{r}+\overline{H}\right)R^{*}\right\Vert _{*}
\qquad \mbox{(rotational invariance)}
 \nonumber\\
 & =
 \left\Vert \left[\begin{array}{cc}
L_{11}\overline{M}_{r,11}R_{11}\\
 & 0_{n-r}
\end{array}\right]+L\overline{H}R^{*}\right\Vert _{*}
,
\qquad \mbox{(see \eqref{eq:M decompose})}
\end{align}
where the third line uses \eqref{eq:H decompose} and then \eqref{eq:norm of L}.
By substituting \eqref{eq:upp bnd MC} and \eqref{eq:low bnd MC} back in \eqref{eq:opt cond},  and then using the convexity of   nuclear norm,
we arrive at the following:
\begin{align}
& \left\langle L\overline{H}R^{*},G\right\rangle \nonumber\\
& \le \left\Vert \left[\begin{array}{cc}
L_{11}\overline{M}_{r,11}R_{11}\\
 & 0_{n-r}
\end{array}\right]+L\overline{H}R^{*}\right\Vert _{*}
\nonumber\\
& \qquad \qquad - \left\Vert \left[\begin{array}{cc}
L_{11}\overline{M}_{r,11}R_{11}
\nonumber\\
 & 0_{n-r}
\end{array}\right]\right\Vert _{*}
\le 0.
\\
& \qquad\forall G\in\partial\left\Vert \left[\begin{array}{cc}
L_{11}\overline{M}_{r,11}R_{11}\\
 & 0_{n-r}
\end{array}\right]\right\Vert _{*}.
\label{eq:opt cnd 2}
\end{align}
Above, $\partial\|A\|_{*}$ stands for the sub-differential of the
nuclear norm at $A$ (e.g., \cite[equation 2.9]{recht2010guaranteed}). In order to fully characterize the sub-differential, we take the following steps.
 First, from \eqref{eq:M decompose}, recall that $\mbox{rank}(\overline{M}_{r,11})=\mbox{rank}(\overline{M}_{r})=\mbox{rank}(M_{r})=r$. Second, assume that $\lambda\cdot \rho\ne0$ so that $\mbox{rank}(L_{11}\overline{M}_{11}R_{11})=\mbox{rank}(\overline{M}_{11})=r$
too (see \eqref{eq:def of Wu} for the definitions of $L_{11},R_{11}$).  Third, consider the SVD
\[
L_{11}\overline{M}_{r,11}R_{11}=\overline{U}_{r}\,\overline{\Lambda}_{r}\,\overline{V}_{r}^{*},\qquad\overline{U}_{r},\overline{V}_{r}\in\mathbb{R}^{r\times r},
\]
and define the \emph{sign matrix} $S\in\mathbb{R}^{n\times n}$ as
\begin{equation}
S:=\left[\begin{array}{cc}
S_{11}\\
 & 0_{n-r}
\end{array}\right]:=\left[\begin{array}{cc}
\overline{U}_{r}\,\overline{V}_{r}{}^{*}\\
 & 0_{n-r}
\end{array}\right].\label{eq:Sw def}
\end{equation}
Finally, the sub-differential in \eqref{eq:opt cnd 2} is specified as
\begin{align}
 & \partial\left\Vert \left[\begin{array}{cc}
L_{11}\overline{M}_{r,11}R_{11}\\
 & 0_{n-r}
\end{array}\right]\right\Vert _{*}\nonumber \\
 & =\Big\{ G\in\mathbb{R}^{n\times n}\, :  \nonumber\\
&  \qquad G=\left[\begin{array}{cc}
S_{11}\in\mathbb{R}^{r\times r}, \\
 & G_{22}\in\mathbb{R}^{(n-r)\times (n-r)}
\end{array}\right]
\nonumber\\
& \qquad \mbox{and } \|G_{22}\|\le1 \Big\} \nonumber \\
 & =\Big\{ G\in\mathbb{R}^{n\times n}\,: \nonumber\\
 & \qquad \mathcal{P}_{\overline{\ST}}(G)=S=\left[\begin{array}{cc}
S_{11}\\
 & 0_{n-r}
\end{array}\right],
\nonumber\\
& \qquad \mbox{and } \left\Vert \mathcal{P}_{\overline{\ST}^{\perp}}(G)\right\Vert \le1\Big\} .
\qquad \mbox{(see \eqref{eq:T bar})}
\label{eq:sub diff}
\end{align}
For the record, \eqref{eq:Sw def} also implies that
\begin{equation*}
\mbox{rank}(S)= \mbox{rank}(S_{11})= r,\qquad
\|S\|=\|S_{11}\|=1,
\end{equation*}
\begin{equation}
\|S\|_{F}=\|S_{11}\|_F=\sqrt{r}.\label{eq:sign matrix props}
\end{equation}
 With the characterization of the sub-differential in (\ref{eq:sub diff}),
we  rewrite (\ref{eq:opt cnd 2}) as
\[
\left\langle L\overline{H}R^{*},S+\left[\begin{array}{cc}
0_{r}\\
 & G_{22}
\end{array}\right]\right\rangle \le 0 ,
\]
for any $G_{22}\in\mathbb{R}^{(n-r)\times (n-r)}$
 such that
$\|G_{22}\|\le1$,
from which it follows that
\begin{align}
0 & \ge\left\langle L\overline{H}R^{*},S\right\rangle +\sup_{\|G_{22}\|\le 1}\left\langle L\overline{H}R^{*},\left[\begin{array}{cc}
0_{r}
\\
 & G_{22}
\end{array}\right]\right\rangle\nonumber \\
& =
\left\langle L\overline{H}R^*,S\right\rangle
+ \sup_{\|G\|\le 1}
\left\langle \mathcal{P}_{\overline{\ST}^\perp}\left( L\overline{H}R^* \right),G \right\rangle
\qquad \mbox{(see \eqref{eq:T bar})}\nonumber
\\
& =\left\langle L\overline{H}R^{*},S\right\rangle +\left\Vert \mathcal{P}_{\overline{\ST}^{\perp}}(L\overline{H}R^{*})\right\Vert _{*}
\nonumber\\
& =\left\langle \overline{H},L^{*}SR\right\rangle +\left\Vert \mathcal{P}_{\overline{\ST}^{\perp}}\left(L\overline{H}R^{*}\right)\right\Vert _{*}\nonumber \\
 & =\left\langle \overline{H},\left[\begin{array}{ccc}
L_{11}S_{11}R_{11} & L_{11}S_{11}R_{12}\\
L_{12}^{*}S_{11}R_{11} & L_{12}^{*}S_{11}R_{12}\\
 &  & 0_{n-2r}
\end{array}\right]\right\rangle
\nonumber\\
& \qquad +\left\Vert \left[\begin{array}{ccc}
0_{r}\\
 & L_{22}\overline{H}_{22}R_{22} & L_{22}\overline{H}_{23}\\
 & \overline{H}_{32}R_{22} & \overline{H}_{33}
\end{array}\right]\right\Vert _{*}
\nonumber \\
 & =\left\langle \overline{H},\left[\begin{array}{ccc}
L_{11}S_{11}R_{11} & L_{11}S_{11}R_{12}\\
L_{12}^{*}S_{11}R_{11} & 0_{r}\\
 &  & 0_{n-2r}
\end{array}\right]\right\rangle
\nonumber\\
& \qquad +\left\langle \overline{H}_{22},L_{12}^{*}S_{11}R_{12}\right\rangle
\nonumber\\
& \qquad +\left\Vert \left[\begin{array}{ccc}
0_{r}\\
 & L_{22}\overline{H}_{22}R_{22} & L_{22}\overline{H}_{23}\\
 & \overline{H}_{32}R_{22} & \overline{H}_{33}
\end{array}\right]\right\Vert _{*}\nonumber \\
 & =:\left\langle \overline{H},\overline{S}'\right\rangle +\left\langle \overline{H}_{22},L_{12}^{*}S_{11}R_{12}\right\rangle +\left\Vert L'\cdot\mathcal{P}_{\overline{\ST}^{\perp}}(\overline{H})\cdot R'\right\Vert _{*},\label{eq:before dual cert}
\end{align}
where the third line uses the duality of nuclear and spectral norms. The fourth idednity applies \eqref{eq:def of Wu} and \eqref{eq:PT PTperp}.
Above,  we also conveniently defined $\overline{S}',L'\in\mathbb{R}^{n\times n}$
as
\begin{equation*}
\overline{S}':=\left[\begin{array}{ccc}
L_{11}S_{11}R_{11} & L_{11}S_{11}R_{12}\\
L_{12}^{*}S_{11}R_{11} & 0_{r}\\
 &  & 0_{n-2r}
\end{array}\right],
\end{equation*}
\begin{equation}
L':=\left[\begin{array}{ccc}
0_{r}\\
 & L_{22}\\
 &  & I_{n-2r}
\end{array}\right].\label{eq:def of L'}
\end{equation}
We define $R'\in\mathbb{R}^{n\times n}$ similarly. The key feature
in (\ref{eq:before dual cert}) is that $\overline{S}'\in \overline{\ST}$, namely $\overline{S}'=\mathcal{P}_{\overline{\ST}}(\overline{S}')$.
Before going any further, let us record the following properties of $\overline{S}',L',R'$ for future reference:
\begin{align}
& \left\Vert \overline{S}'\right\Vert _{F}
\nonumber\\
& =\left\Vert \left[\begin{array}{ccc}
L_{11}S_{11}R_{11} & L_{11}S_{11}R_{12}\\
L_{12}^{*}S_{11}R_{11} & 0_{r}\\
 &  & 0_{n-2r}
\end{array}\right]\right\Vert _{F}
\quad \mbox{(see \eqref{eq:def of L'})}
\nonumber \\
& \le \left\Vert \left[\begin{array}{ccc}
L_{11}S_{11}R_{11} & L_{11}S_{11}R_{12}\\
L_{12}^{*}S_{11}R_{11} & L_{12}^*S_{11}R_{12}\\
 &  & 0_{n-2r}
\end{array}\right]\right\Vert _{F}
\nonumber\\
 & \le\left\Vert \left[\begin{array}{cc}
L_{11} & L_{12}\end{array}\right]\right\Vert \cdot
\left\Vert S_{11}\right\Vert _{F}
\cdot \left\Vert \left[\begin{array}{cc}
R_{11} & R_{12}\end{array}\right]\right\Vert
\nonumber \\
 & =\left\Vert \left[\begin{array}{cc}
L_{11} & L_{12}\end{array}\right]\right\Vert \cdot \|S\|_{F}
\cdot
\left\Vert
 \left[\begin{array}{cc}
R_{11} & R_{12}\end{array}\right]\right\Vert
\qquad \mbox{(see \eqref{eq:Sw def})}
\nonumber \\
 & =\sqrt{r}\left\Vert \left[\begin{array}{cc}
L_{11} & L_{12}\end{array}\right]\right\Vert \cdot\left\Vert \left[\begin{array}{cc}
R_{11} & R_{12}\end{array}\right]\right\Vert
\qquad \mbox{(see \eqref{eq:sign matrix props})}
\nonumber \\
 & \le \sqrt{r}\cdot \sqrt{\frac{\lambda^{4}\cos^{2}u_1+\sin^{2}u_1}{\lambda^{2}\cos^{2}u_1+\sin^{2}u_1}}
 \cdot \sqrt{\frac{\rho^{4}\cos^{2}v_1+\sin^{2}v_1}{\rho^{2}\cos^{2}v_1+\sin^{2}v_1}}
 \nonumber\\
&
 =:\sqrt{r}\cdot \Cr{4}(u_1,v_1,\lambda,\rho),
 \label{eq:frob of S' bar}
\end{align}
\begin{equation}
\left\| L'\right\| \le \|L\| =1. \qquad
\mbox{(see \eqref{eq:def of Wu}, (\ref{eq:norm of L}), and (\ref{eq:def of L'}))}
\label{eq:prop of L'}
\end{equation}
The second inequality in \eqref{eq:frob of S' bar} uses the fact that $\|AB\|_F\le \|A\|\cdot \|B\|_F$ for conforming matrices $A,B$. The last inequality there uses Lemma \ref{lem:props of L}.
Let us now continue the line of argument in (\ref{eq:before dual cert})
by writing that
\begin{align}
0 & \ge\left\langle \overline{H},\overline{S}'\right\rangle +\left\langle \overline{H}_{22},L_{12}^{*}S_{11}R_{12}\right\rangle +\left\Vert L'\mathcal{P}_{\overline{\ST}^{\perp}}(\overline{H}) R'\right\Vert _{*}
 \nonumber \\
 & \ge\left\langle \overline{H},\overline{S}'\right\rangle -\left\Vert \mathcal{P}_{\overline{\ST}^{\perp}}(\overline{H})\right\Vert _{*}\cdot\|L_{12}\|\|S_{11}\|\|R_{12}\| \nonumber\\
& \quad  +\left\Vert \mathcal{P}_{\overline{\ST}^{\perp}}(\overline{H})\right\Vert _{*}-\left\Vert \mathcal{P}_{\overline{\ST}^{\perp}}(\overline{H})-L'\mathcal{P}_{\overline{\ST}^{\perp}}(\overline{H})R'\right\Vert _{*}
 \nonumber \\
 & =\left\langle \overline{H},\overline{S}'\right\rangle +\left(1-\|L_{12}\|\|R_{12}\|\right)\left\Vert \mathcal{P}_{\overline{\ST}^{\perp}}(\overline{H})\right\Vert _{*}
\nonumber\\
& \quad  -\left\Vert \mathcal{P}_{\overline{\ST}^{\perp}}(I_{n})\cdot\mathcal{P}_{\overline{\ST}^{\perp}}(\overline{H})\cdot\mathcal{P}_{\overline{\ST}^{\perp}}(I_{n})-L' \mathcal{P}_{\overline{\ST}^{\perp}}(\overline{H}) R'\right\Vert _{*}
 \nonumber \\
 & \ge\left\langle \overline{H},\overline{S}'\right\rangle +\left(1-\|L_{12}\|\|R_{12}\|\right)\left\Vert \mathcal{P}_{\overline{\ST}^{\perp}}(\overline{H})\right\Vert _{*}
\nonumber\\
& \qquad
 -\left\Vert \mathcal{P}_{\overline{\ST}^{\perp}}(I_{n})
  -L'\right\Vert \left\Vert \mathcal{P}_{\overline{\ST}^{\perp}}(\overline{H})\right\Vert _{*} \nonumber\\
& \qquad  -\|L'\|\left\Vert \mathcal{P}_{\overline{\ST}^{\perp}}(\overline{H})\right\Vert _{*}\left\Vert \mathcal{P}_{\overline{\ST}^{\perp}}(I_{n})-R'\right\Vert
 \nonumber \\
 & \ge\left\langle \overline{H},\overline{S}'\right\rangle +\left(1-\|L_{12}\|\|R_{12}\|\right)\left\Vert \mathcal{P}_{\overline{\ST}^{\perp}}(\overline{H})\right\Vert _{*}
\nonumber\\
& \quad  -\left\Vert I_{r}-L_{22}\right\Vert \left\Vert \mathcal{P}_{\overline{\ST}^{\perp}}(\overline{H})\right\Vert _{*}-\left\Vert \mathcal{P}_{\overline{\ST}^{\perp}}(\overline{H})\right\Vert _{*}\left\Vert I_{r}-R_{22}\right\Vert
 \nonumber \\
 & =\left\langle \overline{H},\overline{S}'\right\rangle +\left(1-\|L_{12}\|\|R_{12}\|-\|I_{r}-L_{22}\| \right.\nonumber\\
& \qquad \qquad \qquad \qquad \left.  -\|I_{r}  -R_{22}\|\right)
\cdot  \left\Vert \mathcal{P}_{\overline{\ST}^{\perp}}(\overline{H})\right\Vert _{*}\nonumber\\
 &
 \ge
 \left\langle \overline{H},\overline{S}'\right\rangle  +\left(1-
\frac{3\sqrt{1-\lambda^{2}}\sin u_1}{2\sqrt{\lambda^{2}\cos^{2}u_1+\sin^{2}u_1}} \right. \nonumber\\
& \qquad \qquad \qquad \left.
-\frac{3\sqrt{1-\rho^{2}}\sin v_1}{2\sqrt{\rho^{2}\cos^{2}v_1+\sin^{2}v_1}}
 \right)\left\Vert \mathcal{P}_{\overline{\ST}^{\perp}}(\overline{H})\right\Vert _{*}
 \nonumber\\
 & =:
 \left\langle \overline{H},\overline{S}'\right\rangle +\left(1-
\Cr{3}\left(u_1,v_1,\lambda,\rho \right)
 \right)\left\Vert \mathcal{P}_{\overline{\ST}^{\perp}}(\overline{H})\right\Vert _{*}
.
 \label{eq:before dual cert 3}
\end{align}
Above, the first inequality applies \eqref{eq:before dual cert}. The second inequality uses the Holder's inequality, and
the fact that $\overline{H}_{22}$ is
a submatrix of $\mathcal{P}_{\overline{\ST}^{\perp}}(\overline{H})$ (see \eqref{eq:T bar}),
which yields $\|\overline{H}_{22}\|_{*}\le\|\mathcal{P}_{\overline{\ST}^{\perp}}(\overline{H})\|_{*}$.
An application of the
triangle inequality also appears in the second inequality. The first identity above uses \eqref{eq:sign matrix props} and then \eqref{eq:PT PTperp}. In the third inequality, we deployed the polarization identity  \eqref{eq:polarization},
followed
with an application of the triangle inequality, and then applied the fact that $\|AB\|_*\le \|A\|\cdot  \|B\|_*$ for conforming matrices $A,B$. In the fourth inequality, we used \eqref{eq:def of L'} and \eqref{eq:prop of L'}.  In the last inequality above, we used Lemma \ref{lem:props of L} and the observation that  $ab+a+b\le \frac{3}{2}(a+b)$ for    $a,b\in[0, 1]$.
At this point, we introduce the dual certificate. Validating the claims below are postponed until Appendix  \ref{sec:dual cert}.
\begin{lem}
\emph{\textbf{(Dual certificate)}} \label{cri:dual cri}Let the subspace $\overline{\ST}\subset\mathbb{R}^{n\times n}$
be the support of $\overline{M}_{r}$,  as defined in (\ref{eq:T bar}).
Assume that $\min_{i,j}p_{ij}\ge l$, with $l^{-1}=1/l(n)=\mbox{poly}(n)$, so that  $l^{-1}$ is bounded above by a polynomial in $n$ (of finite degree).
 Recall also the operators $\overline{\mathcal{R}}_{p}(\cdot)$ and $\overline{\mathcal{P}}_p(\cdot)$
from (\ref{eq:def of RUV}) and (\ref{eq:P bar}), respectively. Then, as long as
\begin{align*}
& \max\left[\log\left(\Cr{4}\cdot n\right),1\right]\cdot
\frac{(\mu_{i}+\nu_{j})r\log n}{n} \nonumber\\
& \qquad \cdot\max\left[\Cr{5} \left(1+{\max_{i}\frac{\breve{\mu}_{i}}{\mu_{i}}}+{\max_{j}\frac{\breve{\nu}_{j}}{\nu_{j}}}\right),1\right]
\lesssim p_{ij} \le 1
,
\end{align*}
for all $i,j\in [1:n]$, the following statements are all true. First,
\begin{equation}
\left\Vert \left(\mathcal{P}_{\overline{\ST}}-\mathcal{P}_{\overline{\ST}}\circ\overline{\mathcal{R}}_{p}\circ\mathcal{P}_{\overline{\ST}}\right)(\cdot)\right\Vert _{F\rightarrow F}\le\frac{1}{2},\label{eq:aux 12}
\end{equation}
except with a probability of  $o(n^{-19})$. Moreover, there exists
$\overline{\Lambda}\in\mathbb{R}^{n\times n}$ such that
\begin{equation}
\left\Vert \overline{S}'-\mathcal{P}_{\overline{\ST}}(\overline{\Lambda})\right\Vert _{F}\le\frac{l}{4\sqrt{2}},\label{eq:aux 13}
\end{equation}
\begin{equation}
\left\Vert \mathcal{P}_{\overline{\ST}^{\perp}}(\overline{\Lambda})\right\Vert \le\frac{1}{2},\label{eq:aux 14}
\end{equation}
\begin{equation}
\overline{\Lambda}=\overline{\mathcal{P}}_{p}(\overline{\Lambda}).\label{eq:aux 16}
\end{equation}
 Here,
\begin{align*}
\Cr{4} & = \Cr{4}\l(u_1,v_1,\lambda,\rho \r) \nonumber\\
& := \sqrt{\frac{\lambda^{4}\cos^{2}u_1+\sin^{2}u_1}{\lambda^{2}\cos^{2}u_1+\sin^{2}u_1}}
 \cdot \sqrt{\frac{\rho^{4}\cos^{2}v_1+\sin^{2}v_1}{\rho^{2}\cos^{2}v_1+\sin^{2}v_1}},
 \end{align*}
 \begin{align*}
 \Cr{5} & = \Cr{5}\l(u_1,v_1,\lambda,\rho \r)  \\
&  := \l(
\sqrt{\frac{\lambda^2 \cos^2 u_1+\sin^2 u_1}{\rho^2 \cos^2 v_1+\sin^2 v_1}} + \sqrt{\frac{\rho^2\cos^2 v_1+\sin^2 v_1}{\lambda^2 \cos^2 u_1+\sin^2 u_1}} \r) \\
& \qquad \cdot
\l(
 \sqrt{\lambda^4 \cos^2 u_1 +\sin^2 u_1}
 +\sqrt{\rho^4 \cos^2 v_1+\sin^2 v_1}
\r).
 \end{align*}
\end{lem}
Under Lemma \ref{cri:dual cri}, in particular,
there exists $\overline{\Lambda}$ that satisfies (\ref{eq:aux 13}-\ref{eq:aux 16}).
This allows us to continue the line of argument in (\ref{eq:before dual cert})
by writing that
\begin{align}
 0  & \ge\left\langle \overline{H},\overline{S}'\right\rangle +\left(1-\Cr{3} \right)\left\Vert \mathcal{P}_{\overline{\ST}^{\perp}}(\overline{H})\right\Vert _{*}
\qquad \mbox{(see \eqref{eq:before dual cert})}
 \nonumber \\
 & =\left\langle \overline{H},\mathcal{P}_{\overline{\ST}}(\overline{\Lambda})\right\rangle +\left\langle \overline{H},\overline{S}'-\mathcal{P}_{\overline{\ST}}(\overline{\Lambda})\right\rangle
 \nonumber\\
& \qquad  +\left(1-\Cr{3}\right)\left\Vert \mathcal{P}_{\overline{\ST}^{\perp}}(\overline{H})\right\Vert _{*}\nonumber \\
 & =\left\langle \overline{H},\overline{\Lambda}\right\rangle -\left\langle \overline{H},\mathcal{P}_{\overline{\ST}^{\perp}}(\overline{\Lambda})\right\rangle +\left\langle \overline{H},\overline{S}'-\mathcal{P}_{\overline{\ST}}(\overline{\Lambda})\right\rangle \nonumber\\
& \qquad
 +\left(1-\Cr{3}\right)\left\Vert \mathcal{P}_{\overline{\ST}^{\perp}}(\overline{H})\right\Vert _{*}\nonumber \\
 & = -\left\langle \overline{H},\mathcal{P}_{\overline{\ST}^{\perp}}(\overline{\Lambda})\right\rangle +\left\langle \overline{H},\overline{S}'-\mathcal{P}_{\overline{\ST}}(\overline{\Lambda})\right\rangle
 \nonumber\\
& \qquad  +\left(1-\Cr{3}\right)\left\Vert \mathcal{P}_{\overline{\ST}^{\perp}}(\overline{H})\right\Vert _{*}
 \nonumber \\
 & \ge -\left\Vert \mathcal{P}_{\overline{\ST}^{\perp}}(\overline{H})\right\Vert _{*}\left\Vert \mathcal{P}_{\overline{\ST}^{\perp}}(\overline{\Lambda})\right\Vert
 -\left\Vert \mathcal{P}_{\overline{\ST}}(\overline{H})\right\Vert _{F}
 \nonumber\\
& \qquad \cdot  \left\Vert \overline{S}'-\mathcal{P}_{\overline{\ST}}(\overline{\Lambda})\right\Vert _{F}
  +\left(1-\Cr{3}\right)\left\Vert \mathcal{P}_{\overline{\ST}^{\perp}}(\overline{H})\right\Vert _{*}
 \nonumber \\
 & \ge-\frac{1}{2}\left\Vert \mathcal{P}_{\overline{\ST}^{\perp}}(\overline{H})\right\Vert _{*}-\frac{l}{4\sqrt{2}}\left\Vert \mathcal{P}_{\overline{\ST}}(\overline{H})\right\Vert _{F}
\nonumber\\
& \qquad +\left(1-\Cr{3}\right)\left\Vert \mathcal{P}_{\overline{\ST}^{\perp}}(\overline{H})\right\Vert _{*}
 \nonumber \\
 & =\left(\frac{1}{2}-\Cr{3}\right)\left\Vert \mathcal{P}_{\overline{\ST}^{\perp}}(\overline{H})\right\Vert _{*}-\frac{l}{4\sqrt{2}}\left\Vert \mathcal{P}_{\overline{\ST}}(\overline{H})\right\Vert _{F},\label{eq:post lambda cert}
\end{align}
or, equivalently,
\begin{equation}
\left(\frac{1}{2}-\Cr{3}\right)\left\Vert \mathcal{P}_{\overline{\ST}^{\perp}}(\overline{H})\right\Vert _{*}\le\frac{l}{4\sqrt{2}}\left\Vert \mathcal{P}_{\overline{\ST}}(\overline{H})\right\Vert _{F}.\label{eq:post lambda cert 2}
\end{equation}
The bound above is nontrivial if $\Cr{3}=\Cr{3}(u_1,v_1,\lambda,\rho)<\frac{1}{2}$.
In  \eqref{eq:post lambda cert}, we  used the Holder's inequality. Also, the third inequality in \eqref{eq:post lambda cert} uses \eqref{eq:aux 13} and \eqref{eq:aux 14}.  Lastly, the third identity in (\ref{eq:post lambda cert}) holds because $\left\langle \overline{H},\overline{\Lambda}\right\rangle =0$.
To see why this is the case, first set $\overline{H}=B_{L}^*HB_{R}$
(see (\ref{eq:H decompose})). Then note that
\begin{align}
\left\Vert \overline{\mathcal{R}}_{p}(\overline{H})\right\Vert_{F} & =\left\Vert \mathcal{R}_{p}(H)\right\Vert _{F}
\qquad \mbox{(see \eqref{eq:aux 9})}
\nonumber \\
 & = \left\Vert \mathcal{R}_{p}(\widehat{M}-M)\right\Vert _{F}
 \nonumber \\
  & =0,
 \label{eq:feas 2}
\end{align}
where the last line uses the feasibility of  $M$ and  $\widehat{M}$  in Program \eqref{eq:p1}.
Therefore,
\begin{align*}
\left\langle \overline{H},\overline{\Lambda}\right\rangle  & =\left\langle \overline{H},\overline{\mathcal{P}}_{p}(\overline{\Lambda})\right\rangle
\qquad \mbox{(see \eqref{eq:aux 16})}
\\
 & =\left\langle \overline{\mathcal{P}}_{p}(\overline{H}),\overline{\Lambda}\right\rangle
\qquad \left( \overline{\mathcal{P}}_p(\cdot) \mbox{ is self-adjoint} \right)
 \\
 & =0,
\qquad \mbox{(see \eqref{eq:feas 2})}
\end{align*}
thereby verifying the fourth line of (\ref{eq:post lambda cert 2}).
On the other hand, to find a matching upper bound for \eqref{eq:post lambda cert 2}, we reason as follows. First, note that
\begin{align}
\left\Vert \overline{\mathcal{R}}_{p}\left(\mathcal{P}_{\overline{\ST}}(\overline{H})\right)\right\Vert _{F}
& =\left\Vert \overline{\mathcal{R}}_{p}\left(\mathcal{P}_{\overline{\ST}^{\perp}}(\overline{H})\right)\right\Vert _{F}
\qquad \mbox{(see \eqref{eq:feas 2})}
\nonumber\\
&
\le\left\Vert \overline{\mathcal{R}}_{p}\left(\cdot\right)\right\Vert _{F\rightarrow F}\cdot \left\Vert \mathcal{P}_{\overline{\ST}^{\perp}}(\overline{H})\right\Vert _{F}
\nonumber \\
 & \le\frac{1}{l}\left\Vert \mathcal{P}_{\overline{\ST}^{\perp}}(\overline{H})\right\Vert _{F}.
 \qquad \mbox{(see (\ref{eq:aux 10}))}
\label{eq:feas arg 0}
\end{align}
Under
 Lemma \ref{cri:dual cri}, $\overline{\mathcal{R}}_{p}(\cdot)$
acts as a near-isometry on the subspace $\overline{\ST}$, which allows us to find a matching lower bound for \eqref{eq:feas arg 0}:
\begin{align}
& \left\Vert \overline{\mathcal{R}}_{p}\left(\mathcal{P}_{\overline{\ST}}(\overline{H})\right)\right\Vert _{F}^{2}
\nonumber\\
& =\left\langle \mathcal{P}_{\overline{\ST}}(\overline{H}),\l(\overline{\mathcal{R}}_{p}
\circ \overline{\mathcal{R}}_p\r)
\left(\mathcal{P}_{\overline{\ST}}(\overline{H})\right)\right\rangle \nonumber \\
 & \ge\left\langle \mathcal{P}_{\overline{\ST}}(\overline{H}),\overline{\mathcal{R}}_{p}\left(\mathcal{P}_{\overline{\ST}}(\overline{H})\right)\right\rangle
\qquad \mbox{(see \eqref{eq:aux 11})}
  \nonumber \\
 & =\left\langle \mathcal{P}_{\overline{\ST}}(\overline{H}),\mathcal{P}_{\overline{\ST}}(\overline{H})\right\rangle
\nonumber\\
& \qquad  +\left\langle \mathcal{P}_{\overline{\ST}}(\overline{H}),\left(\mathcal{P}_{\overline{\ST}}\circ\overline{\mathcal{R}}_{p}\circ\mathcal{P}_{\overline{\ST}}-\mathcal{P}_{\overline{\ST}}\right)\circ\mathcal{P}_{\overline{\ST}}(\overline{H})\right\rangle \nonumber \\
 & \ge\left\Vert \mathcal{P}_{\overline{\ST}}(\overline{H})\right\Vert _{F}^{2}-\left\Vert \mathcal{P}_{\overline{\ST}}\circ\overline{\mathcal{R}}_{p}\circ\mathcal{P}_{\overline{\ST}}-\mathcal{P}_{\overline{\ST}}\right\Vert _{F\rightarrow F}
 \cdot \left\Vert \mathcal{P}_{\overline{\ST}}(\overline{H})\right\Vert _{F}^{2}\nonumber \\
 & \ge\frac{1}{2}\left\Vert \mathcal{P}_{\overline{\ST}}(\overline{H})\right\Vert _{F}^{2}.
\qquad \mbox{(see (\ref{eq:aux 12}))}
 \label{eq:feas arg 2}
\end{align}
Comparing \eqref{eq:feas arg 0} to  \eqref{eq:feas arg 2} yields that
\begin{equation}
\left\Vert \mathcal{P}_{\overline{\ST}}(\overline{H})\right\Vert _{F}\le\frac{\sqrt{2}}{l}\left\Vert \mathcal{P}_{\overline{\ST}^{\perp}}(\overline{H})\right\Vert _{F}.\label{eq:l2}
\end{equation}
The inequality above, when put together with (\ref{eq:post lambda cert 2}),
leads us to
\begin{align*}
\left(\frac{1}{2}-\Cr{3}\right)\left\Vert \mathcal{P}_{\overline{\ST}^{\perp}}(\overline{H})\right\Vert _{*} & \le\frac{l}{4\sqrt{2}}\left\Vert \mathcal{P}_{\overline{T}}(\overline{H})\right\Vert _{F},
\qquad \mbox{(see \eqref{eq:post lambda cert 2})}
 \end{align*}
which, as long as $\Cr{3}=\Cr{3}(u_1,v_1,\lambda,\rho)\le \frac{1}{8}$, immediately yields
that
\begin{equation}
 \mathcal{P}_{\overline{\ST}^{\perp}}(\overline{H})= 0.
\label{eq:final pre}
\end{equation}
We extend the error bound above to $\overline{H}$
by noting that
\begin{align*}
\left\Vert \overline{H}\right\Vert _{F} & \le\left\Vert \mathcal{P}_{\overline{\ST}}(\overline{H})\right\Vert _{F}+\left\Vert \mathcal{P}_{\overline{\ST}^{\perp}}(\overline{H})\right\Vert _{F}\\
 & \le\left(\frac{\sqrt{2}}{l}+1\right)\left\Vert \mathcal{P}_{\overline{\ST}^{\perp}}(\overline{H})\right\Vert _{F}
\qquad \mbox{(see (\ref{eq:l2}))}
 \\
 & = 0.
 \qquad \mbox{(see \eqref{eq:final pre})}
\end{align*}
Since $\|H\|_{F}=\|\overline{H}\|_{F}$
(see (\ref{eq:H decompose})), we find that $\widehat{M}=M$. Extending to nearly low-rank matrices ($M_{r^+}\ne 0$) and accounting for noise ($e> 0$) is a straightforward generalization of Theorem 7 in \cite{candes2010matrix}, matching \cite[Proposition 2]{eftekhari2016mc} nearly verbatim. Such an argument would lead us to:
\begin{align*}
\l\| \widehat{M}-M\r\|_F
& \lesssim \frac{\sqrt{h}}{l}\l\|Q_{\widetilde{\SU}_r,\lambda}  M_{r^+} Q_{\widetilde{\SV}_r,\rho}\r\|_*+ \frac{e \sqrt{n}h^{\frac{3}{2}}}{l}
\qquad
\nonumber\\
& \le \frac{\sqrt{h}}{l}\l\|Q_{\widetilde{\SU}_r,\lambda}  \r\| \cdot \l\| M_{r^+} \r\|_* \cdot  \l\| Q_{\widetilde{\SV}_r,\rho}\r\|+ \frac{e \sqrt{n}h^{\frac{3}{2}}}{l}
\nonumber\\
& = \frac{\sqrt{h}}{l} \l\| M_{r^+} \r\|_*+ \frac{e \sqrt{n}h^{\frac{3}{2}}}{l},
\qquad \mbox{(see \eqref{eq:norm of L})}
\end{align*}
where, in the first line, $\{p_{ij}\}\subset [l,h]$ and $e$   is the noise level. The second inequality uses the fact that  $\|AB\|_* \le \|A\|\cdot \|B\|_*$ for conforming matrices $A,B$.
We therefore arrive at the following result.
\begin{thm}
\textbf{\emph{(Leveraged matrix completion with prior knowledge)}}
For integer $r$ and matrix $M\in\mathbb{R}^{n\times n}$, let $M_r\in\mathbb{R}^{n\times n}$ be a rank-$r$ truncation of $M$, and let $M_{r^+}=M-M_r$ be the residual. Let $\{\mu_i,\nu_i\}_{i=1}^n$ be the leverage scores of $M_r$  (see (\ref{eq:mus}) and (\ref{eq:nus})).
Suppose that the $r$-dimensional subspaces $\SUT_r$ and $\SVT_r$ represent our prior knowledge about the column and row spaces of $M_r$, respectively, and let
\begin{equation*}
u = \angle \l[\SU_r,\SUT_r \r],
\qquad
v = \angle \l[\SV_r,\SVT_r \r],
\end{equation*}
denote the largest principal angles.  Take $\{\breve{\mu}_i,\breve{\nu}_i\}_{i=1}^n$ to be the leverage scores of subspaces $\breve{\SU}= \mbox{span}([\SU_r,\widetilde{\SU}_{r}])$
and $\breve{\SV}= \mbox{span}([\SV_r,\widetilde{\SV}_{r}])$, respectively. Moreover, consider the probabilities $\{p_{ij}\}_{i,j=1}^n \subset [l,h]$ for $0<l\le h\le 1$, and assume that $l^{-1}=1/l(n)$ is bounded above by a  polynomial in $n$ of finite degree. Recalling (\ref{eq:def of R(Z)}), acquire the (possibly noisy) measurement matrix $Y=\mathcal{R}_p(M+E)$ where $\|\mathcal{R}_p(E)\|_F\le  e$ for noise level $ e\ge 0$. Lastly, for $\lambda,\rho\in(0,1]$, let $\widehat{M}$ be a solution of Program (\ref{eq:p1 appendix}). Then, it holds that
\begin{equation*}
\left\Vert \widehat{M}-M\right\Vert _{F}
\lesssim \frac{\sqrt{h}}{l} \l\| M_{r^+} \r\|_*+ \frac{e \sqrt{n}h^{\frac{3}{2}}}{l}. ,
\end{equation*}
except with a probability of $o(n^{-19})$, and provided that
\begin{align*}
& \max\left[\log\left(\Cr{4}\cdot n\right),1\right]\cdot
\frac{(\mu_{i}+\nu_{j})r\log n}{n}
\nonumber\\
& \qquad \cdot\max\left[\Cr{5} \left(1+{\max_{i}\frac{\breve{\mu}_{i}}{\mu_{i}}}+{\max_{j}\frac{\breve{\nu}_{j}}{\nu_{j}}}\right),1\right]
\lesssim p_{ij} \le 1
,
\end{align*}
\begin{equation}
\Cr{3} \le \frac{1}{8},
\label{eq:p req thm proof}
\end{equation}
for all $i,j\in [1:n]$.
Above, we set
\begin{align*}
\Cr{4} & = \Cr{4}\l(u,v,\lambda,\rho \r) \\
& := \sqrt{\frac{\lambda^{4}\cos^{2}u+\sin^{2}u}{\lambda^{2}\cos^{2}u+\sin^{2}u}}
 \cdot \sqrt{\frac{\rho^{4}\cos^{2}v+\sin^{2}v}{\rho^{2}\cos^{2}v+\sin^{2}v}},
 \end{align*}
 \begin{align*}
\Cr{5} & = \Cr{5}\l(u,v,\lambda,\rho \r)  \\
& := \l(
\sqrt{\frac{\lambda^2 \cos^2 u+\sin^2 u}{\rho^2 \cos^2 v+\sin^2 v}} + \sqrt{\frac{\rho^2\cos^2 v+\sin^2 v}{\lambda^2 \cos^2 u+\sin^2 u}} \r)
\\
& \qquad \cdot
\l(
 \sqrt{\lambda^4 \cos^2 u +\sin^2 u}
 +\sqrt{\rho^4 \cos^2 v+\sin^2 v}
\r),
 \end{align*}
\begin{align*}
\Cr{3} & = \Cr{3}\l(u,v,\lambda,\rho \r) \\
& :=
\frac{3\sqrt{1-\lambda^{2}}\sin u}{2\sqrt{\lambda^{2}\cos^{2}u+\sin^{2}u}}
+\frac{3\sqrt{1-\rho^{2}}\sin v}{2\sqrt{\rho^{2}\cos^{2}v+\sin^{2}v}}.
\end{align*}
\end{thm}
In particular, suppose we replace the leverage scores $\{\mu_i,\nu_i\}_i$ with their upper bound $\eta_r(M)$ in Section \ref{sec:analysis of mc w pi} (see \eqref{eq:relation btw coh and lev}), i.e., we set $\mu_i=\nu_i=\eta(M_r)$ for all $i$. Then, all lemmas in Section \ref{sec:meas operator section} still hold (confer \cite{chen2013completing})  and so does the rest of the analysis in Section  \ref{sec:analysis of mc w pi}.
This leads to Theorem \ref{thm:main result} after noting that $\eta(\breve{U}\breve{V}^*)$ upper bounds $\{\breve{\mu}_i,\breve{\nu}_i\}_i$ for all $i$ (see \eqref{eq:mu nu tildes}).

\bibliographystyle{unsrt}
\bibliography{References}

\appendices

\section{Proof of Lemma \ref{lem:canonical}}
\label{sec:Proof of lemma canonical}
Let us focus on the column spaces first. Let $U_r,\wt{U}_r\in\mathbb{R}^{n\times r}$ be  orthonormal bases for the subspaces  $\SU_r,\SUT_r$, respectively.  Without loss of generality, assume that
\begin{equation}
U_{r}^{*}\widetilde{U}_{r} = \cos u\in \mathbb{R}^{r\times r}.
\end{equation}
(Otherwise, take the SVD of $U_{r}^{*}\widetilde{U}_{r}$ and redefine $U_r$ and $\widetilde{U}_r$ accordingly.)
To simplify the exposition, assume also that all $\sin u$ is invertible, namely  all principal angles $\{u_i\}_{i=1}^r$ are nonzero.
We set
\begin{equation}
U'_{r}:=-(I_{n}-U_{r}U_{r}^{*})\widetilde{U}_{r}\l(\sin(u) \r)^{-1}\in\mathbb{R}^{n\times r}.
\label{eq:Uprime}
\end{equation}
Then, it is easy to verify that $U'_{r}$ has orthonormal columns
and that $U_{r}^{'*}\widetilde{U}_{r}=-\sin u\in\mathbb{R}^{r\times r}$.
Furthermore, we take $U''_{n-2r}\in\mathbb{R}^{n\times (n-2r)}$ with orthonormal
columns such that
\begin{equation}
\mbox{span}\l(U''_{n-2r} \r) = \mbox{span}\l(\l[
\begin{array}{cc}
U_r & U'_r
\end{array}
\r]\r)^{\perp}.
\end{equation}
Likewise, we set
\[
\widetilde{U}'_{r}:=(I_{n}-\widetilde{U}_{r}\widetilde{U}_{r}^{*})U_{r}\l(\sin(u)\r)^{-1}\in\mathbb{R}^{n\times r},
\]
which, we may again verify, has orthonormal columns and satisfies $U_{r}^{*}\widetilde{U}'_{r}=\sin u\in\mathbb{R}^{r\times r}$.
It is similarly confirmed that $U_{r}^{'*}\widetilde{U}'_{r}=\cos u\in\mathbb{R}^{r\times r}$.
Lastly, we observe that
\begin{align*}
& \mbox{span}\left(\left[\begin{array}{cc}
U_{r} & \widetilde{U}_{r}\end{array}\right]\right) \nonumber\\
& =\mbox{span}\left(\left[\begin{array}{cc}
U_{r} & (I_{n}-U_{r}U_{r}^{*})\widetilde{U}_{r}\end{array}\right]\right)\nonumber\\
& =\mbox{span}\left(\left[\begin{array}{cc}
U_{r} & -(I_{n}-U_{r}U_{r}^{*})\widetilde{U}_{r}\l(\sin(u)\r)^{-1}\end{array}\right]\right)
\nonumber\\
& =\mbox{span}\left(\left[\begin{array}{cc}
U_{r} & U'_{r}\end{array}\right]\right).
\qquad \mbox{(see \eqref{eq:Uprime})}
\end{align*}
A similar argument shows that $\mbox{span}([U_{r}\,\,\widetilde{U}_{r}])=\mbox{span}([\widetilde{U}_{r}\,\,\widetilde{U}'_{r}])$ and, overall, we find that
\begin{align*}
\mbox{span}\l(\l[
\begin{array}{cc}
U_r & U'_r
\end{array}
\r]\r)
& =\mbox{span}
\l(\l[
\begin{array}{cc}
\wt{U}_r & \wt{U}'_r
\end{array}
\r]\r) \\
& =\mbox{span}
\l(\l[
\begin{array}{cc}
U_r & \wt{U}_r
\end{array}
\r]\r),
\end{align*}
which completes the proof of Lemma \ref{lem:canonical}.

\section{Proof of Lemma \ref{lem:props of L}}
\label{sec:proof of Lemma prop of L}
Below, we derive the claimed inequalities  directly from the definition of $L$ in \eqref{eq:def of Wu}, one by one:
\begin{align*}
& \left\Vert L_{11}\right\Vert \\
& = \left\| \Delta_L \right\|
\qquad \mbox{(see \eqref{eq:def of Wu})}
\nonumber\\
& = \max_i \sqrt{\lambda^2 \cos^2 u_i + \sin^2 u_i} \nonumber\\
& =\sqrt{\lambda^{2}\cos^{2}u_1+\sin^{2}u_1},
\qquad \l(u_1 = \max_i u_i\mbox{ and } \lambda\in (0,1] \r)
\end{align*}
\begin{align*}
\|L_{12}\|
& \le  \max_{i} \frac{\l(1-\lambda^2\r)\sin u_i \cdot \cos u_i}{\sqrt{\lambda^2 \cos^2 u_i +\sin^2 u_i}}
\qquad \mbox{(see \eqref{eq:def of Wu})}
\\
& \le \max_i \frac{\left(1-\lambda^{2}\right)\sin u_i}{\sqrt{\lambda^{2}\cos^{2}u_i+\sin^{2}u_i}}
\\
& = \frac{\left(1-\lambda^{2}\right)\sin u_1}{\sqrt{\lambda^{2}\cos^{2}u_1+\sin^{2}u_1}}
\qquad \l(u_1= \max_i u_i \r)
\\
& \le 1-\lambda^2,
\end{align*}
\begin{align*}
&\l\|I_{r}-L_{22}\r\|\\
& = \l\| I_r - \lambda\Delta_L^{-1} \r\|
\qquad \mbox{(see \eqref{eq:def of Wu})}
\\
& = = \max_i \l| 1- \frac{\lambda }{\sqrt{\lambda^2 \cos^2 u_i +\sin^2 u_i}} \r|
\nonumber\\
& = \max_i \l( 1- \frac{\lambda }{\sqrt{\lambda^2 \cos^2 u_i +\sin^2 u_i}} \r)
\qquad \l(\lambda \in (0,1] \r)
\\
& =  1- \frac{\lambda }{\sqrt{\lambda^2 \cos^2 u_1 +\sin^2 u_1}}
\qquad \l(u_1 = \max_i u_i \r)
\\
& \le\frac{\sqrt{1-\lambda^{2}}\sin u_1}
{\sqrt{\lambda^{2}\cos^{2}u_1+\sin^{2}u_1}}
\,\,
\left(\sqrt{a+b}-\sqrt{a}\le \sqrt{b},\, a,b\ge 0 \right)
\\
& \le \sqrt{1-\lambda^2},
\end{align*}
\begin{align*}
& \left\Vert \left[\begin{array}{cc}
L_{11} & L_{12}\end{array}\right]\right\Vert ^{2} \\
& =\max_{i}\left\Vert \left[\begin{array}{cc}
\sqrt{\lambda^{2}+(1-\lambda^{2})\sin^{2}u_i} & \frac{(1-\lambda^{2})\sin u_{i}\cos u_{i}}{\sqrt{\lambda^{2}+(1-\lambda^{2})\sin^{2}u_{i}}}\end{array}\right]\right\Vert _{2}^{2}
\nonumber \\
 & =\max_{i}\frac{\lambda^{4}+(1-\lambda^{4})\sin^{2}u_{i}}{\lambda^{2}+(1-\lambda^{2})\sin^{2}u_{i}}\nonumber \\
 & =\frac{\lambda^{4}+(1-\lambda^{4})\sin^{2}u_{1}}{\lambda^{2}+(1-\lambda^{2})\sin^{2}u_{1}}\le 1.
 \qquad
 \l( \lambda \in(0,1] \r)
\end{align*}
\begin{align*}
& \left\Vert \left[\begin{array}{cc}
L_{22}-I_{r} & L_{12}\end{array}\right]\right\Vert ^{2} \\
& =\max_{i}\left\Vert \left[\begin{array}{cc}
\frac{\lambda}{\sqrt{\lambda^{2}+(1-\lambda^{2})\sin^{2}u_{i}}}-1 & \frac{(1-\lambda^{2})\sin u_{i}\cos u_{i}}{\sqrt{\lambda^{2}+(1-\lambda^{2})\sin^{2}u_{i}}}\end{array}\right]\right\Vert _{2}^{2}
\nonumber \\
 & =\max_{i} \Bigg[ \frac{\left(\lambda-\sqrt{\lambda^{2}+(1-\lambda^{2})\sin^{2}u_{i}}\right)^{2}}{\lambda^{2}+(1-\lambda^{2})\sin^{2}u_{i}} \nonumber \\
& \qquad\qquad \qquad \qquad  \frac{(1-\lambda^{2})^{2}\sin^{2}u_{i}\cos^{2}u_{i}}{\lambda^{2}+(1-\lambda^{2})\sin^{2}u_{i}} \Bigg] \\
 & \le\max_{i}\frac{(1-\lambda^{2})\sin^{2}u_{i}+(1-\lambda^{2})^{2}\sin^{2}u_{i}\cos^{2}u_{i}}{\lambda^{2}+(1-\lambda^{2})\sin^{2}u_{i}}
 \nonumber \\
 & \le\max_{i}\frac{(1-\lambda^{2})\sin^{2}u_{i}+(1-\lambda^{2})\sin^{2}u_{i}}{\lambda^{2}+(1-\lambda^{2})\sin^{2}u_{i}}\nonumber \\
 & =\frac{2(1-\lambda^{2})\sin^{2}u_{1}}{\lambda^{2}+(1-\lambda^{2})\sin^{2}u_{1}}.
\end{align*}
To derive the last bound above, we used the inequality $\sqrt{a}-\sqrt{b}\le \sqrt{a-b} $ for $a\ge b\ge 0$.
Similar bounds hold for $R$ and its  blocks, $R_{11},R_{12},R_{22}$. This completes the proof of Lemma \ref{lem:props of L}.

\section{Proof of Lemma \ref{lem:props of Rbar}}
\label{sec:proof of props of Rbar}
With\textbf{ $\overline{Z}=B_{L}^{*}ZB_{R}\in\mathbb{R}^{n\times n}$},
note that
\begin{align}
\left\Vert \overline{\mathcal{R}}_{p}(\overline{Z})\right\Vert _{F}^{2}
 & =\left\langle \overline{\mathcal{R}}_{p}(\overline{Z}),\overline{\mathcal{R}}_{p}(\overline{Z})\right\rangle
 \nonumber \\
 & =\left\langle \mathcal{R}_{p}(Z),\mathcal{R}_{p}(Z)\right\rangle
\qquad \mbox{(see \eqref{eq:def of RUV})}
  \nonumber \\
 & =\left\Vert \mathcal{R}_{p}(Z)\right\Vert _{F}^{2},
 \label{eq:app 10}
\end{align}
as claimed in \eqref{eq:aux 9}, and \eqref{eq:aux 8.9} is proved similarly.
Additionally, if  $\{p_{ij}\}\subset[l,h]$ with $0<l\le h\le 1$, we observe that
\begin{align*}
&\left\langle \overline{Z},\overline{\mathcal{R}}_{p}(\overline{Z})\right\rangle \nonumber\\
& =\left\langle Z,\mathcal{R}_{p}(Z)\right\rangle
\qquad \mbox{(see \eqref{eq:aux 8.9})}
 \\
 & =\sum_{i,j}\frac{\epsilon_{ij}}{p_{ij}}\left|Z[i,j]\right|^{2}
\qquad \mbox{(see \eqref{eq:def of R(Z)})}
 \\
 & \le\sum_{i,j}\frac{\epsilon_{ij}}{p_{ij}^{2}}\left|Z[i,j]\right|^{2}
\qquad \left( p_{i,j}\le 1, \,\,\forall i,j \right)
 \\
 & =\left\langle \mathcal{R}_p(Z),\mathcal{R}_{p}(Z)\right\rangle
 \qquad \left( \epsilon_{ij}^2=\epsilon_{ij} \right)
  \\
  & =\left\langle \overline{\mathcal{R}}_p\left(\overline{Z}\right),\overline{\mathcal{R}}_{p}\left (\overline{Z}\right)\right\rangle
\qquad \mbox{(see \eqref{eq:app 10})}
 \\
 & =\left\langle \overline{Z},
 \l( \overline{\mathcal{R}}_p\circ \overline{\mathcal{R}}_{p}\r) \l(\overline{Z}\r)\right\rangle,
  \qquad \left( \overline{\mathcal{R}}_p(\cdot) \mbox{ is self-adjoint} \right)
\end{align*}
for every $\overline{Z}=B_{L}^{*}ZB_{R}\in\mathbb{R}^{n\times n}$,
thereby establishing that \eqref{eq:aux 11}.
 Also note that
\begin{align}\label{eq:mid 1e3}
\left\Vert \overline{\mathcal{R}}_{p}(\overline{Z})\right\Vert _{F} & =\left\Vert \mathcal{R}_{p}(Z)\right\Vert _{F}
\qquad \mbox{(see \eqref{eq:app 10})}
\nonumber
 \\
 & =\sqrt{\sum_{i,j}\frac{\epsilon_{ij}}{p_{ij}^{2}}\left|Z[i,j]\right|^{2}}
  \qquad \mbox{(see \eqref{eq:def of R(Z)})}
\nonumber
 \\
 & \le\frac{\|Z\|_F}{l},
 \qquad \left( p_{ij}\ge l,\,\,\forall i,j \right)
\end{align}
for any $\overline{Z}=B_{L}^{*}ZB_{R}\in\mathbb{R}^{n\times n}$. This proves \eqref{eq:aux 10}. Lastly, to verify \eqref{eq:P n R}, we write that
\begin{align}
\l\| \overline{\mathcal{R}}_p\l( \overline{Z}\r)\r\|_F
& = \sqrt{\sum_{i,j}\frac{\epsilon_{ij}}{p_{ij}^2} \l|Z[i,j]\r|^2 }
\qquad \mbox{(see \eqref{eq:mid 1e3})}
\nonumber\\
& \ge h^{-1} \sqrt{\sum_{i,j}{\epsilon_{ij}}\l|Z[i,j]\r|^2 } \nonumber\\
& = h^{-1} \l\| \mathcal{P}_p\l( Z\r) \r\|_F
\qquad \mbox{(see \eqref{eq:P})}
 \nonumber\\
& = h^{-1} \l\| \overline{\mathcal{P}}_p\l( \overline{Z}\r) \r\|_F.
\qquad \mbox{(similar to \eqref{eq:aux 9})}
\end{align}
This completes the proof of Lemma \ref{lem:props of Rbar}.

\section{Proof of Lemma \ref{cri:dual cri} (Constructing the Dual Certificate)}
\label{sec:dual cert}

Recall \eqref{eq:def of L'} and conveniently define
\begin{equation}
S':=B_{L}\overline{S}'B_{R}^{*}\in\mathbb{R}^{n\times n},\qquad\Lambda:=B_{L}\overline{\Lambda}B_{R}^{*}\in\mathbb{R}^{n\times n}.\label{eq:aux 2}
\end{equation}
Because of \eqref{eq:T bar} and \eqref{eq:def of L'}, we note that $\overline{S}'\in \overline{T}$ and  that consequently
\begin{equation}
S'\in\ST ,
\qquad
\mbox{(see \eqref{eq:aux 2} and \eqref{eq:T n Tbar})}
\label{eq:S' is in T}
\end{equation}
so that  $\mathcal{P}_{\ST}(S')=S'$.  After making this transformation and recalling \eqref{eq:connection btw projections}, \eqref{eq:def of RUV}, and \eqref{eq:P bar},  it is easily verified
that it suffices to prove that
\begin{equation}
\left\Vert \left(\mathcal{P}_{\ST}-\mathcal{P}_{\ST}\circ\mathcal{R}_{p}\circ\mathcal{P}_{\ST}\right)(\cdot)\right\Vert _{F\rightarrow F}\le\frac{1}{2},\label{eq:Lambda cnd 0-1}
\end{equation}
with high probability, and to prove the existence of $\Lambda\in\mathbb{R}^{n\times n}$ such that
\begin{equation}
\left\Vert S'-\mathcal{P}_{\ST}(\Lambda)\right\Vert _{F}\le\frac{l}{4\sqrt{2}},
\label{eq:Lambda cnd 1-1}
\end{equation}
\begin{equation}
\left\Vert \mathcal{P}_{\ST^{\perp}}\left(\Lambda\right)\right\Vert \le\frac{1}{2},\label{eq:Lambda cnd 2-1}
\end{equation}
\begin{equation}
\Lambda
= \mathcal{P}_{p}(\Lambda)
.\label{eq:lambda cnd 3-1}
\end{equation}
According to Lemma \ref{lem:near isometry 1}, (\ref{eq:Lambda cnd 0-1})
holds if the sampling probabilities $\{p_{ij}\}$ are sufficiently large (see \eqref{eq:no of samples}) and except with a probability of at most $n^{-20}$.

It remains to construct an admissible $\Lambda$. To that end, we use the golfing
scheme as follows \cite{gross2011recovering,chen2013completing}. Instead of $\mathcal{R}_{p}(\cdot)$, we equivalently measure
$M$ through $K$ independent applications of $\mathcal{R}_{q}(\cdot)$
(with probabilities $\{q_{ij}\}$ and integer $K$ to be set later).
Given $K$, the two measurement schemes are equivalent (in distribution) if
\begin{equation}
(1-q_{ij})^{K}=1-p_{ij},\qquad
i,j\in[1: n].
\label{eq:qpK}
\end{equation}
We in fact  assume that the set of
observed entries at hand is generated through $K$ independent applications
of $\mathcal{R}_{q}(\cdot)$ (rather than an application of $\mathcal{R}_{p}(\cdot)$).
Next, we set  $\Lambda^{0}=0$, and define
$\Lambda^{k}$, $k\in [1:K]$, as follows:
\begin{equation}
\Lambda^{k}:=\sum_{k=1}^{k'}\mathcal{R}_{q}\left(W^{k'-1}\right),\label{eq:const 1}
\end{equation}
\begin{equation}
W^{k'}:=S'-\mathcal{P}_{\ST}(\Lambda^{k'}).\label{eq:const 2}
\end{equation}
We then set $\Lambda=\Lambda^{K}$; it  readily follows that $\Lambda$  satisfies \eqref{eq:lambda cnd 3-1}, once we recall \eqref{eq:P}. We turn
our attention to verifying (\ref{eq:Lambda cnd 1-1}) next. For
every $k\in [1:K]$, note that
\begin{align}
W^{k} & =S'-\mathcal{P}_{\ST}(\Lambda^{k})
\nonumber\\
 & =S'-\mathcal{P}_{\ST}(\Lambda^{k-1})-(\mathcal{P}_{\ST}\circ\mathcal{R}_{q})(W^{k-1})
\qquad \mbox{(see \eqref{eq:const 1})}
\nonumber\\
 & =S'-\mathcal{P}_{\ST}(\Lambda^{k-1})-(\mathcal{P}_{\ST}\circ\mathcal{R}_{q}\circ\mathcal{P}_{\ST})(W^{k-1})
\nonumber \\
 & =W^{k-1}-(\mathcal{P}_{\ST}\circ\mathcal{R}_{q}\circ\mathcal{P}_{\ST})(W^{k-1})
\qquad \mbox{(see \eqref{eq:const 2})}
\nonumber \\
 & =\left(\mathcal{P}_{\ST}-\mathcal{P}_{\ST}\circ\mathcal{R}_{q}\circ\mathcal{P}_{\ST}\right)\left(W^{k-1}\right),
 \qquad \left(W^{k-1}\in \ST \right)
 \label{eq:leg 100}
\end{align}
where the third line above uses the fact that $W^{k-1}\in \ST
$  from \eqref{eq:const 2} and \eqref{eq:S' is in T}. Under Lemma \ref{lem:near isometry 1}, it then follows that
\begin{align}
\left\Vert W^{k}\right\Vert _{F} & \le\left\Vert \left(\mathcal{P}_{\ST}-\mathcal{P}_{\ST}\circ\mathcal{R}_{q}\circ\mathcal{P}_{\ST}\right)(\cdot)\right\Vert _{F\rightarrow F}\left\Vert W^{k-1}\right\Vert _{F}
\nonumber
\\
 & \le\frac{1}{2}\left\Vert W^{k-1}\right\Vert _{F},
 \label{eq:1/2 drop}
\end{align}
as long as
\begin{equation}
\frac{\left(\mu_{i}+\nu_{j}\right)r\log n}{n} \lesssim q_{ij} \le 1
 ,
\qquad \forall i,j\in[1:n],
\quad \mbox{ (see \eqref{eq:no of samples})}
\label{eq:leg 101}
\end{equation}
and except with a probability of at most $n^{-20}$. It immediately follows
that
\begin{align}
\left\Vert  W^{K}\right\Vert _{F} & \le\left\Vert \l( \mathcal{P}_{\ST}-\mathcal{P}_{\ST}\circ\mathcal{R}_{q}\circ\mathcal{P}_{\ST} \r) (\cdot) \right\Vert _{F\rightarrow F}^{K}\cdot \left\Vert W^{0}\right\Vert _{F}\nonumber \\
 & \le\left(\frac{1}{2}\right)^{K}\left\Vert S'\right\Vert _{F}
 \nonumber \\
 & =\left(\frac{1}{2}\right)^{K}\left\Vert B_{L}\overline{S}'B_{R}^{*}\right\Vert _{F}
\qquad \mbox{(see \eqref{eq:aux 2})}
 \nonumber \\
 & =\left(\frac{1}{2}\right)^{K}\left\Vert \overline{S}'\right\Vert _{F}
\quad \l(B_L,B_R \mbox{ are orthonormal bases} \r)
 \nonumber \\
 & \le\left(\frac{1}{2}\right)^{K}\Cr{4} \sqrt{r},
\quad \mbox{(see (\ref{eq:frob of S' bar}))}
 \label{eq:needed later 3}
\end{align}
except with a probability of at most $Kn^{-20}$ (invoking the union bound).
 The second line above uses \eqref{eq:1/2 drop} and then \eqref{eq:const 2}  for $k=0$.
From \eqref{eq:const 2} and \eqref{eq:needed later 3}, it now follows that
\[
\left\Vert S'-\mathcal{P}_{\ST}\left(\Lambda\right)\right\Vert _{F}=\left\Vert S'-\mathcal{P}_{\ST}\left(\Lambda^{K}\right)\right\Vert _{F}=\left\Vert W^{K}\right\Vert _{F}\le\frac{l}{4\sqrt{2}},
\]
if we take
\[
K \gtrsim \max\l[ \log\l( \frac{8 \Cr{4}\sqrt{r}}{l} \r) , 1 \r].
\]
Assume that $l^{-1}$ is polynomial in $n$, namely that $l^{-1}=l^{-1}(n)$ is bounded above by a polynomial in $n$ of finite degree.  We therefore established that $\Lambda=\Lambda^{K}$, as constructed above,
satisfies (\ref{eq:Lambda cnd 1-1}) with $K\approx \log(\beta n)$ and except with a probability of
at most
\begin{equation}
Kn^{-20}=
O\l(\log \l(\Cr{4} \cdot n\r)\r)\cdot n^{-20} =
o(n^{-19}).
\label{eq:poly takes control}
\end{equation}
It remains to verify that $\Lambda$  also meets the remaining
requirements in (\ref{eq:Lambda cnd 2-1}).
Introducing a factor $\Delta>0$
to be set later, we observe that
\begin{align}
& \left\Vert \mathcal{P}_{\ST^{\perp}}\left(\Lambda\right)\right\Vert \nonumber\\
&  = \left\Vert \mathcal{P}_{\ST^{\perp}}\left(\Lambda^{K}\right)\right\Vert
\nonumber\\
& \le\sum_{k=1}^{K}\left\Vert \l( \mathcal{P}_{\ST^{\perp}}\circ\mathcal{R}_{q}
\r)
\left(W^{k-1}\right)\right\Vert
\qquad \mbox{(see \eqref{eq:const 1})}
\nonumber \\
 & =\sum_{k=1}^{K}\left\Vert
 \l( \mathcal{P}_{\ST^{\perp}}\circ\left(\mathcal{I}-\mathcal{R}_{q}\right)
 \r)
 \left(W^{k-1}\right)\right\Vert
\qquad  \l(W^{k-1}\in \ST \r)
  \nonumber \\
 & \le\sum_{k=1}^{K}\left\Vert \left(\mathcal{I}-\mathcal{R}_{q}\right)\left(W^{k-1}\right)\right\Vert \nonumber \\
 & \le\frac{1}{\Delta}\left[\sum_{k=1}^{K}\left\Vert W^{k-1}\right\Vert _{\mu(\infty)}+\left\Vert W^{k-1}\right\Vert _{\mu(\infty,2)}\right],
\quad \mbox{(Lemma \ref{lem:any Z})}
 \label{eq:leg 10}
\end{align}
as long as \begin{equation}
 \frac{\Delta^2(\mu_{i}+\nu_{j})r\log n}{n}
\lesssim q_{ij} \le  1 ,
\qquad i,j\in [1:n],
\label{eq:the one with Delta}
\end{equation}
and except for a probability
of $Kn^{-20}=o(n^{-19})$ since $l^{-1}=\mbox{poly}(n)$ (as described in \eqref{eq:poly takes control}).
Consider the weighted infinity norm
in the last line of \eqref{eq:leg 10}. Under Lemma \ref{lem:inf bound}, we note
that
\begin{align}
& \left\Vert W^{k-1}\right\Vert _{\mu(\infty)}\nonumber\\
 & =\left\Vert \left(\mathcal{P}_{\ST}-\mathcal{P}_{\ST}\circ\mathcal{R}_{q}\circ\mathcal{P}_{\ST}\right)\left(W^{k-2}\right)\right\Vert _{\mu(\infty)}
\qquad \mbox{(see \eqref{eq:leg 100})}
\nonumber \\
 & \le\frac{1}{2}\left\Vert W^{k-2}\right\Vert _{\mu(\infty)}
\qquad \mbox{(see Lemma \ref{lem:inf bound})}
 \nonumber \\
 & \le\left(\frac{1}{2}\right)^{k-1}\left\Vert W^{0}\right\Vert _{\mu(\infty)},\label{eq:leg 20}
\end{align}
as long as \eqref{eq:leg 101} holds
 and except for a probability of at most $(k-1)n^{-20}=o(n^{-19})$, since $k\le K$. Next, we consider the second norm in the last
line of (\ref{eq:leg 10}). Appealing to  Lemma \ref{lem:inf two bnd}, we observe that
\begin{align*}
& \left\Vert W^{k-1}\right\Vert _{\mu(\infty,2)} \nonumber\\
& =\left\Vert \left(\mathcal{P}_{\ST}-\mathcal{P}_{\ST}\circ\mathcal{R}_{q}\circ\mathcal{P}_{\ST}\right)\left(W^{k-2}\right)\right\Vert _{\mu(\infty,2)}
\qquad \mbox{(see \eqref{eq:leg 100})}
\\
 & \le\frac{1}{2}\left\Vert W^{k-2}\right\Vert _{\mu(\infty)}+\frac{1}{2}\left\Vert W^{k-2}\right\Vert _{\mu(\infty,2)}
\qquad \mbox{(see Lemma \ref{lem:inf two bnd})}
 \\
 & \le\left(\frac{1}{2}\right)^{k-1}\left\Vert W^{0}\right\Vert _{\mu(\infty)}+\frac{1}{2}\left\Vert W^{k-2}\right\Vert _{\mu(\infty,2)},
 \quad \mbox{(see \eqref{eq:leg 20})}
\end{align*}
as long as \eqref{eq:leg 101} holds and except for a probability of at most $k n^{-20}=o(n^{-19})$.
From the above recursion, it immediately  follows that
\begin{align}
\label{eq:leg 102}
& \left\Vert W^{k-1}\right\Vert _{\mu(\infty,2)} \nonumber\\
& \le (k-1)\left(\frac{1}{2}\right)^{k-1}\left\Vert W^{0}\right\Vert _{\mu(\infty)}+\left(\frac{1}{2}\right)^{k-1}\left\Vert W^{0}\right\Vert _{\mu(\infty,2)}.
 \end{align}
Substituting the last two estimates back into (\ref{eq:leg 10}), we arrive
at
\begin{align}
 & \left\Vert \mathcal{P}_{\ST^{\perp}}\left(\Lambda^{K}\right)\right\Vert \nonumber \\
 & \le\frac{1}{\Delta}\left[\sum_{k=1}^{K}\left\Vert W^{k-1}\right\Vert _{\mu(\infty)}+\left\Vert W^{k-1}\right\Vert _{\mu(\infty,2)}\right]
\qquad \mbox{(see \eqref{eq:leg 10})}
 \nonumber \\
 & \le\frac{1}{\Delta}\Bigg[\sum_{k=1}^{K}\left(\frac{1}{2}\right)^{k-1}\left\Vert W^0\right\Vert _{\mu(\infty)} \nonumber\\
 &  +\left((k-1)\left(\frac{1}{2}\right)^{k-1}\left\Vert W^0\right\Vert _{\mu(\infty)}+\left(\frac{1}{2}\right)^{k-1}\left\Vert W^0\right\Vert _{\mu(\infty,2)}\right)\Bigg]
 \nonumber \\
 & \le
\frac{4}{\Delta}\left\Vert W^0\right\Vert _{\mu(\infty)}+\frac{2}{\Delta}\left\Vert W^0\right\Vert _{\mu(\infty,2)}
 \nonumber \\
 & =\frac{4}{\Delta}\left\Vert S'\right\Vert _{\mu(\infty)}+\frac{2}{\Delta}\left\Vert S'\right\Vert _{\mu(\infty,2)}.
\qquad \mbox{(see \eqref{eq:const 2})}
 \label{eq:leg 19}
\end{align}
The second inequality above uses \eqref{eq:leg 20} and \eqref{eq:leg 102}.
Both norms in the last line above are bounded in Appendix \ref{sec:Estimates-of-S' norms}.
\begin{lem}\label{lem:bnd on Sp}
For $S'\in\mathbb{R}^{n\times n}$ defined in (\ref{eq:aux 2}) (also see (\ref{eq:def of L'})), it holds that
\begin{equation*}
\l\| S' \r\|_{\mu(\infty)} \le \Cr{5}
\left(1+\sqrt{2\max_{i}\frac{\breve{\mu}_{i}}{\mu_{i}}}+\sqrt{2\max_{j}\frac{\breve{\nu}_{j}}{\nu_{j}}}\right),
\end{equation*}
\begin{equation*}
\l\| S' \r\|_{\mu(\infty,2)} \le 2\Cr{5}
\left(1+\sqrt{2\max_{i}\frac{\breve{\mu}_{i}}{\mu_{i}}}+\sqrt{2\max_{j}\frac{\breve{\nu}_{j}}{\nu_{j}}}\right),
\end{equation*}
where
\begin{align}
\Cr{5} & = \Cr{5}\l(u_1,v_1,\lambda,\rho \r)  \nonumber\\
&
= \l(
\sqrt{\frac{\lambda^2 \cos^2 u_1+\sin^2 u_1}{\rho^2 \cos^2 v_1+\sin^2 v_1}} + \sqrt{\frac{\rho^2\cos^2 v_1+\sin^2 v_1}{\lambda^2 \cos^2 u_1+\sin^2 u_1}} \r) \nonumber\\
& \qquad \cdot
\l(
\sqrt{\lambda^4 \cos^2 u_1 +\sin^2 u_1}+
\sqrt{\rho^4 \cos^2 v_1+\sin^2 v_1}
\r).
\end{align}
\end{lem}
In light of Lemma \ref{lem:bnd on Sp}, we accordingly update (\ref{eq:leg 19})
to read
\begin{align}
& \l\| \mathcal{P}_{\ST}^\perp (\Lambda) \r\| \nonumber\\
& =\left\Vert \mathcal{P}_{{\ST}^{\perp}}\left({\Lambda}^{K}\right)\right\Vert \nonumber\\
 & \le\frac{4}{\Delta}\left\Vert {S}'\right\Vert _{\mu(\infty)}+\frac{2}{\Delta}\left\Vert {S}'\right\Vert _{\mu(\infty,2)}
\qquad \mbox{(see \eqref{eq:leg 19})}
 \nonumber\\
 & \le\frac{8}{\Delta}\cdot  \Cr{5}\left(1+\sqrt{2\max_{i}\frac{\breve{\mu}_{i}}{\mu_{i}}}+\sqrt{2\max_{j}\frac{\breve{\nu}_{j}}{\nu_{j}}}\right)
\quad \mbox{(Lemma \ref{lem:bnd on Sp})}
 \nonumber\\
 & \le\frac{1}{2},
 \label{eq:1/2 bnd proof}
\end{align}
where we took
\[
\Delta=16 \Cr{5}  \l(
1+\sqrt{2\max_{i}\frac{\breve{\mu}_{i}}{\mu_{i}}}+\sqrt{2\max_{j}\frac{\breve{\nu}_{j}}{\nu_{j}}}
\r).
\]
After recalling  (\ref{eq:the one with Delta}), we observe that  \eqref{eq:1/2 bnd proof} (equivalently, \eqref{eq:Lambda cnd 2-1})  holds provided that
\begin{equation}
\Cr{5}^2 \l(1+\max_i \frac{\breve{\mu}_i}{\mu_i} +\max_j \frac{\breve{\nu}_j}{\nu_j}\r)
\frac{\l(\mu_i+\nu_j \r) r \log n}{n}
\lesssim
q_{ij}
\le 1,\label{eq:q so far}
\end{equation}
and except with a probability of  $o(n^{-20})$.
Overall, we conclude that the dual certificate exists as long as
\begin{align*}
& \max\l[  \Cr{5}^2\l(1+{\max_i \frac{\breve{\mu}_i}{\mu_i}} + {\max_j \frac{\breve{\nu}_j}{\nu_j}}\r),1  \r]
\cdot
\frac{(\mu_i+\nu_j)r\log n}{n} \\
& \qquad \qquad \qquad  \lesssim  q_{ij} \le 1, \qquad i,j\in [1:n],
\end{align*}
\[
K=\max\left[\log\left(\Cr{4}\cdot n\right),1\right].
\]
Lastly, recall the relation between $\{q_{ij}\},K$ and $\{p_{ij}\}$ in \eqref{eq:qpK}:
\begin{align*}
p_{ij} & =1-(1-q_{ij})^{K}
\qquad \mbox{(see \eqref{eq:qpK})}
\\
 & \gtrsim K\cdot q_{ij}\\
 & =\max\left[\log\left(\Cr{4}\cdot n\right),1\right]\cdot
\frac{\l(\mu_i+\nu_j\r)r\log n}{n} \\
& \qquad  \cdot\max\left[\Cr{5} \left(1+\sqrt{\max_{i}\frac{\widetilde{\mu}_{i}}{\mu_{i}}}+\sqrt{\max_{j}\frac{\widetilde{\nu}_{j}}{\nu_{j}}}\right),1\right].
\end{align*}
The second line holds if $\{q_{ij}\}$ are sufficiently small, i.e.,
when $n$ is sufficiently large. This completes the proof of Lemma \ref{cri:dual cri}.

\section{Proof of Lemma \ref{lem:bnd on Sp} \label{sec:Estimates-of-S' norms}}

Here, we will estimate the norms of $S'$ (see \eqref{eq:aux 2} and \eqref{eq:def of L'}).
 We write that
\begin{align*}
&\left\Vert S'\right\Vert _{\mu(\infty)} \nonumber\\
& =\left\Vert \left(\frac{\mu r}{n}\right)^{-\frac{1}{2}}\cdot S' \cdot\left(\frac{\nu r}{n}\right)^{-\frac{1}{2}}\right\Vert _{\infty}
\qquad \mbox{(see \eqref{eq:mu inf norm})}
\nonumber\\
& =\left\Vert \left(\frac{\mu r}{n}\right)^{-\frac{1}{2}}\cdot B_{L}\overline{S}'B_{R}^{*}\cdot\left(\frac{\nu r}{n}\right)^{-\frac{1}{2}}\right\Vert _{\infty}
\qquad \mbox{(see \eqref{eq:aux 2})}
\nonumber \\
 & =\Bigg\Vert \left(\frac{\mu r}{n}\right)^{-\frac{1}{2}}\cdot B_{L}\left[\begin{array}{ccc}
L_{11}S_{11}R_{11} & L_{11}S_{11}R_{12}\\
L_{12}^{*}S_{11}R_{11} & 0_{r}\\
 &  & 0_{n-2r}
\end{array}\right] \nonumber\\
& \qquad \qquad  \cdot B_{R}^{*}\cdot\left(\frac{\nu r}{n}\right)^{-\frac{1}{2}}\Bigg\Vert _{\infty},
\end{align*}
 where, in the last line, we used \eqref{eq:def of L'}. It follows that
 \begin{align*}
 &\left\Vert S'\right\Vert _{\mu(\infty)} \nonumber\\
 & \le\left\Vert \left(\frac{\mu r}{n}\right)^{-\frac{1}{2}}\cdot U_{r}\cdot L_{11}S_{11}R_{11}\cdot V_{r}^{*}\cdot\left(\frac{\nu r}{n}\right)^{-\frac{1}{2}}\right\Vert _{\infty} \nonumber\\
&   +\left\Vert \left(\frac{\mu r}{n}\right)^{-\frac{1}{2}}\cdot U_{r}\cdot L_{11}S_{11}R_{12}\cdot \l(V_{r}'\r)^{*}\cdot\left(\frac{\nu r}{n}\right)^{-\frac{1}{2}}\right\Vert _{\infty}\nonumber \\
 & +\left\Vert \left(\frac{\mu r}{n}\right)^{-\frac{1}{2}}\cdot U'_{r}\cdot L_{12}^{*}S_{11}R_{11}\cdot V_{r}^{*}\cdot\left(\frac{\nu r}{n}\right)^{-\frac{1}{2}}\right\Vert _{\infty}
\, \mbox{(see \eqref{eq:def of BL})}
 \nonumber \\
 & \le\left\Vert \left(\frac{\mu r}{n}\right)^{-\frac{1}{2}}U_{r}\right\Vert _{2\rightarrow\infty}\cdot\left\Vert L_{11}S_{11}R_{11}\right\Vert \cdot\left\Vert  \left(\frac{\nu r}{n}\right)^{-\frac{1}{2}}V_{r}\right\Vert _{2\rightarrow\infty}\nonumber \\
 & +\left\Vert \left(\frac{\mu r}{n}\right)^{-\frac{1}{2}}U_{r}\right\Vert _{2\rightarrow\infty}\cdot\left\Vert L_{11}S_{11}R_{12}\right\Vert \cdot\left\Vert \left(\frac{\nu r}{n}\right)^{-\frac{1}{2}}V'_{r}\right\Vert _{2\rightarrow\infty}\nonumber \\
 & +\left\Vert \left(\frac{\mu r}{n}\right)^{-\frac{1}{2}}U'_{r}\right\Vert _{2\rightarrow\infty}\cdot\left\Vert L_{12}^{*}S_{11}R_{11}\right\Vert \nonumber\\
&  \quad \cdot\left\Vert \left(\frac{\nu r}{n}\right)^{-\frac{1}{2}}V_{r}'\right\Vert _{2\rightarrow\infty},
\end{align*}
where the last inequality uses the fact that $
\|AB\|_{\infty} \le \|A\|_{2\rightarrow\infty} \cdot \|B^*\|_{2\rightarrow\infty}$ for conforming matrices $A,B$. We continue by writing that
\begin{align*}
& \l\|S'\r\|_{\mu(\infty)} \nonumber\\
 & \le\left\Vert L_{11}\right\Vert \left\Vert S_{11}\right\Vert \left\Vert R_{11}\right\Vert +\left\Vert L_{11}\right\Vert \left\Vert S_{11}\right\Vert \left\Vert R_{12}\right\Vert \sqrt{2\max_{j}\frac{\breve{\nu}_{j}}{\nu_{j}}}\nonumber\\
 & \qquad +\sqrt{2\max_{i}\frac{\breve{\mu}_{i}}{\mu_{i}}}\left\Vert L_{12}\right\Vert \left\Vert S_{11}\right\Vert \left\Vert R_{11}\right\Vert
\qquad \mbox{(see \eqref{eq:useful eqs})}
  \nonumber \\
 & =\left\Vert L_{11}\right\Vert \left\Vert R_{11}\right\Vert +\left\Vert L_{11}\right\Vert \left\Vert R_{12}\right\Vert \sqrt{2\max_{j}\frac{\breve{\nu}_{j}}{\nu_{j}}} \nonumber\\
& \qquad  +\sqrt{2\max_{i}\frac{\breve{\mu}_{i}}{\mu_{i}}}\left\Vert L_{12}\right\Vert \left\Vert R_{11}\right\Vert,
\qquad \mbox{(see \eqref{eq:sign matrix props})}
\end{align*}
which itself leads to
\begin{align*}
& \l\| S'\r\|_{\mu(\infty)} \nonumber\\
&
\le \left\Vert L_{11} \r\Vert
\l(  \left\Vert R_{11}\right\Vert + \l\|R_{12} \r\| \sqrt{2\max_j \frac{\breve{\nu}_j}{\nu_j}} \r)
\nonumber\\
& \qquad + \l\| R_{11} \r\| \l( \l\|L_{11}\r\| + \l\| L_{12} \r\|\sqrt{2\max_i \frac{\breve{\mu}_i}{{\mu}_i}} \r)\nonumber\\
& \le
  \left\Vert L_{11} \r\Vert
 \cdot
  \left\Vert
\l[
\begin{array}{cc}
R_{11} & R_{12}
\end{array}
\r]
\r\|
\cdot \l(1+\sqrt{2\max_j \frac{\breve{\nu}_j}{\nu_j}}\r)
\nonumber\\
& \quad+  \l\| R_{11}\r\|
\cdot
 \left\Vert
\l[
\begin{array}{cc}
L_{11} & L_{12}
\end{array}
\r]
\r\|
\cdot \l(1+\sqrt{2\max_i \frac{\breve{\mu}_i}{\mu_i}} \r).
\end{align*}
After invoking Lemma \ref{lem:props of L}, we continue simplifying the last line above by writing that
\begin{align}
& \l\| S' \r\|_{\mu(\infty)} \nonumber\\
&
\le
\sqrt{\lambda^2 \cos^2 u_1+\sin^2 u_1}
\cdot \sqrt{\frac{\rho^{4} \cos^2 v_1 +\sin^{2}v_{1}}{\rho^{2}\cos^2 v_1+\sin^{2}v_{1}}} \nonumber\\
& \cdot \l(1+\sqrt{2\max_j \frac{\breve{\nu}_j}{\nu_j}}\r)
+
\sqrt{\rho^2 \cos^2 v_1+\sin^2 v_1} \nonumber\\
& \qquad
\cdot
\sqrt{\frac{\lambda^{4} \cos^2 u_1 +\sin^{2}u_{1}}{\lambda^{2}\cos^2 u_1+\sin^{2}u_{1}}}  \cdot \l(1+\sqrt{2\max_i \frac{\breve{\mu}_i}{\mu_i}} \r)
\nonumber\\
& \le
\l(
\sqrt{\frac{\lambda^2 \cos^2 u_1+\sin^2 u_1}{\rho^2 \cos^2 v_1+\sin^2 v_1}} + \sqrt{\frac{\rho^2\cos^2 v_1+\sin^2 v_1}{\lambda^2 \cos^2 u_1+\sin^2 u_1}} \r)
\nonumber\\
& \qquad \cdot
\l(
\sqrt{\rho^4 \cos^2 v_1+\sin^2 v_1} + \sqrt{\lambda^4 \cos^2 u_1 +\sin^2 u_1}
\r)\nonumber\\
& \qquad
\cdot \l(1+\sqrt{2\max_j \frac{\breve{\nu}_j}{\nu_j}}+\sqrt{2\max_i \frac{\breve{\mu}_i}{\mu_i}} \r)
\nonumber\\
& =: \Cr{5}\l(u_1,v_1,\lambda,\rho  \r)  \cdot \l(1+\sqrt{2\max_j \frac{\breve{\nu}_j}{\nu_j}}+\sqrt{2\max_i \frac{\breve{\mu}_i}{\mu_i}} \r),
 \label{eq:leg 20-1}
\end{align}
where the last inequality above uses the fact that $ac+bd\le (a+b)(c+d)$ whenever $a,b,c,d\ge 0$.
As for $\|S'\|_{\mu(2,\infty)}$, we begin with writing that
\begin{align*}
& \left\Vert \left(\frac{\mu r}{n}\right)^{-\frac{1}{2}}S'\right\Vert _{2\rightarrow\infty}
\nonumber\\
 & =\left\Vert \left(\frac{\mu r}{n}\right)^{-\frac{1}{2}}\cdot B_{L}\overline{S}'B_{R}^*\right\Vert _{2\rightarrow\infty}
\qquad \mbox{(see \eqref{eq:aux 2})}
\nonumber \\
 & =\left\Vert \left(\frac{\mu r}{n}\right)^{-\frac{1}{2}}\cdot B_{L}\overline{S}'\right\Vert _{2\rightarrow\infty}.
 \qquad \mbox{(rotational invariance)}
\end{align*}
It follows that
\begin{align*}
& \left\Vert \left(\frac{\mu r}{n}\right)^{-\frac{1}{2}}S'\right\Vert _{2\rightarrow\infty} \nonumber\\
 & \le\left\Vert \left(\frac{\mu r}{n}\right)^{-\frac{1}{2}}\cdot U_{r}\cdot L_{11}S_{11}R_{11}\right\Vert _{2\rightarrow\infty}
\nonumber\\
&  +\left\Vert \left(\frac{\mu r}{n}\right)^{-\frac{1}{2}}\cdot U_{r}\cdot L_{11}S_{11}R_{12}\right\Vert _{2\rightarrow\infty}
 \nonumber \\
 & +\left\Vert \left(\frac{\mu r}{n}\right)^{-\frac{1}{2}}\cdot U'_{r}\cdot L_{12}^{*}S_{11}R_{11}\right\Vert _{2\rightarrow\infty}
\,\, \mbox{(see \eqref{eq:def of BL} and \eqref{eq:def of L'})}
 \nonumber \\
 & \le\left\Vert \left(\frac{\mu r}{n}\right)^{-\frac{1}{2}}U_{r}\right\Vert _{2\rightarrow\infty}\left\Vert L_{11}\right\Vert \left\Vert S_{11}\right\Vert \left\Vert R_{11}\right\Vert \nonumber\\
& \qquad  +\left\Vert \left(\frac{\mu r}{n}\right)^{-\frac{1}{2}}U_r\right\Vert _{2\rightarrow\infty}\left\Vert L_{11}\right\Vert \left\Vert S_{11}\right\Vert \left\Vert R_{12}\right\Vert
 \nonumber \\
 & \qquad+\left\Vert \left(\frac{\mu r}{n}\right)^{-\frac{1}{2}}U'_r\right\Vert _{2\rightarrow\infty}\left\Vert L_{12}\right\Vert \left\Vert S_{11}\right\Vert \left\Vert R_{11}\right\Vert,
\end{align*}
where the last inequality applies the fact that $\|AB\|_{2\rightarrow \infty} \le \|A\|_{2\rightarrow\infty} \cdot \|B\|$ for conforming matrices $A,B$. We continue by simplifying the last inequality and write that
\begin{align}
& \left\Vert \left(\frac{\mu r}{n}\right)^{-\frac{1}{2}}S'\right\Vert _{2\rightarrow\infty} \nonumber\\
 & \le \left\Vert L_{11}\right\Vert \left\Vert R_{11}\right\Vert +\left\Vert L_{11}\right\Vert \left\Vert R_{12}\right\Vert +\sqrt{2\max_{i}\frac{\breve{\mu}_{i}}{\mu_{i}}}\left\Vert L_{12}\right\Vert \left\Vert R_{11}\right\Vert
  \nonumber \\
& \le  \left\|L_{11}\r\|  \l( \l\|R_{11} \r\|+ \l\|R_{12}\r\| \r)
\nonumber\\
& \qquad  +  \sqrt{2\max_i\frac{\breve{\mu}_i}{\mu_i}}
\l\| R_{11}\r\| \l(\l\|L_{11} \r\|+\l\| L_{12}\r\| \r) \nonumber\\
 &
 \le
  2\left\|L_{11}\r\|  \cdot \l\|
  \l[
  \begin{array}{cc}
  R_{11} & R_{12}
  \end{array}
  \r]
  \r\| \nonumber\\
& \qquad   +  2\sqrt{2\max_i\frac{\breve{\mu}_i}{\mu_i}}
\l\| R_{11}\r\| \cdot
\l\| \l[
\begin{array}{cc}
L_{11} & L_{12}
\end{array}
\r]
\r\|
\nonumber\\
& \le 2 \sqrt{\lambda^2 \cos^2 u_1 + \sin^2 u_1} \cdot \sqrt{\frac{\rho^4 \cos^2 v_1+\sin^2 v_1}{\rho^2 \cos^2 v_1 + \sin^2 v_1}} \nonumber\\
& \qquad
+
2 \sqrt{2\max_i \frac{\breve{\mu}_i}{\mu_i}} \cdot   \sqrt{\rho^2 \cos^2 v_1 + \sin^2 v_1}
\nonumber\\
&\qquad  \cdot \sqrt{\frac{\lambda^4 \cos^2 u_1+\sin^2 u_1}{\lambda^2 \cos^2 u_1 + \sin^2 u_1}}
\qquad \mbox{(see Lemma \ref{lem:props of L})}
\nonumber\\
& \le 2 \Cr{5} \cdot \l(1+ \sqrt{2\max_i \frac{\breve{\mu}_i}{\mu_i}}  \r),
\qquad \mbox{(see \eqref{eq:leg 20-1})}
\label{eq:leg 22}
\end{align}
where the first inequality uses \eqref{eq:useful eqs} and \eqref{eq:sign matrix props}.
Similarly, it holds that
\begin{equation}
\left\Vert \left(\frac{\nu r}{n}\right)^{-\frac{1}{2}}\l({S}'\r)^{*}\right\Vert _{2\rightarrow\infty}\le
2 \Cr{5} \left(1+\sqrt{2\max_{j}\frac{\breve{\nu}_{j}}{\nu_{j}}}\right),
\label{eq:}
\end{equation}
so that, recalling \eqref{eq:mu inf 2 norm}, we find that
\begin{align}
\left\Vert {S}'\right\Vert _{\mu(2,\infty)} & = \left\Vert \left(\frac{\mu r}{n}\right)^{-\frac{1}{2}}{S}'\right\Vert _{2\rightarrow\infty}\vee\left\Vert \left(\frac{\nu r}{n}\right)^{-\frac{1}{2}}\l(S'\r)^{*}\right\Vert _{2\rightarrow\infty}
\nonumber \\
 & \le
2 \Cr{5} \left(1+\sqrt{2\max_{i}\frac{\breve{\mu}_{i}}{\mu_{i}}}+\sqrt{2\max_{j}\frac{\breve{\nu}_{j}}{\nu_{j}}}\right),\label{eq:leg 23}
\end{align}
which completes the proof of Lemma \ref{lem:bnd on Sp}.

\section*{Biographies}

Armin Eftekhari received his PhD from Colorado School of Mines in
2015 under the supervision of Michael Wakin. Before joining the Alan
Turing Institute as a Research Fellow in 2016, he was a Peter
O'Donnell, Jr. Postdoctoral Fellow at the University of Texas at
Austin.

Dehui Yang received the B. Eng. degree in Telecommunications Engineering from Zhejiang University of Technology, Hangzhou, China. He completed his Ph.D. degree in Electrical Engineering at the Colorado School of Mines, Golden, CO, USA, in 2018. His research interests include signal processing using low-dimensional models, machine learning, and large-scale optimization.

Michael B. Wakin (Student member 2001, Member 2007, Senior member 2013) is an Associate Professor in the Department of Electrical Engineering at the Colorado School of Mines. Dr. Wakin received a B.S. in electrical engineering and a B.A. in mathematics in 2000 (summa cum laude), an M.S. in electrical engineering in 2002, and a Ph.D. in electrical engineering in 2007, all from Rice University. He was an NSF Mathematical Sciences Postdoctoral Research Fellow at Caltech from 2006-2007, an Assistant Professor at the University of Michigan from 2007-2008, and a Ben L. Fryrear Associate Professor at Mines from 2015-2017. His research interests include sparse, geometric, and manifold-based models for signal processing and compressive sensing.

In 2007, Dr. Wakin shared the Hershel M. Rich Invention Award from Rice University for the design of a single-pixel camera based on compressive sensing. In 2008, Dr. Wakin received the DARPA Young Faculty Award for his research in compressive multi-signal processing for environments such as sensor and camera networks. In 2012, Dr. Wakin received the NSF CAREER Award for research into dimensionality reduction techniques for structured data sets. In 2014, Dr. Wakin received the Excellence in Research Award for his research as a junior faculty member at Mines. Dr. Wakin is a recipient of the Best Paper Award from the IEEE Signal Processing Society and has served as an Associate Editor for IEEE Signal Processing Letters.

\end{document}